\documentclass[fontsize=10pt,a4paper]{scrartcl}
\usepackage[a4paper, top=4cm, bottom=4cm]{geometry}
\usepackage[utf8x]{inputenc}
\usepackage[T1]{fontenc}
\usepackage{tikz}
\usetikzlibrary{matrix,arrows}
\usepackage{enumitem}
\usepackage{hyperref}
\usepackage{amsmath}
\usepackage{amsthm}
\usepackage{amssymb}
\usepackage{amsfonts}
\usepackage{mathrsfs}
\usepackage{bbm}
\usepackage{abstract}
\usepackage{graphicx}
\usepackage{placeins}
\usepackage{subfigure}
\usepackage{nicefrac}
\usepackage{booktabs}
\usepackage{bookmark}
\usepackage{multirow}
\usepackage{dsfont}
\usepackage{longtable}
\usepackage{braket}
\usepackage{tikz}
\usepackage{scalerel}
\usepackage{listings}
\usepackage{color}
\usepackage[all]{xy}
\usepackage{tensor}
\usepackage{faktor}
\usepackage{lipsum}
\usepackage{comment}

\newcommand{\sj}[6]{\begin{Bmatrix}
  #1 & #2 & #3 \\
  #4 & #5 & #6 
 \end{Bmatrix}}
 
 \newcommand{\updownarrows}{\uparrow\mathrel{\mspace{-1mu}}\downarrow}
\newcommand{\downuparrows}{\downarrow\mathrel{\mspace{-1mu}}\uparrow}
\renewcommand{\upuparrows}{\uparrow\uparrow}
\renewcommand{\downdownarrows}{\downarrow\downarrow}

\theoremstyle{definition}
\newtheorem{definition}{Definition}[section] 
 
\theoremstyle{definition}
\newtheorem{remark}[definition]{Remark}

\theoremstyle{definition}

\title{$\mathcal{N}=1$ Supergravity with LQG methods\\
and \\
quantization of the SUSY constraint}
\author{
  \large Konstantin Eder\thanks{Email: konstantin.eder@gravity.fau.de}, Hanno Sahlmann\thanks{Email: hanno.sahlmann@gravity.fau.de} \\
		\large Institute for Quantum Gravity (IQG)\\
 	\large Friedrich-Alexander-Universit\"at Erlangen-N\"urnberg (FAU)\\\\
}

\begin{document}

\maketitle

\begin{abstract}
In this paper, the classical and quantum theory of $\mathcal{N}=1$ supergravity in four spacetime dimensions will be studied in the framework of loop quantum gravity. We discuss the canonical analysis of the supergravity Holst action as first introduced by Tsuda. In this way, we also derive a compact expression of the supersymmetry constraint, which plays a crucial role in canonical supergravity theories, akin to the role of the Hamiltonian constraint in non-supersymmetric generally covariant theories. 

The resulting theory is then quantized using loop quantum gravity methods. In particular, we propose and discuss a quantization of the supersymmetry constraint and derive explicit expressions of the action of the resulting operator. This is important as it is the first step on the way of analyzing the Dirac algebra generated by supersymmetry and Hamiltonian constraint in the quantum theory and for finding physical states. We also discuss some qualitative properties of such solutions of the SUSY constraint. 
\end{abstract}

\newpage 
\section{Introduction}
The study of supergravity theories in the framework of loop quantum gravity (LQG) already has a long history. About ten years after the discovery of supergravity in 1976 by Freedman, Ferrara and van Nieuwenhuizen \cite{Freedman:1976xh}, Jacobson \cite{Jacobson:1987cj} introduced a chiral variant of the real $\mathcal{N}=1$ Poincaré supergravity action using Ashtekar's self-dual connection variables. Soon after, Fülöp \cite{Fulop:1993wi} extended this theory to anti-de Sitter supergravity including a cosmological constant where he also pointed out some interesting remnant supersymmetric structure in the resulting Poisson algebra between the Gauss and left supersymmetry (SUSY) constraint. This paved the way towards a new approach to nonperturbative supergravity in which parts of SUSY were kept manifest. In particular, this was more intensively studied by Gambini and Pullin et al. \cite{Gambini:1995db} as well as Ling and Smolin \cite{Ling:1999gn,Ling:2000ss}, where the notion of super spin networks first appered. Later it was also considered by Livine and Oeckl \cite{Livine:2003hn} in the spinfoam approach to quantum gravity.\\

Canonical supergravity with \emph{real} Asthekar-Barbero variables was first time considered by Tsuda \cite{Tsuda:1999bg} where a generalization of the chiral $\mathcal{N}=1$ supergravity action to arbitrary real Barbero-Immirzi parameters was found. In parallel, Sawaguchi \cite{Sawaguchi:2001wi} constructed the phase space in terms of real Ashtekar-Barbero variables performing a canonical transformation of the ADM phase space. However, these considerations did not include a full consistent treatment of half-densitized fermionic fields as proposed by Thiemann in \cite{Thiemann:1997rq} in order to solve the reality conditions to be satisfied by the Rarita-Schwinger field. Generalizations in the classical setting have been studied for instance in \cite{Kaul:2007gz}, where Holst actions for extended $D=4$ supergravity theories have been constructed.\\  
Finally, these considerations have been extended to higher spacetime dimensions by Bodendorfer at el. \cite{Bodendorfer:2011pb,Bodendorfer:2011pc} based on a new method discovered by the same authors in \cite{Bodendorfer:2011nv} allowing them to construct Ashekar-Barbero type variables in case of more general spacetime dimensions going beyond the limitations of the variables usually applied in LQG. This, among other things, has the advantage of being able to apply LQG methods to the maximal $\mathcal{N}=1$, $D=11$ supergravity which is thought to be the low energy limit of $M$-theory, a non-perturbative unification of all existing $D=10$ superstring theories. 
Since we are not working in higher dimensions, we use the standard Ashtekar connection, shifted by some torsion terms. These are slightly different variables for the gravitational field than \cite{Bodendorfer:2011pb,Bodendorfer:2011pc}. However, \cite{Bodendorfer:2011pb} uses half densitized variables for the Rarita-Schwinger field, and it introduces an ingenious technique for dealing with its Majorana-nature, which we will also employ.   
\\
\\
In this work, we will be mainly interested in the $\mathcal{N}=1$, $D=4$ case, in particular, in the implementation of the SUSY constraint in the quantum theory. In the chiral approach, Jacobson studied the classical Poisson algebra generated by the left and right supersymmetry constraints which maintain the right balance between fermionic and bosonic degrees of freedom. In particular, it was shown that the Poisson bracket among the SUSY constraints generates the Hamiltonian constraint which is in fact a generic feature in canonical supergravity theories. Similar results obtained in \cite{Sawaguchi:2001wi} using real Ashtekar variables supported this hypothesis showing that, on the constrained surface of gauge and diffeomorphism invariant states, the Poisson bracket between the SUSY constraints is indeed proportional to the Hamiltonian constraint.\\
This has interesting consequences implying that the SUSY constraint is superior to the Hamiltonian constraint in the sense that the solutions of the SUSY constraint immediately are solution of the latter. Hence, in case of presence local supersymmetry, the SUSY constraint plays a similar role as the Hamiltonian constraint in ordinary field theories. This is precisely what makes its study in LQG particularly interesting. However, an explicit implementation of the SUSY constraint in the quantum theory not has been considered so far in the literature. In fact, the SUSY constraint turns out to have a different structure than the Hamiltonian constraint which also requires special care for its regularization. As a result, its implementation in the quantum theory leads to an operator which has a different structure than the Hamiltonian constraint operator. It would be interesting to check by computing the commutators, in which sense these operators can be related to each other. This may also fix some of the quantization ambiguities. In fact, for a certain subclass of symmetry reduced models, we have explicitly shown in \cite{Eder:2020okh} that such a strong relationship can indeed be maintained in the quantum theory. It would be of great interest to see whether these results can be extended to the full theory.\\
\\
The structure of this paper is as follows: In section \ref{sec:majo}, we will review very briefly some important aspects about Clifford algebras and Majorana spinors. We will use this opportunity to fix our notation and conventions as well as to collect important identities used in the main text. In section \ref{sec:holst} We will subsequently discuss the canonical analysis of the Holst action of $D=4$, $\mathcal{N}=1$ supergravity as introduced in \cite{Tsuda:1999bg} filling in some details concerning half-densitized fermion fields. We will finally derive a compact expression of the supersymmetry constraint that will be used for the implementation in the quantum theory. The quantization of the Rarita-Schwinger field will be discussed in detail in section \ref{section 5.1} following the proposal of \cite{Bodendorfer:2011pb} performing an appropriate extension of the canonical phase space. In this way, we will also use this occasion to point out some interesting mathematical structure underlying the usual quantization scheme of fermion fields in LQG also discussed in more detail in \cite{Eder:2020erq} in the context of the manifestly supersymmetric approach to quantum supergravity.\\
Finally, in section \ref{section 5.2}, we will turn to the quantization of the SUSY constraint in the quantum theory. In particular, an explicit expression of the quantum SUSY constraint will be derived using a specific adapted regularization scheme. In this way, we will also find some explicit formulas for its action on spin network functions which may be of particular interest in order to find relations to the standard quantization scheme of Hamiltonian constraint. In section \ref{section 5.3}, possible solutions of the SUSY constraint will be discussed on a qualitative level showing that general solutions may indeed be supersymmetric in the sense that they need to contain both fermionic and bosonic degrees of freedom.\\ 

Unless otherwise stated, we work in signature $(-+++)$. The gravitational coupling constant is denoted by $\kappa=8 \pi G$, the Barbero-Immirzi parameter by $\beta$. Indices $I,J\ldots=0,\ldots,3$ are local Lorentz indices, $i,j,\ldots=1,2,3$ their spatial part. 4D Majorana spinor indices are denoted by $\alpha,\beta,\ldots$.

\section{Some notes on Clifford algebras and Majorana spinors}
\label{sec:majo}
In this section, we will only recall some essential aspects of Clifford algebras and Majorana spinors. Therefore, we will mainly follow the mathematical exposition in \cite{Hamilton:2017}, although our conventions are those in \cite{Freedman:2012zz}.\\
Let  $(\mathbb{R}^{s,t},\eta)$ be the inner product space where $\eta$ is a symmetric bilinear form of signature $(s,t)$, i.e., with respect to the standard basis $\{e_{I}\}$ of $\mathbb{R}^{s,t}$, $I=0,\ldots,s+t=:n$, one has
\begin{equation}
    \eta(e_I,e_I)=\left\{\begin{array}{lr}
        -1, & \text{for } I=1,\ldots,s\\
        +1, & \text{for } I=s+1,\ldots,t
        \end{array}\right.
\end{equation}
and $\eta(e_I,e_J)=0$ for $I\neq J$. The \emph{Clifford algebra} $\mathrm{Cl}(\mathbb{R}^{s,t},\eta)$ is an associative algebra over the reals with unit $1$ generated by $n$ elements $\gamma_I\in\mathrm{Cl}(\mathbb{R}^{s,t},\eta)$ satisfying
\begin{equation}
    \left\{\gamma_{I},\gamma_{J}\right\}=2\eta_{IJ}
    \label{eq:1.1}
\end{equation}
It follows that $\mathrm{Cl}(\mathbb{R}^{s,t},\eta)$ is real vector space of dimension $\mathrm{dim}\,\mathrm{Cl}(\mathbb{R}^{s,t},\eta)=2^n$ spanned by the unit $1$ together with elements of the form
\begin{equation}
    \gamma_{I_{1}I_2\cdots I_{k}}:=\gamma_{[I_1}\gamma_{I_2}\cdot\ldots\cdot\gamma_{I_k]}
    \label{eq:1.2}
\end{equation}
for $k=1,\ldots,n$, where the bracket denotes antisymmetrization.\\
The Clifford algebra has the structure of a graded algebra via the decomposition $\mathrm{Cl}(\mathbb{R}^{s,t},\eta)=\mathrm{Cl}(\mathbb{R}^{s,t},\eta)_0\oplus\mathrm{Cl}(\mathbb{R}^{s,t},\eta)_1$ where $\mathrm{Cl}(\mathbb{R}^{s,t},\eta)_i$ for $i=0$ or $1$ is the subalgebra generated by elements of the form (\ref{eq:1.2}) containing an even resp. odd number of elements $\gamma_I$. The even part $\mathrm{Cl}(\mathbb{R}^{s,t},\eta)_0$  contains a subset $\mathrm{Spin}^+(s,t)$ which turns out to have the structure of a Lie group. In particular, it follows that this Lie group defines a universal covering of the orthochronous pseudo-orthogonal group $\mathrm{SO}^+(s,t)$ together with a covering map 
\begin{equation}
    \lambda^+:\,\mathrm{Spin}^+(s,t)\rightarrow\mathrm{SO}^+(s,t)
\end{equation}
In case of Minkowski spacetime in $D=4$, $\mathrm{Spin}^+(1,3)$ is isomorphic to $\mathrm{SL}(2,\mathbb{C})$. The Lie algebra $\mathfrak{spin}^+(s,t)$ of $\mathrm{Spin}^+(s,t)$ is generated by the elements
\begin{equation}
    M_{IJ}:=\frac{1}{2}\gamma_{IJ}
    \label{eq:1.3}
\end{equation}
In this article, we are mainly concerned about four spacetime dimensions. In fact, most of the computations do not require a specific representation of the Clifford algebra. However, in section \ref{section 5.1}, it will be worthwhile to choose a representation in which the gamma matrices are explicitly real. An explicit realization for such a type of representation for arbitrary even spacetime dimensions can be found for instance in \cite{Freedman:2012zz}. \\
For section \ref{section 5.2}, it will prove particularly beneficial to work instead in a \emph{chiral representation} or \emph{Weyl representation}. This will also play a prominent role in the context of self-dual variables as discussed in \cite{Eder:2020erq,Eder:2020okh}. In this representation, the gamma matrices take the form
\begin{equation}
    \gamma_{I}=\begin{pmatrix}
0 & \sigma_{I}\\
\bar{\sigma}_{I} & 0
\end{pmatrix}\quad\text{and}\quad\gamma_{*}=\begin{pmatrix}
\mathds{1} & 0\\
0&-\mathds{1}
\end{pmatrix}
\label{eq:1.4}
\end{equation}
with $\gamma_*:=i\gamma_0\gamma_1\gamma_2\gamma_3$ the highest rank Clifford algebra element also commonly denoted by $\gamma_5$ and $\sigma_{I}:=(-\mathds{1},\sigma_i)$ and $\bar{\sigma}_{I}:=(\mathds{1},\sigma_i)$ where $\sigma_i$, $i=1,\ldots,3$ denote the ordinary Pauli matrices satisfying the product relation
\begin{equation}
    \sigma_i\sigma_j=\delta_{ij}\mathds{1}+i\tensor{\epsilon}{_{ij}^k}\sigma_k
    \label{eq:1.5}
\end{equation}
The generators (\ref{eq:1.3}) of $\mathfrak{spin}^{+}(1,3)$ then take the form
\begin{equation}
    M_{IJ}=\frac{1}{2}\gamma_{IJ}=\frac{1}{4}\begin{pmatrix}
\sigma_{I}\bar{\sigma}_{J}-\sigma_{J}\bar{\sigma}_{I} & 0\\
0 & \bar{\sigma}_{I}\sigma_{J}-\bar{\sigma}_{J}\sigma_{I}
\end{pmatrix}
\label{eq:1.6}
\end{equation}
Moreover, they satisfy well-known Lie algebra relations
\begin{equation}
[M_{IJ},M_{KL}]=\eta_{JK}M_{IL}-\eta_{IK}M_{JL}-\eta_{JL}M_{IK}+\eta_{IL}M_{JK}
\label{eq:1.7}
\end{equation}
A useful formula which interrelates elements of the form (\ref{eq:1.2}) with different degree is given by the following
\begin{equation}
    \tensor{\gamma}{^{I_1 I_2\ldots I_r}}\gamma_{*}=\frac{i}{(4-r)!}\tensor{\epsilon}{^{I_r I_{r-1}\ldots I_1 J_1\ldots J_{4-r}}}\tensor{\gamma}{_{J_1\ldots J_{4-r}}}
    \label{eq:1.8b}
\end{equation}
for $0\leq r\leq 4$, which will often be needed in the main text. Here, $\tensor{\epsilon}{^{IJKL}}=-\tensor{\epsilon}{_{IJKL}}$ denotes the completely antisymmetric symbol in $D=4$ with the convention $\tensor{\epsilon}{^{0123}}=1$.\\
\\
Finally, let us briefly say something about Majorana representations and Majorana spinors. Let $\kappa:\,\mathrm{Spin}^+(s,t)\rightarrow\mathrm{GL}(\Delta_n)$ be the complex Dirac representation (for a detailed account on complex Dirac representations in arbitrary spactime dimensions see for instance \cite{Hamilton:2017} and references therein). A Majorana representation is then defined as an induced represenation on a real subspace of the complex vector space $\Delta_n$. More precisely,
\begin{definition} The complex spinor representation $\kappa$ is called \emph{Majorana} if it admits a \emph{real structure} $\sigma$, i.e. a complex antilinear map $\sigma:\,\Delta_n\rightarrow\Delta_n$ such that $\sigma$ is $\mathrm{Spin}^+(s,t)$-equivariant
\begin{equation}
    \sigma\circ\kappa(g)=\kappa(g)\circ\sigma
\end{equation}
$\forall g\in\mathrm{Spin}^+(s,t)$ and $\sigma$ is involutive $\sigma^2=\mathrm{id}_{\Delta_n}$.\\
The real structure defines a proper real $\mathrm{Spin}^+(s,t)$-invariant subspace
\begin{equation}
    \Delta_{\mathbb{R}}:=\{\psi\in\Delta_n|\,\sigma(\psi)=\psi\}
\end{equation}
of $\Delta_n$ of real dimension $\mathrm{dim}_{\mathbb{R}}\,\Delta_{\mathbb{R}}=\mathrm{dim}_{\mathbb{C}}\,\Delta_n$. Moreover, due to $\mathrm{Spin}^+(s,t)$-equivariance, it induces a real sub representation
\begin{equation}
    \kappa_{\mathbb{R}}:\,\mathrm{Spin}^+(s,t)\rightarrow\mathrm{GL}(\Delta_{\mathbb{R}})
\end{equation}
of the complex Dirac representation of $\mathrm{Spin}^+(s,t)$ on $\Delta_{\mathbb{R}}$ called the \emph{Majorana representation} of $\mathrm{Spin}^+(s,t)$.
\end{definition}
Choosing a basis of $\Delta_n$, one can write the condition $\psi=\sigma(\psi)$ equivalently in the form
\begin{equation}
    \psi^*=B\psi
    \label{eq:1.8}
\end{equation} 
with $B$ a complex matrix satisfying $B^*B=\mathds{1}$, which is also often referred to as the \emph{Majorana condition} in the literature. This matrix is related to the charge conjugation matrix $C$ via $B=it_0C\gamma^0$ where $t_0\in\{\pm 1\}$ depends on the signature and the dimension of the spacetime.\\
In case of Minkowski spacetime in four spactime dimenions, one usually sets $t_0=1$ in which case the charge conjugation matrix is given by $C=i\gamma^3\gamma^1$ and therefore, in the chiral representation,
\begin{equation}
    B=\gamma^0\gamma^1\gamma^3=\begin{pmatrix}
0 &	-i\sigma_2\\
i\sigma_2 & 0
\end{pmatrix}
\end{equation}
For a Dirac fermion $\psi=(\chi,\phi)^T$, the Majorana condition (\ref{eq:1.8}) then reads
\begin{equation}
\psi^*=B\psi\quad\Leftrightarrow\quad\chi=-i\sigma_2\phi^*\text{ or }\phi=i\sigma_2\chi^*
\end{equation}

\section{Holst action for Supergravity in $D=4$ and its $3+1$ decomposition}
\label{sec:holst}
Supergravity in $D=4$ with $\mathcal{N}=1$ fermionic generator in the SUSY algebra can be described, in case of a vanishing cosmological constant, as a super Cartan geometry modeled on a super Klein geometry $(\mathrm{ISO}(\mathbb{R}^{1,3|4}),\mathrm{Spin}^+(1,3))$ with $\mathrm{ISO}(\mathbb{R}^{1,3|4})$ the super Poincaré group with super Lie algebra
\begin{equation}
    \mathfrak{iso}(\mathbb{R}^{1,3|4})=\mathbb{R}^{1,3}\rtimes\mathfrak{spin}^+(1,3)\oplus\Delta_{\mathbb{R}}
\end{equation}
The super Cartan connection $\mathcal{A}=e^I P_I+\frac{1}{2}\omega^{IJ}M_{IJ}+\psi^{\alpha}Q_{\alpha}$ splits into the spin connection $\omega\in\Omega^1(P,\mathfrak{spin}^+(1,3))$, the soldering form $e\in\Omega^1_{hor}(P,\mathbb{R}^{1,3})$ as well as the Rarita-Schwinger field $\psi\in\Omega^1_{hor}(P,\Delta_{\mathbb{R}})$ with $P$ the underlying spin structure\footnote{The spin structure arises as the body of the principal super fiber bundle corresponding to the super Cartan geometry.}.\\
\\
For the purpose of describing supergravity in the context of LQG, we take the Holst action of $\mathcal{N}=1$ supergravity as stated in \cite{Tsuda:1999bg} which, adapted to our conventions and written in a coordinate free from, reads\footnote{for convenience, $\kappa$ will be absorbed in the Rarita-Schwinger field.}
\begin{align}
S(e,\omega,\psi)=\frac{1}{4\kappa}\int_{M}\Sigma^{IJ}\wedge(P\circ F(\omega))^{KL}\epsilon_{IJKL}+
2\kappa e^I\wedge\bar{\psi}\wedge\gamma_I\frac{\mathds{1}+i\beta\gamma_{*}}{\beta}D^{(\omega)}\psi
\label{eq:2.1}
\end{align}
where $\kappa=8\pi G$ and $D^{(\omega)}\psi:=\mathrm{d}\psi+\kappa_{\mathbb{R}*}(\omega)\wedge\psi$ denotes the exterior covariant derivative of $\psi$ and
\begin{equation}
    (P\circ F(\omega))^{IJ}:=\tensor{P}{^{IJ}_{KL}}F(\omega)^{KL}\quad\text{with}\quad\tensor{P}{^{IJ}_{KL}}:=\delta^I_{[K}\delta^J_{L]}-\frac{1}{2\beta}\tensor{\epsilon}{^{IJ}_{KL}}
\end{equation}
with $\beta$ the \emph{Barbero Immirzi parameter}. Moreover, $F(\omega):=\mathrm{d}\omega+\omega\wedge\omega$ is the associated curvature of $\omega$ and
\begin{equation}
   \Sigma:=e\wedge e\in\Omega^2_{hor}(P,\mathfrak{spin}^+(1,3))
   \label{eq:2.2}
\end{equation}
One needs to ensure that the equations of motion resulting from (\ref{eq:2.1}) are independent on the choice of the Barbero-Immirzi parameter and, at second order, are equivalent to those of ordinary $\mathcal{N}=1$ supergravity.\\
Therefore, one has to the vary (\ref{eq:2.1}) with respect to the spin connection $\omega$. As this is rarely done explicitly in the literature, let us perform the variation for a general matter contribution. That is, we consider an action $S$ of the form $S=S_H+S_{H-matter}$, where $S_H$ is the standard Holst action of pure gravity and $S_{H-matter}$ is some Holst-like modification of the matter contribution such that the resulting equations of motion remain unchanged.\\
First, let us consider the Holst term
\begin{equation}
S_H=\frac{1}{4\kappa}\int_{M}\Sigma^{IJ}\wedge(P\circ F(\omega))^{KL}\epsilon_{IJKL}=:\frac{1}{4\kappa}\int_{M}\braket{\Sigma\wedge P\circ F(\omega)}
\label{eq:2.3}
\end{equation}
where $\braket{\,\cdot\,\wedge\,\cdot\,}:\,\Omega^2(P,\mathfrak{spin}^+(1,3))\times\Omega^2(P,\mathfrak{spin}^+(1,3))\rightarrow\mathbb{R}$ is the extension of the Adjoint invariant bilinear form on $\mathfrak{spin}^+(1,3)$ to $\mathfrak{spin}^+(1,3)$-valued forms on $P$. Let us then consider a variation of connection $\omega+\delta\omega$. The variation of $F(\omega)$ is then given by $\delta F(\omega)=D^{(\omega)}\delta\omega$. Since $P\circ D^{(\omega)}\delta\omega=D^{(\omega)}(P\circ\delta\omega)$ and $\braket{\Sigma\wedge D^{(\omega)}(P\circ\delta\omega)}=-\braket{D^{(\omega)}\Sigma\wedge P\circ\delta\omega}$ up to a total derivative \cite{Baum}, this yields
\begin{equation}
\delta S_H=\frac{1}{4\kappa}\int_{M}\braket{D^{(\omega)}\Sigma\wedge P\circ\delta\omega}=-\frac{1}{4\kappa}\int_{M}D^{(\omega)}\Sigma^{IJ}\wedge(P\circ \delta\omega)^{KL}\epsilon_{IJKL}
\label{eq:2.4}
\end{equation}
Using (\ref{eq:1.7}), it follows
\begin{align}
    D^{(\omega)}\Sigma^{IJ}&=\mathrm{d}(e^I\wedge e^J)+\frac{1}{4}\omega^{IJ}\wedge\Sigma^{KL}\otimes[M_{IJ},M_{KL}]^{IJ}\nonumber\\
    &=\mathrm{d}e^I\wedge e^J-e^I\wedge\mathrm{d}e^J+\tensor{\omega}{^I_K}\wedge\Sigma^{KJ}+\tensor{\omega}{^J_K}\wedge\tensor{\Sigma}{^{IK}}\nonumber\\
    &=T^I\wedge e^J-e^I\wedge T^{J}
    \label{eq:2.5}
\end{align}
with $T^I=\mathrm{d}e^I+\tensor{\omega}{^I_K}\wedge e^K$ the associated torsion 2-form. Inserting (\ref{eq:2.5}) into (\ref{eq:2.4}), this yields  
\begin{align}
\delta S_H&=-\frac{1}{2\kappa}\int_{M}T^I\wedge e^J\wedge(P\circ \delta\omega)^{KL}\epsilon_{IJKL}\nonumber\\
&=-\frac{1}{4\kappa}\int_{M}\epsilon^{MNJO}\epsilon_{IJKL}T^I_{\mu\nu}e^{\mu}_Me ^{\nu}_N(P\circ \delta\omega_{\rho})^{KL}e^{\rho}_O\,\mathrm{d}\mathrm{vol}_M\nonumber\\
&=-\frac{1}{4\kappa}\int_{M}3!\delta^{[M}_I\delta^N_K\delta^{O]}_LT^I_{\mu\nu}e^{\mu}_Me ^{\nu}_N(P\circ \delta\omega_{\rho})^{KL}e^{\rho}_O\,\mathrm{d}\mathrm{vol}_M\nonumber\\
&=-\frac{1}{2\kappa}\int_{M}\tensor{P}{^{KL}_{IJ}}(2T^{\rho}_{\rho\mu}e^{\mu}_Ke^{\nu}_L+T^{\nu}_{\mu\rho}e^{\mu}_{K}e^{\rho}_{L})\delta\omega_{\nu}^{IJ}\,\mathrm{d}\mathrm{vol}_M
 \label{eq:2.6}
\end{align}
Hence, including the matter contribution, we find for the variation of the total action
\begin{align}
    \delta S_{\text{$H$-$SG$}}=\int_{M}{-\frac{1}{2\kappa}\tensor{P}{^{KL}_{IJ}}(2T^{\rho}_{\rho\mu}e^{\mu}_Ke^{\nu}_L+T^{\nu}_{\mu\rho}e^{\mu}_{K}e^{\rho}_{L})\delta\omega_{\nu}^{IJ}+\frac{\delta S_{H-matter}}{\delta\omega_{\nu}^{IJ}}\delta\omega_{\nu}^{IJ}}\,\mathrm{d}\mathrm{vol}_M
     \label{eq:2.7}
\end{align}
which vanishes if and only if
\begin{align}
    \tensor{P}{^{KL}_{IJ}}(2T_{\rho K}^{\rho}e_L^{\nu}+T_{KL}^{\nu})=2\kappa e^{-1}\frac{\delta S_{H-matter}}{\delta\omega_{\nu}^{IJ}}
     \label{eq:2.8}
\end{align}
Applying the inverse
\begin{align}
    \tensor{(P^{-1})}{_{IJ}^{KL}}=\frac{\beta^2}{1+\beta^2}\left(\delta^K_{[I}\delta^L_{J]}+\frac{1}{2\beta}\tensor{\epsilon}{_{IJ}^{KL}}\right)
     \label{eq:2.9}
\end{align} 
on both sides of (\ref{eq:2.8}), this gives
\begin{align}
    2T_{\rho I}^{\rho}e_J^{\nu}+T_{IJ}^{\nu}=2\kappa e^{-1}\tensor{(P^{-1})}{_{IJ}^{KL}}\frac{\delta S_{H-matter}}{\delta\omega_{\nu}^{KL}}
\end{align}
This is the most general formula for the equations of motion of the spin connection for arbitrary matter contributions resulting from the variation of the Holst action. In case of $\mathcal{N}=1$ supergravity, we have
\begin{align}
    \frac{\delta S_{H-matter}}{\delta\omega_{\nu}^{KL}}=-\frac{1}{4}\tensor{\epsilon}{^{\mu\nu\rho\sigma}}\bar{\psi}_{\mu}\gamma_{\sigma}\frac{\mathds{1}+i\beta\gamma_{*}}{2\beta}\gamma_{KL}\psi_{\rho}
\end{align}
so that
\begin{align}
    \tensor{(P^{-1})}{_{IJ}^{KL}}\frac{\delta S_{H-matter}}{\delta\omega_{\nu}^{KL}}&=\frac{\beta^2}{4(1+\beta^2)}\tensor{\epsilon}{^{\mu\nu\rho\sigma}}\bar{\psi}_{\mu}\gamma_{\sigma}\frac{\mathds{1}+i\beta\gamma_{*}}{2\beta}\left(\gamma_{IJ}+\frac{1}{2\beta}\tensor{\epsilon}{_{IJ}^{KL}}\gamma_{KL}\right)\psi_{\rho}
\end{align}
Since $\tensor{\epsilon}{_{IJ}^{KL}}\gamma_{KL}=2i\gamma_{IJ}\gamma_{*}$ by (\ref{eq:1.8b}), this implies
\begin{align}
    \tensor{(P^{-1})}{_{IJ}^{KL}}\frac{\delta S_{H-matter}}{\delta\omega_{\nu}^{KL}}&=\frac{\beta^2}{4(1+\beta^2)}\tensor{\epsilon}{^{\mu\nu\rho\sigma}}\bar{\psi}_{\mu}\gamma_{\sigma}\frac{\mathds{1}+i\beta\gamma_{*}}{2\beta}\left(\gamma_{IJ}+\frac{i}{\beta}\gamma_{IJ}\gamma_{*}\right)\psi_{\rho}\nonumber\\
    &=\frac{\beta^2}{4(1+\beta^2)}i\tensor{\epsilon}{^{\mu\nu\rho\sigma}}\bar{\psi}_{\mu}\gamma_{\sigma}\gamma_{IJ}\gamma_{*}\frac{\mathds{1}+i\beta\gamma_{*}}{2\beta}\frac{\mathds{1}-i\beta\gamma_{*}}{\beta}\psi_{\rho}\nonumber\\
    &=\frac{i}{8}\tensor{\epsilon}{^{\mu\nu\rho\sigma}}\bar{\psi}_{\mu}\gamma_{\sigma}\gamma_{IJ}\gamma_{*}\psi_{\rho}
\end{align}
Finally, using $\epsilon^{\mu\nu\rho\sigma}\gamma_{\sigma}=ie\gamma^{\mu\nu\rho}\gamma_{*}$, we find
\begin{align}
     2\tensor{T}{_{\rho I}^{\rho}}e_J^{\nu}+\tensor{T}{_{IJ}^{\nu}}=\frac{\kappa}{4}\bar{\psi}_{\mu}\gamma^{\mu\nu\rho}\gamma_{IJ}\psi_{\rho}
\end{align}
which are exactly the equations of motion of $\omega$ of $\mathcal{N}=1$ supergravity, in particular, completely independent of the Barbero-Immirzi parameter. These can be equivalently be written in the form \cite{Freedman:2012zz}
\begin{equation}
    T^{\rho}_{\mu\nu}=\frac{\kappa}{2}\bar{\psi}_{\mu}\gamma^{\rho}\psi_{\nu}
    \label{eq:2.10}
\end{equation}
In view of the decomposition of the action (\ref{eq:2.1}), let us rewrite in a coordinate dependent form which gives
\begin{equation}
S_{\text{$H$-$SG$}}=\int_{M}{\mathrm{d}^4x\,\frac{e}{2\kappa}e_I^{\mu}e_J^{\nu}\left(\tensor*{F(\omega)}{*_{\mu\nu}^{IJ}}-\frac{1}{2\beta}\tensor{\epsilon}{^{IJ}_{KL}}\tensor*{F(\omega)}{*_{\mu\nu}^{KL}}\right)+\tensor{\epsilon}{^{\mu\nu\rho\sigma}}\bar{\psi}_{\mu}\gamma_{\sigma}\frac{\mathds{1}+i\beta\gamma_{*}}{2\beta}D^{(\omega)}_{\nu}\psi_{\rho}}
\label{eq:2.11}
\end{equation}
As shown above, variation of (\ref{eq:2.11}) yields the same equation of motion as the standard action of $\mathcal{N}=1$ supergravity. It was then shown explicitly in \cite{Kaul:2007gz}, inserting the unique solution of (\ref{eq:2.10}) into (\ref{eq:2.11}), that the terms proportional to $\beta^{-1}$ together become purely topological. Hence, the Holst action coincides with the ordinary one provided $\omega$ satisfies its field equations.\\
\\
The 3+1-split of the action (\ref{eq:2.11}) follows the standard procedure. Since $M$ is supposed to be globally hyperbolic, it is diffeomorphic to a foliation $\mathbb{R}\times\Sigma$, where $\Sigma$ is a spacelike Cauchy surface. Let $\phi: \mathbb{R}\times\Sigma\rightarrow M$ denote such a diffeomorphism. Then, for a specific time $t\in\mathbb{R}$, we define the time slice $\Sigma_t$ via $\Sigma_t=\phi_t(\Sigma)$, where $\phi_t:=\phi(t,\cdot)$ describing the evolution of $\Sigma$ in $M$. Furthermore, the flow of the time slices induces a global timelike vector field $\partial_t$ which, on smooth functions $f\in C^{\infty}(M)$, acts via
\begin{equation}
\partial_t (f)=\frac{\mathrm{d}}{\mathrm{d}t}(f\circ\phi_t)
\label{eq:3.0.1}
\end{equation}
We choose a unit normal vector field $n$ which is normal to the time slices such that there exists a lapse function $N$ shift vector field $\vec{N}$  with $\vec{N}$ tangential to the foliation, such that
\begin{equation}
\partial_t=Nn+\vec{N}
\label{eq:3.0.1.1}
\end{equation}
As the canonical analysis of the purely bosonic term in (\ref{eq:2.11}) is very well-known, let us only comment on some mains steps. The decomposition of the curvature tensor w.r.t. to the unit normal (co)vector field yields  
\begin{equation}
\frac{e}{2}e_I^{\mu}e_J^{\nu}\tensor{P}{^{IJ}_{KL}}\tensor*{F(\omega)}{*_{\mu\nu}^{KL}}=\frac{e}{2}e_i^{a}e_j^{b}\tensor{P}{^{ij}_{KL}}\tensor*{F}{*_{ab}^{KL}}+ee_I^{\mu}e_J^{\nu}n^{\rho}\tensor{P}{^{IJ}_{KL}}\tensor*{F}{*_{\rho[\mu}^{KL}}n_{\nu]}
\label{eq:2.12}
\end{equation}
with $\tensor*{F}{*_{\mu\nu}^{KL}}=2\partial_{[\mu}\omega_{\nu]}^{KL}+2\tensor*{\omega}{*^K_{[\mu}_{|M|}}\tensor*{\omega}{*_{\nu]}^{ML}}$. Using $n^{\rho}\partial_{[\rho}\omega_{a]}=\frac{1}{2N}\left(L_{\partial_t}\omega_a-2N^b\partial_{[b}\omega_{a]}-\partial_a\omega_t\right)$, where $L_{\partial_t}\omega_a$, $a=1,\ldots,3$, denotes the Lie derivative of a spatial component of $\omega$ w.r.t. the global timelike vector field $\partial_t$, the last term in (\ref{eq:2.12}) becomes
\begin{align}
ee_I^{\mu}e_J^{\nu}n^{\rho}\tensor{P}{^{IJ}_{KL}}\tensor*{F}{*_{\rho[\mu}^{KL}}n_{\nu]}=&-N\sqrt{q}n^{\rho}e_i^{\mu}\left(\tensor*{F}{*_{\rho\mu}^{i0}}-\frac{1}{2\beta}\tensor{\epsilon}{^{i0}_{kl}}\tensor*{F}{*_{\rho\mu}^{kl}}\right)=N\sqrt{q}e_i^{a}\tensor{P}{^{0i}_{KL}}n^{\rho}\tensor*{F}{*_{\rho a}^{KL}}\nonumber\\
=&N\sqrt{q}e_i^{a}\tensor{P}{^{0i}_{KL}}\left(2n^{\rho}\partial_{[\rho}\omega_{a]}^{KL}+2n^{\rho}\tensor*{\omega}{*^K_{[\rho}_{|M|}}\tensor*{\omega}{*_{a]}^{ML}}\right)\nonumber\\
=&\frac{1}{\beta}\sqrt{q}e_i^{a}L_{\partial_t}\left(\beta\tensor*{\omega}{*^{0i}_{a}}-\frac{1}{2}\tensor{\epsilon}{^i_{kl}}\tensor*{\omega}{*^{kl}_a}\right)-\sqrt{q}e_i^{a}\tensor{P}{^{0i}_{KL}}\partial_a\tensor*{\omega}{*^{KL}_t}-2N^b\sqrt{q}e_i^{a}\tensor{P}{^{0i}_{KL}}\partial_{[b}\tensor{\omega}{*^{KL}_{a]}}\nonumber\\
&+2\sqrt{q}e_i^{a}\tensor{P}{^{0i}_{KL}}n^{\rho}\tensor*{\omega}{*^K_{[\rho}_{|M|}}\tensor*{\omega}{*_{a]}^{ML}}\nonumber\\
=&\frac{1}{\beta}E_i^{a}L_{\partial_t}A_a^i-\frac{1}{\beta}E_i^{a}\partial_a A_t^i+2E_i^{a}\tensor{P}{^{0i}_{KL}}\tensor*{\omega}{*^K_{t}_{M}}\tensor*{\omega}{*_{a}^{ML}}\nonumber\\
&-N^bE_i^{a}\tensor{P}{^{0i}_{KL}}\left(2\partial_{[b}\tensor{\omega}{*^{KL}_{a]}}+2\tensor*{\omega}{*^K_{[b}_{|M|}}\tensor*{\omega}{*_{a]}^{ML}}\right)
\label{eq:2.13}
\end{align}
where
\begin{equation}
A_a^i=\Gamma_a^i+\beta K_a^i\quad\text{and}\quad E^a_i=\sqrt{q}e_i^a
\label{eq:2.14}
\end{equation}
is the usual Ashtekar connection and dual electric field, respectively, where we set $\Gamma_a^i:=-\frac{1}{2}\tensor{\epsilon}{^i_{kl}}\omega^{kl}_a$ and  $K_a^i:=\tensor{\omega}{_a^{0i}}$ for the 3D spin connection on $\Sigma$ and extrinsic curvature, respectively. These satisfy the non-vanishing Poisson brackets
\begin{equation}
\left\{A_a^i(x),E_j^b(y)\right\}=\kappa\beta\delta^i_j\delta^b_a\delta^{(3)}(x,y)
\label{eq:2.15}
\end{equation}
Furthermore, in (\ref{eq:2.13}), we introduced the Lagrange multiplier $A_t^i:=-\frac{1}{2}\tensor{\epsilon}{^i_{kl}}\omega^{kl}_t+\beta\tensor{\omega}{_t^{0i}}=:\Gamma^i_t+\beta K_t^i$. Since
\begin{align}
2\tensor{P}{^{0i}_{KL}}=\frac{1}{\beta}A_t^m\tensor{\epsilon}{_{mn}^i}A_a^n-\frac{1+\beta^2}{\beta}K_t^m\tensor{\epsilon}{_{mn}^i}K_a^n
\end{align}
the two mid terms in (\ref{eq:2.13}) can be combined to give, after integration by parts and dropping a boundary term, 
\begin{align}
\frac{1}{\beta}A_t^i\partial_aE_i^{a}+2E^a_i\tensor{P}{^{0i}_{KL}}\tensor*{\omega}{*^K_{t}_{M}}\tensor*{\omega}{*_{a}^{ML}}&=A_t^i\frac{1}{\beta}\left(\partial_aE_i^{a}+\tensor{\epsilon}{_{ik}^l}A_a^kE^a_l\right)-\frac{1+\beta^2}{\beta}K_t^m\tensor{\epsilon}{_{mn}^i}K_a^nE^a_i\nonumber\\
&=A_t^i\frac{1}{\beta}D_a^{(A)}E_i^{a}-\frac{1+\beta^2}{\beta}K_t^m\tensor{\epsilon}{_{mn}^i}K_a^n E^a_i
\label{eq:2.16}
\end{align}
For the last term in (\ref{eq:2.13}) proportional to the shift vector field, it follows
\begin{align}
&-N^aE_i^{b}\tensor{P}{^{0i}_{KL}}\left(2\partial_{[a}\tensor{\omega}{*^{KL}_{b]}}+2\tensor*{\omega}{*^K_{[a}_{|M|}}\tensor*{\omega}{*_{b]}^{ML}}\right)\nonumber\\
=&N^a\frac{1}{\beta}E_i^{b}\left(F(A)^i_{ab}+(1+\beta^2)\tensor{\epsilon}{^i_{kl}}K^k_aK^l_b\right)
\label{eq:2.17}
\end{align}
with $F(A)^i=\mathrm{d}A^i+\frac{1}{2}\tensor{\epsilon}{^i_{jk}}A^j\wedge A^k$ the curvature of the Ashtekar-Barbero connection. Finally. let us comment on the first term appearing in the decomposition (\ref{eq:2.12}). Since $e=N\sqrt{q}$, this can be written in the form
\begin{align}
    \frac{e}{2}e_i^{a}e_j^{b}\tensor{P}{^{ij}_{KL}}\tensor*{{F(\omega)}}{*_{ab}^{KL}}&=\frac{N\sqrt{q}}{2}e_i^{a}e_j^{b}(\tensor*{{F(\omega)}}{*_{ab}^{ij}}+\frac{1}{\beta}\tensor{\epsilon}{^{ij}_k}\tensor*{{F(\omega)}}{*_{ab}^{0k}})\nonumber\\
    &=\frac{NE^a_iE^b_j}{2\sqrt{q}}(\tensor*{{F(\Gamma)}}{*_{ab}^{ij}}+2\tensor*{\omega}{*^{0i}_{[a}}\tensor*{\omega}{*_{b]}^{0j}}+\frac{1}{\beta}\tensor{\epsilon}{^{ij}_k}\tensor*{{F(\omega)}}{*_{ab}^{0k}})
\label{eq:2.18}
\end{align}
with $F(\Gamma)$ the curvature of the 3D spin connection $\Gamma$. Using 
\begin{align}
  \tensor*{{F(\Gamma)}}{*^i_{ab}}&=\tensor*{{F(A)}}{*^i_{ab}}-2\beta D^{(\Gamma)}_{[a}K^i_{b]}-\beta^2\tensor{\epsilon}{^i_{jk}}K^j_a K^k_b
  \label{eq:2.19}
\end{align}
it follows that (\ref{eq:2.18}) can be written in the form 
\begin{align}
    \frac{e}{2}e_i^{a}e_j^{b}\tensor{P}{^{ij}_{KL}}\tensor*{{F(\omega)}}{*_{ab}^{KL}}&=-\frac{NE^a_iE^b_j}{2\sqrt{q}}\tensor{\epsilon}{^{ij}_{k}}\left(\tensor*{{F(A)}}{*^k_{ab}}-(1+\beta^2)\tensor{\epsilon}{^k_{mn}}K^m_a K^n_b-\frac{2(1+\beta^2)}{\beta}D^{(\Gamma)}_{[a}K^k_{b]}\right)
    \label{eq:2.20}
\end{align}
Next, let us decompose the fermionic part of the supergravity action (\ref{eq:2.11}). Since $e_t^0=-n^\flat(\partial_t)=N$ and $e_t^i=N^ae^i_a$, we find
\begin{align}
\tensor{\epsilon}{^{\mu\nu\rho\sigma}}\bar{\psi}_{\mu}\gamma_{\sigma}\frac{\mathds{1}+i\beta\gamma_{*}}{2\beta}D^{(\omega)}_{\nu}\psi_{\rho}=&\tensor{\epsilon}{^{abc}}\bar{\psi}_{t}\gamma_{a}\frac{\mathds{1}+i\beta\gamma_{*}}{2\beta}D^{(\omega)}_{b}\psi_{c}\nonumber\\
&-N\tensor{\epsilon}{^{abc}}\bar{\psi}_{a}\gamma_{0}\frac{\mathds{1}+i\beta\gamma_{*}}{2\beta}D^{(\omega)}_{b}\psi_{c}\nonumber\\
&-N^d\tensor{\epsilon}{^{abc}}\bar{\psi}_{a}\gamma_{d}\frac{\mathds{1}+i\beta\gamma_{*}}{2\beta}D^{(\omega)}_{b}\psi_{c}\nonumber\\
&+\tensor{\epsilon}{^{abc}}\bar{\psi}_{a}\gamma_{b}\frac{\mathds{1}+i\beta\gamma_{*}}{2\beta}\left(L_{\partial_t}\psi_c+\frac{1}{4}\tensor*{\omega}{*^{IJ}_{t}}\gamma_{IJ}\psi_c\right)\nonumber\\
&-\tensor{\epsilon}{^{abc}}\bar{\psi}_{a}\gamma_{b}\frac{\mathds{1}+i\beta\gamma_{*}}{2\beta}\left(\partial_c\psi_t+\frac{1}{4}\tensor*{\omega}{*^{IJ}_{c}}\gamma_{IJ}\psi_t\right)
\label{eq:2.21}
\end{align}
Hence, taking the left-derivative of kinematical term apearing in (\ref{eq:2.21}) with respect to $\psi_t$ and noticing that fermionic fields are anticommuting, it follows that the momentum conjugate to $\psi_a$ is given by,
\begin{align}
\pi^a&=-\tensor{\epsilon}{^{abc}}\bar{\psi}_{b}\gamma_{c}\frac{\mathds{1}+i\beta\gamma_{*}}{2\beta}=\frac{i}{2}\tensor{\epsilon}{^{abc}}\bar{\psi}_{b}\gamma_{*}\gamma_{c}\frac{\gamma_{*}+i\beta}{i\beta}
\label{eq:2.22}
\end{align}
These satisfy the non-vanishing Poisson brackets
\begin{align}
\left\{\psi_{a}^{\alpha}(x),\pi^b_{\beta}(y)\right\}=-\delta^a_b\delta^{\alpha}_{\beta}\delta^{(3)}(x,y)
\label{eq:2.23}
\end{align}
In particular, according to (\ref{eq:2.22}), the canonically conjugate momentum $\pi^a$ is related to $\psi_a$ via the reality condition
\begin{equation}
    \Omega^a:=\pi^a-\frac{i}{2}\tensor{\epsilon}{^{abc}}\bar{\psi}_{b}\gamma_{*}\gamma_{c}\mathcal{P}^+_{\beta}=0
    \label{eq:2.100}
\end{equation}
where we set
\begin{align}
\mathcal{P}^{\pm}_{\beta}:=\frac{i\beta\pm\gamma_{*}}{i\beta}
\end{align}
If we consider the last term in (\ref{eq:2.21}), it again follows after integration by parts and dropping a boundary term
\begin{align}
-\tensor{\epsilon}{^{abc}}\bar{\psi}_{a}\gamma_{b}\frac{\mathds{1}+i\beta\gamma_{*}}{2\beta}\left(\partial_c\psi_t+\frac{1}{4}\tensor*{\omega}{*^{IJ}_{c}}\gamma_{IJ}\psi_t\right)=&\tensor{\epsilon}{^{abc}}\partial_c\bar{\psi}_{t}\frac{\mathds{1}+i\beta\gamma_{*}}{2\beta}\gamma_{b}\psi_a\nonumber\\
&-\frac{1}{4}\tensor{\epsilon}{^{abc}}\bar{\psi}_{t}\tensor*{\omega}{*^{IJ}_{c}}\gamma_{IJ}\frac{\mathds{1}+i\beta\gamma_{*}}{2\beta}\gamma_b\psi_a\nonumber\\
=&\bar{\psi}_{t}\frac{\mathds{1}+i\beta\gamma_{*}}{2\beta}\partial_a\left(\tensor{\epsilon}{^{abc}}\gamma_{b}\psi_c\right)\nonumber\\
&+\bar{\psi}_{t}\frac{\mathds{1}+i\beta\gamma_{*}}{2\beta}\frac{1}{4}\tensor{\epsilon}{^{abc}}\tensor*{\omega}{*^{IJ}_{a}}\gamma_{IJ}\gamma_b\psi_c\nonumber\\
=&\bar{\psi}_{t}\frac{\mathds{1}+i\beta\gamma_{*}}{2\beta}D^{(\omega)}_a\left(\tensor{\epsilon}{^{abc}}\gamma_{b}\psi_c\right)
\label{eq:2.24}
\end{align}
Let us rewrite (\ref{eq:2.24}) in terms of the covariant derivative of the Ashtekar connection. Since
\begin{align}
\tensor*{\omega}{*^{IJ}_{a}}\gamma_{IJ}&=\tensor*{\omega}{*^{ij}_{a}}\gamma_{ij}+2\tensor*{\omega}{*^{0i}_{a}}\gamma_{0i}\\
&=2i\Gamma_a^i\gamma_{*}\gamma_{0i}+2K_a^i\gamma_{0i}
\end{align}
we find
\begin{align}
\frac{\mathds{1}+i\beta\gamma_{*}}{2\beta}\tensor*{\omega}{*^{IJ}_{a}}\gamma_{IJ}=&-\frac{\mathds{1}+i\beta\gamma_{*}}{i\beta}\left(\Gamma_a^i\gamma_{*}\gamma_{0i}-iK_a^i\gamma_{0i}\right)\nonumber\\
=&-\frac{1}{i\beta}\left(\Gamma_a^i\gamma_{*}\gamma_{0i}-iK_a^i\gamma_{0i}+i\beta\Gamma_a^i\gamma_{0i}+\beta K_a^i\gamma_{*}\gamma_{0i}\right)\nonumber\\
=&-\frac{1}{i\beta}\left(A_a^i-iK_a^i\gamma_{*}+i\beta\Gamma_a^i\gamma_{*}\right)\gamma_{*}\gamma_{0i}\nonumber\\
=&-\frac{1}{i\beta}\left(A_a^i+i\beta A_a^i\gamma_{*}-i(1+\beta^2)K_a^i\gamma_{*}\right)\gamma_{*}\gamma_{0i}\nonumber\\
=&\frac{\mathds{1}+i\beta\gamma_{*}}{2\beta}2iA_a^i\gamma_{*}\gamma_{0i}+\frac{1+\beta^2}{\beta}K_a^i\gamma_{0i}
\label{eq:2.25}
\end{align}
Hence, this yields
\begin{align}
\frac{\mathds{1}+i\beta\gamma_{*}}{2\beta}D_a^{(\omega)}\psi_b=&\frac{\mathds{1}+i\beta\gamma_{*}}{2\beta}D_a^{(A)}\psi_b+\frac{1+\beta^2}{4\beta}K_a^i\gamma_{0i}\psi_b
\label{eq:2.26}
\end{align}
with
\begin{align}
D_a^{(A)}\psi_b:=\partial_a\psi_b+\frac{i}{2}A_a^i\gamma_{*}\gamma_{0i}\psi_b
\label{eq:2.27}
\end{align}
With respect to the chiral representation of the gamma matrices, one has
\begin{equation}
    \frac{i}{2}\gamma_{*}\gamma_{0i}=\begin{pmatrix}
\tau_{i}&0\\
0 & \tau_{i}
\end{pmatrix}
\label{eq:2.28}
\end{equation}
Hence, in particular, in the chiral representation the covariant derivative acts separately on the respective chiral sub components of the Rarita-Schwinger field. We will use this property later in section \ref{section 5.2}, when we will study the action of SUSY constraint on spin network states. Note that the appearance of the term $\frac{i}{2}\gamma_*\gamma_{0i}$ in the covariant derivative in (\ref{eq:2.27}) is not a coincidence, but follows from the identification of $\mathfrak{su}(2)$ as a Lie subalgebra of $\mathfrak{spin}^+(1,3)$ generated by $M_{jk}=\frac{1}{2}\gamma_{jk}$ such that $A=-\frac{1}{2}\tensor{\epsilon}{_i^{jk}}M_{jk}$ which implies
\begin{equation}
    \kappa_{\mathbb{R}*}(A)=-\frac{1}{2}A^i\tensor{\epsilon}{_i^{jk}} \kappa_{\mathbb{R}*}(M_{jk})=-\frac{1}{4}A^i\tensor{\epsilon}{_i^{jk}} \gamma_{jk}=\frac{i}{2}\gamma_*\gamma_{0i}A^i
\end{equation}
For the derivation of the SUSY constraint, we need to collect the terms in (\ref{eq:2.24}) proportional to $\psi_t$. Using (\ref{eq:2.26}), one finds again by integration by parts and eventually dropping boundary terms
\begin{align}
&\tensor{\epsilon}{^{abc}}\bar{\psi}_{t}\gamma_{a}\frac{\mathds{1}+i\beta\gamma_{*}}{2\beta}D^{(\omega)}_{b}\psi_{c}-\tensor{\epsilon}{^{abc}}\bar{\psi}_{a}\gamma_{b}\frac{\mathds{1}+i\beta\gamma_{*}}{2\beta}\left(\partial_c\psi_t+\frac{1}{4}\tensor*{\omega}{*^{IJ}_{c}}\gamma_{IJ}\psi_t\right)\nonumber\\
=&\bar{\psi}_t\left(\tensor{\epsilon}{^{abc}}\gamma_{a}\frac{\mathds{1}+i\beta\gamma_{*}}{2\beta}D^{(\omega)}_{b}\psi_{c}+\frac{\mathds{1}+i\beta\gamma_{*}}{2\beta}D^{(\omega)}_a\left(\tensor{\epsilon}{^{abc}}\gamma_{b}\psi_c\right)\right)\nonumber\\
=&\bar{\psi}_t\left(\tensor{\epsilon}{^{abc}}\gamma_{a}\frac{\mathds{1}+i\beta\gamma_{*}}{2\beta}D^{(A)}_{b}\psi_{c}+\frac{\mathds{1}+i\beta\gamma_{*}}{2\beta}D^{(A)}_a\left(\tensor{\epsilon}{^{abc}}\gamma_{b}\psi_c\right)-\frac{1+\beta^2}{4\beta}\tensor{\epsilon}{^{abc}}K_b^ie_{a}^j\gamma_0\{\gamma_i,\gamma_j\}\psi_{c}\right)\nonumber\\
=&\bar{\psi}_t\left(\tensor{\epsilon}{^{abc}}\gamma_{a}\frac{\mathds{1}+i\beta\gamma_{*}}{2\beta}D^{(A)}_{b}\psi_{c}+\frac{\mathds{1}+i\beta\gamma_{*}}{2\beta}D^{(A)}_a\left(\tensor{\epsilon}{^{abc}}\gamma_{b}\psi_c\right)-\frac{1+\beta^2}{2\beta}\tensor{\epsilon}{^{abc}}K_{ba}\gamma_0\psi_{c}\right)
\label{eq:2.29}
\end{align}
Hence, the SUSY constraint in the theory takes the form
\begin{align}
S=&\epsilon^{abc}\gamma_a\frac{\mathds{1}+i\beta\gamma_{*}}{2\beta}D^{(A)}_b\psi_c+\frac{\mathds{1}+i\beta\gamma_{*}}{2\beta}D^{(A)}_a\left(\epsilon^{abc}\gamma_b\psi_c\right)\nonumber\\
&-\frac{1+\beta^2}{2\beta}\epsilon^{abc}\gamma_0\psi_cK_{ba}\label{eq:3.01}
\end{align}
For the term proportional to $\omega_t$ in (\ref{eq:2.21}) we compute, using (\ref{eq:2.25}),
\begin{align}
\frac{1}{4}\tensor{\epsilon}{^{abc}}\bar{\psi}_{a}\gamma_{b}\frac{\mathds{1}+i\beta\gamma_{*}}{2\beta}\tensor*{\omega}{*^{IJ}_{t}}\gamma_{IJ}\psi_{c}=&A_t^i\left(-\frac{1}{4}\tensor{\epsilon}{^{abc}}\bar{\psi}_{a}\gamma_{b}\frac{\mathds{1}+i\beta\gamma_{*}}{i\beta}\gamma_{*}\gamma_{0i}\psi_c\right)\nonumber\\
&+\frac{1+\beta^2}{4\beta}K_t^i\tensor{\epsilon}{^{abc}}\bar{\psi}_{a}\gamma_{b}\gamma_{0i}\psi_c\nonumber\\
=&A_t^i\left(-\frac{i}{2}\pi^{a}\gamma_{*}\gamma_{0i}\psi_a\right)+\frac{1+\beta^2}{4\beta}K_t^i\tensor{\epsilon}{^{abc}}\bar{\psi}_{a}\gamma_{b}\gamma_{0i}\psi_c
\label{eq:2.30}
\end{align}
so that, combining with (\ref{eq:2.16}), this yields
\begin{align}
A_t^iG_i=&A_t^i\left(\frac{1}{\kappa\beta}D_a^{(A)}E^a_i-\frac{i}{2}\pi^{a}\gamma_{*}\gamma_{0i}\psi_a\right)
\end{align}
Hence, the Gauss constraint takes the form
\begin{align}
G_i&=\frac{1}{\kappa\beta}D_a^{(A)}E^a_i-\frac{i}{2}\pi^{a}\gamma_{*}\gamma_{0i}\psi_a\nonumber\\
&=\frac{1}{\kappa\beta}D_a^{(A)}E^a_i+\frac{i}{2}\tensor{\epsilon}{^{abc}}\bar{\psi}_a\gamma_{*}\gamma_0\gamma_b\gamma_i\frac{\mathds{1}+i\beta\gamma_{*}}{2\beta}\psi_c
\label{eq:2.101}
\end{align}
As fermion fields anticommute, it follows that 
\begin{align}
\tensor{\epsilon}{^{abc}}\bar{\psi}_{a}\gamma_{0}\gamma_{de}\psi_{c}=\tensor{\epsilon}{^{abc}}\bar{\psi}_{c}\gamma_{de}\gamma_{0}\psi_{a}=-\tensor{\epsilon}{^{abc}}\bar{\psi}_{a}\gamma_{0}\gamma_{de}\psi_{c}=0
\end{align}
Therefore, combining the last term in (\ref{eq:2.30}) with the last term in (\ref{eq:2.16}), this gives
\begin{align}
&-\frac{1+\beta^2}{\beta}K_t^i\left(\frac{1}{\kappa}\tensor{\epsilon}{_{ik}^l}K_a^kE^a_l-\frac{1}{4}\tensor{\epsilon}{^{abc}}\bar{\psi}_{a}\gamma_{b}\gamma_{0i}\psi_c\right)\nonumber\\
=&-\frac{1+\beta^2}{\beta}K_t^i\left(\frac{1}{\kappa}\tensor{\epsilon}{_{ik}^l}K_a^kE^a_l+\frac{1}{4}\tensor{\epsilon}{^{abc}}e_{bi}\bar{\psi}_{a}\gamma_{0}\psi_c\right)
\end{align}
yielding the second class constraint
\begin{align}
\tensor{\epsilon}{_{ik}^l}K_a^kE^a_l+\frac{\kappa}{4}\tensor{\epsilon}{^{abc}}e_{bi}\bar{\psi}_{a}\gamma_{0}\psi_c=0
\label{eq:2.31}
\end{align}
For the vector constraint, we need to collect terms proportional to the shift vector field $N^a$. From (\ref{eq:2.17}), we deduce, using (\ref{eq:2.30}),
\begin{align}
N^d\frac{1}{\kappa\beta}E_i^{b}\left(F(A)^i_{db}+(1+\beta^2)\tensor{\epsilon}{^i_{kl}}K^k_dK^l_b\right)&=N^d\frac{1}{\kappa\beta}E_i^{b}F(A)^i_{db}+\frac{1+\beta^2}{4\kappa\beta}N^dK^k_d\tensor{\epsilon}{_{kl}^i}K^l_bE^b_i\nonumber\\
&=N^d\frac{1}{\kappa\beta}E_i^{b}F(A)^i_{db}-N^d\frac{1+\beta^2}{4\beta}\tensor{\epsilon}{^{abc}}K_{db}\bar{\psi}_{a}\gamma_{0}\psi_c
\end{align}
On the other hand, (\ref{eq:2.21}) yields together with (\ref{eq:2.26})
\begin{align}
-N^d\tensor{\epsilon}{^{abc}}\bar{\psi}_{a}\gamma_{d}\frac{\mathds{1}+i\beta\gamma_{*}}{2\beta}D^{(\omega)}_{b}\psi_{c}=&-N^d\tensor{\epsilon}{^{abc}}\bar{\psi}_{a}\gamma_{d}\frac{\mathds{1}+i\beta\gamma_{*}}{2\beta}D^{(A)}_{b}\psi_{c}\nonumber\\
&-N^d\frac{1+\beta^2}{4\beta}\tensor{\epsilon}{^{abc}}K_b^i\bar{\psi}_{a}\gamma_{d}\gamma_{0i}\psi_{c}\nonumber\\
=&-N^d\tensor{\epsilon}{^{abc}}\bar{\psi}_{a}\gamma_{d}\frac{\mathds{1}+i\beta\gamma_{*}}{2\beta}D^{(A)}_{b}\psi_{c}\nonumber\\
&+N^d\frac{1+\beta^2}{4\beta}\tensor{\epsilon}{^{abc}}K_{bd}\bar{\psi}_{a}\gamma_{0}\psi_{c}
\end{align}
Therefore, the vector constraint is given by 
\begin{align}
H_d:=\frac{1}{\kappa\beta}E_i^{b}F(A)^i_{db}-\tensor{\epsilon}{^{abc}}\bar{\psi}_{a}\gamma_{d}\frac{\mathds{1}+i\beta\gamma_{*}}{2\beta}D^{(A)}_{b}\psi_{c}+\frac{1+\beta^2}{2\beta}\tensor{\epsilon}{^{abc}}K_{[bd]}\bar{\psi}_{a}\gamma_{0}\psi_{c}
\end{align}
Finally, using (\ref{eq:2.20}), we find for the Hamilton constraint of the theory, modulo the second class constraint,
\begin{align}
    H=&\frac{E^a_iE^b_j}{2\kappa\sqrt{q}}\tensor{\epsilon}{^{ij}_{k}}\left(\tensor*{{F(A)}}{*^k_{ab}}-(1+\beta^2)\tensor{\epsilon}{^k_{mn}}K^m_a K^n_b-\frac{2(1+\beta^2)}{\beta}D^{(\Gamma)}_{[a}K^k_{b]}\right)\nonumber\\
    &+\tensor{\epsilon}{^{abc}}\bar{\psi}_{a}\gamma_{0}\frac{\mathds{1}+i\beta\gamma_{*}}{2\beta}D^{(A)}_{b}\psi_{c}+\frac{1+\beta^2}{4\beta}\epsilon^{abc}K_b^i\bar{\psi}_a\gamma_{0i}\psi_c
\end{align}
At this point, we have expressed the constraints discovered so far in terms of $A,E,\psi,\pi, \Gamma$ and $K$. However, while we can further express $K$ as $K(A,\Gamma)$, $\Gamma$ is undetermined as of yet. At the same time we have a further second class constraint, coming from the variation of the action with respect to 
\begin{equation}
^-A^{i}_a=\Gamma^i_a-\beta K^i_a.  
\end{equation}
The 9 components of this constraint, together with the 3 components of \eqref{eq:2.31} should allow us to solve for $\Gamma$ and $K_t$, thus solving the second class constraints. The calculation is tedious already for Dirac fermions coupled to gravity \cite{Bojowald:2007nu}, so we take a shortcut. The precise expression for $K_t$ is not relevant for our purposes and the gravitational contribution to $\Gamma$, the torsion free spin connection, is well known. The fermionic contribution is simply the spatial component of the contortion tensor $C_{\rho IJ}$ which, using \eqref{eq:2.10} is given by
\begin{equation}
C_a^i:=-\epsilon^{ijk}C_{a jk}=-\frac{\kappa}{8|e|}\epsilon^{bcd}e^i_d(\bar{\psi}_b\gamma_a\psi_c+2\bar{\psi}_b\gamma_c\psi_a)
\end{equation}
This is a function of $E, \psi, \pi$. From now on, we always assume that $\Gamma$ and $K$ are determined by the canonical variables in this way. 

\subsection{Introducing half-densitized fermion fields}
As proposed in \cite{Thiemann:1997rq}, in order to solve the reality conditions of fermion fields in canonical quantum gravity, it is worthwhile to go over to half-densitized fermion fields. In case of the Rarita-Schwinger field, this amounts to introducing the new fields
\begin{align}
\phi_i=\sqrt[4]{q}e^a_i\psi_a\quad\text{and}\quad\pi_{\phi}^i=\frac{1}{\sqrt[4]{q}}e_a^i\pi^a
\label{eq:3.1}
\end{align}
As both sides have been rescaled by the spatial metric, it ís clear that this, a priori, does not define a canonical transformation. In fact, as we will see in the following, this requires a redefinition of the Ashtekar connection. Therefore, following the same steps as in \cite{Bodendorfer:2011pb}, we substitute the transformed fields (\ref{eq:3.1}) in the symplectic potential which yields   
\begin{align}
\int_{\mathbb{R}}{\mathrm{d}t\int_{\Sigma}{\mathrm{d}^3x\,\frac{1}{\kappa\beta}E_i^a\dot{A}_a^i-\pi^a\dot{\psi}_a}}&=\int_{\mathbb{R}}{\mathrm{d}t\int_{\Sigma}{\mathrm{d}^3x\,\frac{1}{\kappa\beta}E_i^a\dot{A}_a^i-\frac{1}{\sqrt[4]{q}}E^a_i\pi_{\phi}^iL_{\partial_t}\left(\sqrt[4]{q}E_a^j\phi_j\right)}}\nonumber\\
&=\int_{\mathbb{R}}{\mathrm{d}t\int_{\Sigma}{\mathrm{d}^3x\,\frac{1}{\kappa\beta}E_i^a\dot{A}_a^i-\pi_{\phi}^i\dot{\phi_i}-\pi_{\phi}^iE_i^a\dot{E}_a^j\phi_j}}\nonumber\\
&=\int_{\mathbb{R}}{\mathrm{d}t\int_{\Sigma}{\mathrm{d}^3x\,\frac{1}{\kappa\beta}E_i^a\dot{A}_a^i-\pi_{\phi}^i\dot{\phi_i}+\pi_{\phi}^i\dot{E}_i^aE_a^j\phi_j}}\nonumber\\
&=\int_{\mathbb{R}}{\mathrm{d}t\int_{\Sigma}{\mathrm{d}^3x\,\frac{1}{\kappa\beta}E_i^a\dot{A}_a^i-\pi_{\phi}^i\dot{\phi_i}-E_i^aL_{\partial_t}\left(\pi_{\phi}^iE_a^j\phi_j\right)}}\nonumber\\
&=\int_{\mathbb{R}}{\mathrm{d}t\int_{\Sigma}{\mathrm{d}^3x\,\frac{1}{\kappa}E_i^aL_{\partial_t}\left(A_a^i-\kappa\beta\pi_{\phi}^iE_a^j\phi_j\right)-\pi_{\phi}^i\dot{\phi_i}}}\label{eq:3.2}
\end{align}
where we have dropped a boundary term from the third to the fourth line. Hence, transforming the Ashtekar connection via 
\begin{align}
A_a^i\rightarrow A'^i_a=\Gamma_a^i+\beta K'^i_a
\label{eq:3.3}
\end{align}
with
\begin{align}
K'^i_a&=K_a^i-\kappa\pi_{\phi}^iE_a^l\phi_l=K_a^i+\frac{\kappa}{q}\tensor{\epsilon}{^{dbc}}e_d^ie_b^je_c^ke_a^l\bar{\phi}_{j}\gamma_{k}\frac{\mathds{1}+i\beta\gamma_{*}}{2\beta}\phi_l\nonumber\\
&=K_a^i+\frac{i\kappa}{2\sqrt{q}}\tensor{\epsilon}{^{ijk}}e_a^l\bar{\phi}_{j}\gamma_{k}\frac{\mathds{1}+i\beta\gamma_{*}}{i\beta}\phi_l
\label{eq:3.3.1}
\end{align}
this yields a canonical transformation with the new canonically conjugate pairs $(A'^i_a,E^a_i)$ and $(\phi_i,\pi^i_{\phi})$ and the non-vanishing Poisson brackets
\begin{align}
\{A'^i_a(x),E^b_j(y)\}=\kappa\beta\delta^{(3)}(x,y)\quad\text{and}\quad\{\phi_{i}^{\alpha}(x),\pi_{\phi\,\beta}^j(y)\}=-\delta^j_i\delta^{\alpha}_{\beta}\delta^{(3)}(x,y)
\label{eq:3.3.2}
\end{align}
In the new variables, the reality condition (\ref{eq:2.100}) takes the form
\begin{align}
\Omega^i:=\pi_{\phi}^i-\frac{i}{2}\tensor{\epsilon}{^{ijk}}\bar{\phi}_{j}\gamma_{*}\gamma_{k}\mathcal{P}^{+}_{\beta}=0
\label{eq:3.3.3}
\end{align}
which now, in particular, neither depends on the internal triad nor on the spatial metric simplifying significantly the further canonical analysis. As a next step, we have to reformulate the constraints in the new variables. Since we will mainly be interested in the explicit form of the SUSY constraint, we will only derive the transformed expressions of the Gauss and SUSY constraint in what follows. The remaining constraints can be treated in complete analogy.     

\subsubsection{Gauss constraint}
By (\ref{eq:2.101}), the Gauss constraint takes the form
\begin{align}
G_i&=\frac{1}{\kappa\beta}D_a^{(A)}E^a_i+\frac{i}{2}\tensor{\epsilon}{^{abc}}\bar{\psi}_a\gamma_{*}\gamma_0\gamma_b\gamma_i\frac{\mathds{1}+i\beta\gamma_{*}}{2\beta}\psi_c\nonumber\\
&=\frac{1}{\kappa\beta}D_a^{(A)}E^a_i+\frac{i}{2}\tensor{\epsilon}{^{jmk}}\bar{\phi}_j\gamma_{*}\gamma_0\gamma_m\gamma_i\frac{\mathds{1}+i\beta\gamma_{*}}{2\beta}\phi_k
\label{2.102}
\end{align}
Considering the first part in (\ref{2.102}), we find
\begin{align}
D_a^{(A)}E^a_i=&\partial_aE^a_i+\tensor{\epsilon}{_{im}^{n}}(A'^m_a+\kappa\beta\pi_{\phi}^mE_a^l\phi_l)E^a_n\nonumber\\
=&D_a^{(A')}E^a_i+\frac{i\kappa\beta}{2}\tensor{\epsilon}{_{mi}^{l}}\tensor{\epsilon}{^{mjk}}\bar{\phi}_{j}\gamma_{k}\frac{\mathds{1}+i\beta\gamma_{*}}{i\beta}\phi_l\nonumber\\
=&D_a^{(A')}E^a_i+\frac{i\kappa\beta}{2}\bar{\phi}_{i}\gamma_{k}\frac{\mathds{1}+i\beta\gamma_{*}}{i\beta}\phi^k-\frac{i\kappa\beta}{2}\bar{\phi}_{l}\gamma_{i}\frac{\mathds{1}+i\beta\gamma_{*}}{i\beta}\phi^l\nonumber\\
=&D_a^{(A')}E^a_i+\frac{\kappa}{2}\bar{\phi}_{i}\gamma_{k}\phi^k-\frac{i\kappa\beta}{2}\bar{\phi}_{i}\gamma_{*}\gamma_{k}\phi^k+\frac{i\kappa\beta}{2}\bar{\phi}_{l}\gamma_{*}\gamma_{i}\phi^l
\label{eq:3.4}
\end{align}
Since $\gamma_i\gamma_j=\delta_{ij}+\gamma_{ij}$, one has
\begin{align}
\frac{i}{2}\tensor{\epsilon}{^{jmk}}\bar{\phi}_j\gamma_{*}\gamma_0\gamma_m\gamma_i\frac{\mathds{1}+i\beta\gamma_{*}}{2\beta}\phi_k=&\frac{1}{4}\tensor{\epsilon}{^{jmk}}\bar{\phi}_j\gamma_0\gamma_m\gamma_i\phi_k+\frac{i}{4\beta}\tensor{\epsilon}{^{jmk}}\bar{\phi}_j\gamma_{*}\gamma_0\gamma_m\gamma_i\phi_k\nonumber\\
=&\frac{1}{4}\tensor{\epsilon}{^{jmk}}\bar{\phi}_j\gamma_0\gamma_m\gamma_i\phi_k-\frac{i}{4\beta}\tensor{\epsilon}{^{ijk}}\bar{\phi}_j\gamma_{*}\gamma_0\phi_k\nonumber\\
&+\frac{i}{4\beta}\tensor{\epsilon}{^{jmk}}\bar{\phi}_j\gamma_{*}\gamma_0\gamma_{mi}\phi_k\label{eq:3.5}
\end{align}
By antisymmetry of the fermion fields, it follows
\begin{equation}
\tensor{\epsilon}{^{ijk}}\bar{\phi}_j\gamma_{*}\gamma_0\phi_k=\tensor{\epsilon}{^{ijk}}\bar{\phi}_k\gamma_{*}\gamma_0\phi_j=-\tensor{\epsilon}{^{ijk}}\bar{\phi}_j\gamma_{*}\gamma_0\phi_k=0
\label{eq:3.6}
\end{equation}
so that, using $\gamma_{*}\gamma_{ij}=-i\tensor{\epsilon}{_{ij}^k}\gamma_{0k}$, we find
\begin{align}
\frac{i}{2}\tensor{\epsilon}{^{jmk}}\bar{\phi}_j\gamma_{*}\gamma_0\gamma_m\gamma_i\frac{\mathds{1}+i\beta\gamma_{*}}{2\beta}\phi_k=&\frac{1}{4}\tensor{\epsilon}{^{jmk}}\bar{\phi}_j\gamma_0\gamma_m\gamma_i\phi_k-\frac{1}{4\beta}\bar{\phi}_i\gamma_{k}\phi_k+\frac{1}{4\beta}\bar{\phi}_k\gamma^k\phi_i\nonumber\\
=&\frac{1}{4}\tensor{\epsilon}{^{jmk}}\bar{\phi}_j\gamma_0\gamma_m\gamma_i\phi_k-\frac{1}{2\beta}\bar{\phi}_i\gamma_{k}\phi_k\nonumber\\
=&-\frac{1}{4}\tensor{\epsilon}{_i^{jk}}\bar{\phi}_j\gamma_0\phi_k-\frac{1}{2\beta}\bar{\phi}_i\gamma_{k}\phi_k\label{eq:3.8}
\end{align}
where from the second to the last line we again used (\ref{eq:3.6}). Hence, the Gauss constraint can be written as
\begin{align}
G_i=&D_a^{(A')}E^a_i-\frac{1}{4}\tensor{\epsilon}{_i^{jk}}\bar{\phi}_j\gamma_0\phi_k-\frac{i}{2}\bar{\phi}_{i}\gamma_{*}\gamma_{k}\phi^k+\frac{i}{2}\bar{\phi}_{k}\gamma_{*}\gamma_{i}\phi^k
\label{eq:3.9}
\end{align}
In fact, this can be simplified even further. Therefore, consider 
\begin{align}
\bar{\phi}_j\gamma_{*}\gamma_{k}\gamma_i\gamma^{(j}\phi^{k)}=&\frac{1}{2}\bar{\phi}_j\gamma_{*}\gamma_{k}\gamma_i\gamma^{j}\phi^{k}+\frac{1}{2}\bar{\phi}_j\gamma_{*}\gamma_{k}\gamma_i\gamma^{k}\phi^{j}\nonumber\\
=&\frac{1}{2}\bar{\phi}_j\gamma_{*}\gamma_{k}\gamma_i\gamma^{j}\phi^{k}-\frac{1}{2}\bar{\phi}_j\gamma_{*}\gamma_{i}\phi^{j}
\label{eq:3.10}
\end{align}
which, due to $\gamma_i\gamma^j=2\delta_i^j-\gamma^j\gamma_i$ yields
\begin{align}
\bar{\phi}_j\gamma_{*}\gamma_{k}\gamma_i\gamma^{(j}\phi^{k)}=&\bar{\phi}_i\gamma_{*}\gamma_{k}\phi^{k}-\frac{1}{2}\bar{\phi}^j\gamma_{*}\gamma_{kj}\gamma_i\phi^{k}-\bar{\phi}_k\gamma_{*}\gamma_{i}\phi^{k}\nonumber\\
=&\frac{i}{2}\tensor{\epsilon}{^{klj}}\bar{\phi}_k\gamma_{0}\gamma_{l}\gamma_i\phi_{j}+\bar{\phi}_i\gamma_{*}\gamma_{k}\phi^{k}-\bar{\phi}_k\gamma_{*}\gamma_{i}\phi^{k}\nonumber\\
=&-\frac{i}{2}\tensor{\epsilon}{_i^{kj}}\bar{\phi}_k\gamma_{0}\phi_{j}+\bar{\phi}_i\gamma_{*}\gamma_{k}\phi^{k}-\bar{\phi}_k\gamma_{*}\gamma_{i}\phi^{k}
\label{eq:3.11}
\end{align}
Thus, to summarize, in the new variables, we find that the Gauss constraint can be written in the following compact form
\begin{equation}
G_i=D_a^{(A')}E^a_i-\frac{i}{2}\bar{\phi}_j\gamma_{*}\gamma_{k}\gamma_i\gamma^{(j}\phi^{k)}
\label{eq:3.12}
\end{equation}

\subsubsection{Supersymmetry constraint}
Finally, we want to express the supersymmetry constraint $S$ in the new variables. Therefore, inserting (\ref{eq:3.1}) and (\ref{eq:3.4}) into (\ref{eq:3.01}), the first two terms in (\ref{eq:3.01}) become
\begin{align}
S=&\tensor{\epsilon}{^{abc}}e_a^i\gamma_i\frac{\mathds{1}+i\beta\gamma_{*}}{2\beta}D_b^{(A')}\left(\frac{1}{\sqrt[4]{q}}e^j_c\phi_j\right)+\frac{\mathds{1}+i\beta\gamma_{*}}{2\beta}D_a^{(A')}\left(\frac{1}{\sqrt[4]{q}}\tensor{\epsilon}{^{ijk}}E_i^a\gamma_j\phi_k\right)\nonumber\\
&+\frac{i\kappa\beta}{2\sqrt[4]{q}}\epsilon^{lmn}\epsilon^{ijk}\frac{\mathds{1}-i\beta\gamma_{*}}{2\beta}\gamma_{*}\gamma_0\gamma_m\gamma_i\phi_n\left(\bar{\phi}_{j}\gamma_{k}\frac{\mathds{1}+i\beta\gamma_{*}}{2\beta}\phi_l\right)\nonumber\\
&-\frac{i\kappa\beta}{2\sqrt[4]{q}}\epsilon^{lmn}\epsilon^{ijk}\frac{\mathds{1}+i\beta\gamma_{*}}{2\beta}\gamma_{*}\gamma_0\gamma_i\gamma_m\phi_n\left(\bar{\phi}_{j}\gamma_{k}\frac{\mathds{1}+i\beta\gamma_{*}}{2\beta}\phi_l\right)
\label{eq:3.13}
\end{align}
where the second and last line in (\ref{eq:3.13}) can be summarized as
\begin{align}
&\frac{i\kappa\beta}{2\sqrt[4]{q}}\epsilon^{lmn}\epsilon^{ijk}\left(\bar{\phi}_{j}\gamma_{k}\frac{\mathds{1}+i\beta\gamma_{*}}{2\beta}\phi_l\right)\left[\frac{1}{2\beta}\gamma_{*}\gamma_0[\gamma_m,\gamma_i]-\frac{i}{2}\gamma_0\{\gamma_m,\gamma_i\}\right]\phi_n\nonumber\\
&=\frac{i\kappa}{2\sqrt[4]{q}}\epsilon^{lmn}\epsilon^{ijk}\gamma_{*}\gamma_{0mi}\phi_n\left(\bar{\phi}_{j}\gamma_{k}\frac{\mathds{1}+i\beta\gamma_{*}}{2\beta}\phi_l\right)-\frac{\kappa\beta}{2\sqrt[4]{q}}\tensor{\epsilon}{_i^{ln}}\epsilon^{ijk}\gamma_{0}\phi_n\left(\bar{\phi}_{j}\gamma_{k}\frac{\mathds{1}+i\beta\gamma_{*}}{2\beta}\phi_l\right)\label{eq:3.14}
\end{align}
Since $\gamma_{*}\gamma_{0mi}=-i\tensor{\epsilon}{_{mi}^p}\gamma_p$, the first term in the second line of (\ref{eq:3.14}) takes the form
\begin{align}
\frac{\kappa}{2\sqrt[4]{q}}\epsilon^{lmn}\tensor{\epsilon}{_{mi}^p}\epsilon^{ijk}\gamma_p\phi_n\left(\bar{\phi}_{j}\gamma_{k}\frac{\mathds{1}+i\beta\gamma_{*}}{2\beta}\phi_l\right)=\frac{\kappa}{\sqrt[4]{q}}\epsilon^{ijk}\gamma^{l}\phi_{[l}\left(\bar{\phi}_{i]}\frac{\mathds{1}+i\beta\gamma_{*}}{2\beta}\gamma_{k}\phi_j\right)
\label{eq:3.15}
\end{align}
Next, let us rewrite the '$K$-term' of the supersymmetry constraint (\ref{eq:3.01}) as
\begin{align}
\epsilon^{abc}\gamma_0\psi_cK_{ba}=&\frac{1}{\sqrt[4]{q}}\epsilon^{adc}e_c^ne_{ai}\gamma_0\phi_nK_{b}^ie_d^je_j^b=\frac{1}{\sqrt[4]{q}}\tensor{\epsilon}{_i^{jn}}\gamma_0\phi_nK_{b}^iE_j^b\nonumber\\
=&-\frac{\kappa}{4\sqrt[4]{q}}\tensor{\epsilon}{^{abc}}e_{b}^n\gamma_0\phi_n(\bar{\psi}_{a}\gamma_{0}\psi_c)=\frac{\kappa}{4\sqrt[4]{q}}\tensor{\epsilon}{^{njk}}\gamma_0\phi_n(\bar{\phi}_{j}\gamma_{0}\phi_k)
\label{eq:3.16}
\end{align}
Hence, combining (\ref{eq:3.16}) with second term in the second line of (\ref{eq:3.14}), this yields
\begin{align}
&-\frac{\kappa\beta}{2\sqrt[4]{q}}\tensor{\epsilon}{_i^{ln}}\epsilon^{ijk}\gamma_{0}\phi_n\left(\bar{\phi}_{j}\gamma_{k}\frac{\mathds{1}+i\beta\gamma_{*}}{2\beta}\phi_l\right)-\frac{\kappa}{4\sqrt[4]{q}}\frac{1+\beta^2}{2\beta}\tensor{\epsilon}{^{njk}}\gamma_0\phi_n(\bar{\phi}_{j}\gamma_{0}\phi_k)\nonumber\\
&=\frac{i\kappa\beta}{4\sqrt[4]{q}}\gamma_0\phi^k\left(\bar{\phi}_{l}\gamma_{*}\gamma_k\phi^l\right)+\frac{\kappa\beta}{2\sqrt[4]{q}}\gamma_0\phi^j\left(\bar{\phi}_{j}\gamma^l\frac{\mathds{1}+i\beta\gamma_{*}}{2\beta}\phi_l\right)-\frac{\kappa}{4\sqrt[4]{q}}\frac{1+\beta^2}{2\beta}\tensor{\epsilon}{^{njk}}\gamma_0\phi_n\bar{\phi}_{j}\gamma_{0}\phi_k\nonumber\\
&=\frac{i\kappa\beta}{4\sqrt[4]{q}}\gamma_0\phi^i\left(\bar{\phi}_l\gamma_*\gamma_i\phi^l-\bar{\phi}_i\gamma_*\gamma_l\phi^l+\frac{i}{2}\tensor{\epsilon}{_i^{jk}}\bar{\phi}_j\gamma_0\phi_k\right)+\frac{\kappa}{4\sqrt[4]{q}}\gamma_0\phi^i\left(\bar{\phi}_{i}\gamma^l\phi_l\right)-\frac{\kappa}{8\beta\sqrt[4]{q}}\gamma_0\phi_i\tensor{\epsilon}{^{ijk}}(\bar{\phi}_{j}\gamma_{0}\phi_k)\nonumber\\
&=-\frac{i\kappa\beta}{4\sqrt[4]{q}}\gamma_0\phi^i\left(\bar{\phi}_j\gamma_*\gamma_k\gamma_i\gamma^{(j}\phi^{k)}\right)+\frac{\kappa}{4\sqrt[4]{q}}\gamma_0\phi^i\left(\bar{\phi}_{i}\gamma^l\phi_l\right)-\frac{\kappa}{8\beta\sqrt[4]{q}}\gamma_0\phi_i\tensor{\epsilon}{^{ijk}}(\bar{\phi}_{j}\gamma_{0}\phi_k)
\label{eq:3.17}
\end{align}
where, from the third to the last line, identity (\ref{eq:3.11}) was used. Since
\begin{equation}
\bar{\phi}_{i}\gamma^l\phi_l=-\frac{i}{2}\tensor{\epsilon}{^{jkl}}\bar{\phi}_{j}\gamma_*\gamma_0\gamma_k\gamma_i\phi_l+\bar{\phi}_{j}\gamma_k\gamma_i\gamma^{(j}\phi^{k)}
\label{eq:3.18}
\end{equation}
(this can be shown along the lines of eq. (\ref{eq:3.10}) and (\ref{eq:3.11})) and $\tensor{\epsilon}{^{ijk}}\bar{\phi}_{j}\gamma_0\phi_k=-\tensor{\epsilon}{^{jkl}}\bar{\phi}_{j}\gamma_0\gamma_k\gamma_i\phi_l$, the last line of (\ref{eq:3.17}) finally takes the form
\begin{align}
&-\frac{i\kappa\beta}{4\sqrt[4]{q}}\gamma_0\phi^i\left(\bar{\phi}_j\gamma_*\gamma_k\gamma_i\gamma^{(j}\phi^{k)}\right)+\frac{\kappa}{4\sqrt[4]{q}}\gamma_0\phi^i\left(\bar{\phi}_{i}\gamma^l\phi_l\right)-\frac{\kappa}{8\beta\sqrt[4]{q}}\gamma_0\phi_i\tensor{\epsilon}{^{ijk}}(\bar{\phi}_{j}\gamma_{0}\phi_k)\nonumber\\
&=\frac{\kappa\beta}{2\sqrt[4]{q}}\gamma_0\phi^{i}\left(\bar{\phi}_{j}\gamma_k\frac{\mathds{1}+i\beta\gamma_{*}}{2\beta}\gamma_i\gamma^{(j}\phi^{k)}\right)+\frac{\kappa}{4\sqrt[4]{q}}\gamma_0\phi^{i}\left(\tensor{\epsilon}{^{jkl}}\bar{\phi}_{j}\gamma_0\frac{\mathds{1}+i\beta\gamma_{*}}{2\beta}\gamma_k\gamma_{i}\phi_{l}\right)
\label{eq:3.19}
\end{align}
To summarize, we have found the following form of the supersymmetry constraint in the new variables
\begin{align}
S=&\tensor{\epsilon}{^{abc}}e_a^i\gamma_i\frac{\mathds{1}+i\beta\gamma_{*}}{2\beta}D_b^{(A')}\left(\frac{1}{\sqrt[4]{q}}e^j_c\phi_j\right)+\frac{\mathds{1}+i\beta\gamma_{*}}{2\beta}D_a^{(A')}\left(\frac{1}{\sqrt[4]{q}}\tensor{\epsilon}{^{ijk}}E_i^a\gamma_j\phi_k\right)\nonumber\\
&+\frac{\kappa}{\sqrt[4]{q}}\tensor{\epsilon}{^{ijk}}\gamma^l\phi_{[l}\left(\bar{\phi}_{i]}\frac{\mathds{1}+i\beta\gamma_{*}}{2\beta}\gamma_k\phi_j\right)+\frac{\kappa\beta}{2\sqrt[4]{q}}\gamma_0\phi^{i}\left(\bar{\phi}_{j}\gamma_k\frac{\mathds{1}+i\beta\gamma_{*}}{2\beta}\gamma_i\gamma^{(j}\phi^{k)}\right)\nonumber\\
&+\frac{\kappa}{4\sqrt[4]{q}}\gamma_0\phi^{i}\left(\tensor{\epsilon}{^{jkl}}\bar{\phi}_{j}\gamma_0\frac{\mathds{1}+i\beta\gamma_{*}}{2\beta}\gamma_k\gamma_{i}\phi_{l}\right)
\label{eq:3.20}
\end{align}
With an eye towards quantization of this expression, it is useful to rewrite the second term in (\ref{eq:3.20}) depending on the covariant derivative of the fermion field. In fact, using $\gamma_{*}\gamma_{0i}\gamma_k=-2i\tensor{\epsilon}{_{ik}^l}\gamma_l+\gamma_k\gamma_{*}\gamma_{0i}$, we find 
\begin{align}
D_a^{(A')}\left(\frac{1}{\sqrt[4]{q}}\tensor{\epsilon}{^{abc}}e_b^k\gamma_ke_c^l\phi_l\right)&=\partial_a\left(\frac{1}{\sqrt[4]{q}}\tensor{\epsilon}{^{abc}}e_b^k\gamma_ke_c^l\phi_l\right)+\frac{1}{\sqrt[4]{q}}\tensor{\epsilon}{^{abc}}e_b^ke_c^lA_a^i\frac{i}{2}\gamma_{*}\gamma_{0i}\gamma_k\phi_l\nonumber\\
&=(D_a^{(A')}e_b^k)\frac{1}{\sqrt[4]{q}}\tensor{\epsilon}{^{abc}}\gamma_ke_c^l\phi_l+\tensor{\epsilon}{^{abc}}e_b^k\gamma_kD_a^{(A')}\left(\frac{1}{\sqrt[4]{q}}e_c^l\phi_l\right)
\label{eq:3.21}
\end{align}
so that we can write (\ref{eq:3.20}) equivalently as follows 
\begin{align}
S=&i\tensor{\epsilon}{^{abc}}e_a^i\gamma_i\gamma_{*}D_b^{(A')}\left(\frac{1}{\sqrt[4]{q}}e^j_c\phi_j\right)+\frac{1}{\sqrt[4]{q}}\tensor{\epsilon}{^{abc}}e_c^l\frac{\mathds{1}+i\beta\gamma_{*}}{2\beta}\gamma_k(D_a^{(A')}e_b^k)\phi_l\nonumber\\
&+\frac{\kappa}{\sqrt[4]{q}}\tensor{\epsilon}{^{ijk}}\gamma^l\phi_{[l}\left(\bar{\phi}_{i]}\frac{\mathds{1}+i\beta\gamma_{*}}{2\beta}\gamma_k\phi_j\right)+\frac{\kappa\beta}{2\sqrt[4]{q}}\gamma_0\phi^{i}\left(\bar{\phi}_{j}\gamma_k\frac{\mathds{1}+i\beta\gamma_{*}}{2\beta}\gamma_i\gamma^{(j}\phi^{k)}\right)\nonumber\\
&+\frac{\kappa}{4\sqrt[4]{q}}\gamma_0\phi^{i}\left(\tensor{\epsilon}{^{jkl}}\bar{\phi}_{j}\gamma_0\frac{\mathds{1}+i\beta\gamma_{*}}{2\beta}\gamma_k\gamma_{i}\phi_{l}\right)
\label{eq:3.22}
\end{align}
This is the most compact form of the supersymmetry constraint that we will use for quantization of the theory.
\section{Anti-de Sitter Supergravity}
The canonical analysis of $\mathcal{N}=1$-anti-de Sitter supergravity in the chiral theory has been studied for instance in \cite{Fulop:1993wi,Gambini:1995db,Ling:1999gn}. For sake of completeness, let us briefly discuss it in case of real Barbero-Immirzi parameters.\\
In turns out that the isometry group $\mathrm{SO}(2,3)$ of anti-de Sitter space\footnote{The four-dimensional anti-de Sitter spacetime is an embedded submanifold of the semi-Riemannian manifold $\mathbb{R}^{2,3}$ equipped with the metric $\eta_{AB}=\mathrm{diag}(-+++-)$ defined as
\begin{equation}
\mathrm{AdS}_4:=\{x\in\mathbb{R}^{2,3}|\,\eta_{AB}x^Ax^B=-L^2\}
\end{equation}} $\mathrm{AdS}_4$ can be extended to a super Lie group with $\mathcal{N}$ fermionic generators given by the orthosymplectic Lie group $\mathrm{OSp}(\mathcal{N}|4)$. This leads to a supergravity theory with negative cosmological constant. For $\mathcal{N}=1$, the Holst action then takes the form
\begin{align}
S_{\text{$H$-$ASG$}}=S_{\text{$H$-$SG$}}+\int_{M}{\mathrm{d}^4x\,-e\frac{1}{2L}\bar{\psi}_{\mu}\gamma^{\mu\nu}\psi_{\nu}+\frac{3}{\kappa L^2}e}
\label{eq:4.1}
\end{align}
with $S_{\text{$H$-$SG$}}$ the Holst-action (\ref{eq:2.1}) (or (\ref{eq:2.11})) of $\mathcal{N}=1$ Poincaré-supergravity where $L$ is the so-called anti-de Sitter radius which is related to the cosmological constant via $\Lambda=-\frac{3}{L^2}$. Since these additional terms do not depend on the spin connection, it follows immediately that the variation of (\ref{eq:4.1}) w.r.t. $\omega$ yields the same equations of motions as in the $\Lambda=0$ case and thus, in particular, are again independent of the Barbero-Immirzi parameter.\\
The 3+1-split of the additional terms is straightforward and yields
\begin{align}
    -e\frac{1}{2L}\bar{\psi}_{\mu}\gamma^{\mu\nu}\psi_{\nu}+\frac{3}{\kappa L^2}e=-\frac{1}{2L}N\sqrt{q}\left(2\bar{\psi}_t\gamma^{ta}\psi_a+\bar{\psi}_a\gamma^{ab}\psi_b\right)+N\frac{3}{\kappa L^2}\sqrt{q}
\end{align}
as for anticommuting fermionic fields one has $\bar{\psi}_a\gamma^{at}\psi_t=\bar{\psi}_t\gamma^{ta}\psi_a$. Since $e^t_i=0$ and $e^t_0=\frac{1}{N}$, we find
\begin{align}
    -e\frac{1}{2L}\bar{\psi}_{\mu}\gamma^{\mu\nu}\psi_{\nu}+\frac{3}{\kappa L^2}e=-\frac{1}{L}E^{a}_i\bar{\psi}_t\gamma^{0i}\psi_a+N\left(\frac{1}{2L}\sqrt{q}\bar{\psi}_a\gamma^{ab}\psi_b+\frac{3}{\kappa L^2}\sqrt{q}\right)
    \label{eq:4.3}
\end{align}
The first term in (\ref{eq:4.3}) yields an additional contribution to the SUSY constraint whereas the second term contributes to the Hamiltonian constraint. Hence, it follows that the SUSY constraint in AdS supergravity takes the form 
\begin{align}
S=&\epsilon^{abc}\gamma_a\frac{\mathds{1}+i\beta\gamma_{*}}{2\beta}D^{(A)}_b\psi_c+\frac{\mathds{1}+i\beta\gamma_{*}}{2\beta}D^{(A)}_a\left(\epsilon^{abc}\gamma_b\psi_c\right)\nonumber\\
&-\frac{1+\beta^2}{2\beta}\epsilon^{abc}\gamma_0\psi_cK_{ba}-\frac{1}{L}E^{a}_i\gamma^{0i}\psi_a
\end{align}
which again can be re-expressed in terms of half-densitized fermionic variables.
\section{Quantum theory}
\subsection{Quantization of the Rarita-Schwinger field}\label{section 5.1}
The quantization of the Rarita-Schwinger field is more complicated than for ordinary Dirac fermions due to the form (\ref{eq:3.3.3}) of the reality condition $\Omega^i_{\alpha}$ which, however, has already been drastically simplified using half-densitized fermionic fields since then (\ref{eq:3.3.3}) no longer depends on the triads and the spatial metric.\\
In order to solve this second class constraint, we follow the standard procedure and compute the corresponding Dirac brackets for which we have to compute Poisson brackets of the form $\{\Omega^i_{\alpha},\Omega^j_{\beta}\}$. Using (\ref{eq:3.3.3}) as well as (\ref{eq:3.3.2}), this yields (omitting the delta distribution for convenience)
\begin{align}
    \{\Omega^i_{\alpha},\Omega^j_{\beta}\}&=-\frac{i}{2}\epsilon^{ikl}\{\bar{\phi}_{k\delta},\pi_{\beta}^j\}\tensor{(\gamma_{*}\gamma_l\mathcal{P}_{\beta}^+)}{^{\delta}_{\alpha}}-\frac{i}{2}\epsilon^{jmn}\{\pi^i_{\alpha},\bar{\phi}_{m\delta}\}\tensor{(\gamma_{*}\gamma_n\mathcal{P}_{\beta}^+)}{^{\delta}_{\beta}}\nonumber\\
    &=\frac{i}{2}\epsilon^{ijk}\mathcal{C}_{\beta\delta}\tensor{(\gamma_{*}\gamma_k\mathcal{P}_{\beta}^+)}{^{\delta}_{\alpha}}-\frac{i}{2}\epsilon^{ijk}\mathcal{C}_{\alpha\delta}\tensor{(\gamma_{*}\gamma_k\mathcal{P}_{\beta}^+)}{^{\delta}_{\beta}}\nonumber\\
    &=\frac{i}{2}\epsilon^{ijk}\left[(C\gamma_*\gamma_k\mathcal{P}_{\beta}^+)^T_{\alpha\beta}-(C\gamma_*\gamma_k\mathcal{P}_{\beta}^+)_{\alpha\beta}\right]\nonumber\\
    &=\frac{i}{2}\epsilon^{ijk}(C\gamma_*\gamma_k[\mathcal{P}_{\beta}^++\mathcal{P}_{\beta}^-])=i\epsilon^{ijk}(C\gamma_*\gamma_k)_{\alpha\beta}=:\mathbf{C}_{\alpha\beta}^{ij}
    \label{eq:4.2}
\end{align}
As we see, the operator $\mathcal{P}^{\pm}_{\beta}$ has dropped out completely so that, in particular, (\ref{eq:4.2}) is independent of the Barbero-Immirzi parameter. Finally, since
\begin{equation}
    \{\phi_i^{\alpha},\Omega^j_{\beta}\}=-\delta^j_i\delta^{\alpha}_{\beta}\quad\text{and}\quad\{\Omega^i_{\alpha},\bar{\phi}_{j\beta}\}=-\delta^i_j C_{\alpha\beta}
\end{equation}
it follows that the Dirac brackets for the Rarita-Schwinger field take the form
\begin{equation}
    \{\phi^{\alpha}_i,\bar{\phi}_{j\beta}\}_{\mathrm{DB}}=-\{\phi_i^{\alpha},\Omega^k_{\gamma}\}(\mathbf{C}^{-1})^{\gamma\delta}_{kl}\{\Omega^l_{\delta},\bar{\phi}_{j\beta}\}=-(\tensor{(\mathbf{C}^{-1})_{ij}C)}{^{\alpha}_{\beta}}
    \label{eq:4.4}
\end{equation}
with $\mathbf{C}^{-1}$ the inverse of (\ref{eq:4.2}) which satisfies $(\mathbf{C}^{-1})_{ij}\mathbf{C}^{jk}=\delta_i^k\mathds{1}$. As can be checked by direct computation, this matrix takes the form 
\begin{equation}
    (\mathbf{C}^{-1})_{ij}=-\gamma_0\left(\mathds{1}\delta_{ij}-\frac{1}{2}\gamma_i\gamma_j\right)C^{-1}
    \label{eq:4.5}
\end{equation}
so that the resulting Dirac brackets can be written as
\begin{equation}
    \{\phi^{\alpha}_i(x),\bar{\phi}_{j\beta}(y)\}_{\mathrm{DB}}=\tensor{\left(\left(\mathds{1}\delta_{ij}-\frac{1}{2}\gamma_i\gamma_j\right)\gamma_0\right)}{^{\alpha}_{\beta}}\delta^{(3)}(x,y)
    \label{eq:4.6}
\end{equation}
Note that, since (\ref{eq:3.3.3}) does not depend on the internal triads, the Dirac brackets of the bosonic degrees of freedom $(A^i_a,E_a^i)$ coincide with the original Poisson brackets. In particular, the mixed Dirac brackets between bosonic and fermionic degrees of freedom are still vanishing. For further simplification, we will work in a real representation of the Clifford algebra such that Majorana fermions are explicitly real. In such a representation, the charge conjugation matrix is given by $C=i\gamma^0$ and (\ref{eq:4.6}) yields
\begin{equation}
    \{\phi^{\alpha}_i(x),\phi_{j}^{\beta}(y)\}_{\mathrm{DB}}=\frac{i}{2}\tensor{\left(\mathds{1}\delta_{ij}-\frac{1}{2}\gamma_i\gamma_j\right)}{^{\alpha\beta}}\delta^{(3)}(x,y)
    \label{eq:4.7}
\end{equation}
together with the Majorana condition $\phi_i^*=\phi_i$. Due to the complicated form of the Dirac bracket (\ref{eq:4.7}), the implementation of the Rarita-Schwinger field which simultaneously also allows a direct solution of the Gauss constraint in the quantum theory is by far not straightforward. However, in 
\cite{Bodendorfer:2011pb} a clever way was found to solve all these issues simultaneously by appropriately enlarging the phase space.\\
More precisely, the idea in \cite{Bodendorfer:2011pb} is to decompose $\phi_i$ in its trace part $\sigma:=\gamma^i\phi_i$ and its trace free part $\rho_i:=\phi_i-\frac{1}{3}\gamma_i\sigma$ w.r.t. to the gamma matrices $\gamma_i$ such that $\phi_i=\rho_i+\frac{1}{3}\gamma_i\sigma$. On the enlarged phase space, we then impose the Poisson brackets 
\begin{equation}
    \{\rho_i^{\alpha},\rho_j^{\beta}\}=i\delta_{ij}\delta^{\alpha\beta}\delta^{(3)}(x,y)\quad\text{and}\quad\{\sigma^{\alpha},\sigma^{\beta}\}=-\frac{9i}{2}\delta_{ij}\delta^{\alpha\beta}\delta^{(3)}(x,y)
\end{equation}
with the remaining brackets being zero such that the Dirac bracket (\ref{eq:4.7}) is recovered. Moreover, in order to account for the superfluous degrees of freedom, i.e. the trace freeness of $\rho_i$, one has to add the additional secondary constraint $\Lambda:=\gamma^i\rho_i=0$ \cite{Bodendorfer:2011pb}. Using $\{\Lambda^{\alpha},\Lambda^{\beta}\}=3i\delta^{\alpha\beta}$, this yields the Dirac brackets
\begin{equation}
    \{\rho^{\alpha}_i,\rho_{j}^{\beta}\}_{\mathrm{DB}}=i\left(\delta_{ij}\delta^{\alpha\beta}-\frac{1}{3}\tensor{(\gamma_i\gamma_j)}{^{\alpha\beta}}\right)\delta^{(3)}(x,y)=:i\mathbf{P}^{\alpha\beta}_{ij}\delta^{(3)}(x,y)
    \label{eq:4.9}
\end{equation}
where $\mathbf{P}^{\alpha\beta}_{ij}$ is the projection operator onto the subspace of trace free Rarita-Schwinger fields, i.e., $\rho_i=\mathbf{P}_{ij}\phi^j$. Due to the fact that, in contrast to (\ref{eq:4.7}), this indeed defines a projection now allows for a direct implementation in the quantum theory.\\
Before we do so, following \cite{Thiemann:1997rq}, we first exploit the fact that the $\phi_i$ (resp. $\rho_i$ and $\sigma$) are half densities and introduce new Grassmann-valued variables. For later purposes, in contrast to \cite{Thiemann:1997rq}, in view of the regularization of the supersymmetry constraint, we therefore triangulate the spatial slice $\Sigma$ by disjoint (again up to common faces, edges and vertices) tetrahedra $\Delta_i$ instead of boxes at countably infinite discrete points $x_i\in\Sigma$, $i\in\mathcal{I}$ ($|\mathcal{I}|=\aleph_0$), and coordinate volume $\delta_{i}^3/6$ such that $\Sigma=\bigcup_{i\in\mathcal{I}}\Delta_{i}$. Here, $\delta_{i}>0$ $\forall i\in\mathcal{I}$ are small positive numbers determining the fineness of the triangulation. Then, for each $i\in\mathcal{I}$, we define \cite{Thiemann:1997rq}
\begin{equation}
    \theta^{(\delta_i)}(x_i):=\int_{\Sigma}{\mathrm{d}^3y\,\frac{\chi_{\delta_i}(x_i-y)}{\sqrt{\frac{\delta_i^3}{6}}}\phi(y)}
    \label{eq:4.9.1}
\end{equation}
where $\chi_{\delta_i}(x_i-y)$ is the characteristic function of the tetrahedron $\Delta_i$ centered at $x_i$. These satisfy the bracket relations
\begin{align}
    \{\theta^{(\delta_k)}_i(x_k),\theta^{(\delta_l)}_j(x_l)\}&=\int_{\Sigma}\mathrm{d}^3x\,\frac{\chi_{\delta_k}(x_k-x)}{\sqrt{\frac{\delta_k^3}{6}}}\int_{\Sigma}\mathrm{d}^3y\,\frac{\chi_{\delta_l}(x_l-y)}{\sqrt{\frac{\delta_l^3}{6}}}\{\phi_i(x),\phi_j(y)\}_{\mathrm{DB}}\nonumber\\
    &=\frac{i}{2}\left(\mathds{1}\delta_{ij}-\frac{1}{2}\gamma_i\gamma_j\right)\delta_{kl}\int_{\Sigma}\mathrm{d}^3x\,\frac{\chi_{\delta_k}(x_k-x)}{\delta_k^3/6}\nonumber\\
    &=\frac{i}{2}\left(\mathds{1}\delta_{ij}-\frac{1}{2}\gamma_i\gamma_j\right)\delta_{kl}
\end{align}
We then take the continuum limit $\sup_{i\in\mathcal{I}}\{\delta_i\}\rightarrow 0$ and set $\theta_i(x):=\lim_{\delta_x\rightarrow 0}\,\theta^{(\delta_x)}_i(x)$ $\forall x\in\Sigma$. Furthermore, setting $\theta^{(\rho)}_i(x):=\mathbf{P}_{ij}\theta^j(x)$ as well as $\theta^{(\sigma)}:=\gamma^i\theta_{j}(x)$, this finally yields
\begin{equation}
    \{\theta^{(\rho)}_i(x),\theta^{(\rho)}_{j}(y)\}=i\mathbf{P}_{ij}\delta_{x,y}\quad\text{and}\quad\{\theta^{(\sigma)}(x),\theta^{(\sigma)}(y)\}=-\frac{9i}{2}\mathds{1}\delta_{x,y}
\label{eq:4.8}
\end{equation}
together with the Majorana conditions $\theta^{(\rho)}_i(x)^*=\theta^{(\rho)}_i(x)$ and $\theta^{(\sigma)}(x)^*=\theta^{(\sigma)}(x)$ $\forall x,y\in\Sigma$. Hence, one ends up with an abstract CAR *algebra at any point $x\in\Sigma$. The quantization of the theory can be performed following \cite{Bodendorfer:2011pb}. We will sketch the main idea and also use this opportunity to point out some mathematical structure lying behind this quantization scheme which has a beautiful interpretation in the framework supergeometry and even naturally arises in the chiral approach (see \cite{Eder:2020erq}).\\
For any point $x\in\Sigma$ we choose the superspace $\mathbb{R}_x^{0|N}:=(\{x\},\Lambda_N)$, also called a \emph{superpoint}, with $N$ fermionic generators $\theta^A$, $A=1,\ldots,N$, whose sections $f\in\Lambda_N^{\mathbb{C}}:=\Lambda_N\otimes\mathbb{C}$ of the complexified function sheaf take the form 
\begin{equation}
    f=\sum_{I\in M_N}f_I\theta^I
\end{equation}
with $f_I\in\mathbb{C}$ for all multi-indices $I\in M_N$ of length $0\leq |I|\leq N$. On the superspace one has the standard translation-invariant super scalar product $\mathscr{S}:\,\Lambda_N^{\mathbb{C}}\times\Lambda_N^{\mathbb{C}}\rightarrow \mathbb{C}$ given by the \emph{Berezin integral}
\begin{equation}
    \mathscr{S}(f,g):=\int_{B}{\mathrm{d}\theta^1\cdots\mathrm{d}\theta^N\,\bar{f}g},\quad\forall f,g\in\Lambda_N^{\mathbb{C}}
\label{eq:4.10}
\end{equation}
This gives the space $(\Lambda_N^{\mathbb{C}},\mathscr{S})$ the structure of a \emph{Krein space}, i.e., an indefinite inner product space for which there exists an endomorphism $\mathbf{S}\in\mathrm{End}(\Lambda_N^{\mathbb{C}})$ such that $\mathscr{S}(\,\cdot\,,\mathbf{S}\,\cdot\,)$ defines a positive definite scalar product on $\Lambda_N^{\mathbb{C}}$. The choice of such an endomorphism $\mathbf{S}$ is not unique but is strongly restricted by the implementation of reality conditions. A standard choice of a scalar product is given by identifying $\Lambda_N^{\mathbb{C}}\cong\mathbb{C}^{2^N}$ and setting 
\begin{equation}
   \braket{f,g}:=\sum_{I\in M_N}\bar{f}_Ig_I 
   \label{eq:4.11}
\end{equation}
It follows, even for general super Lie groups, that there always exists an endomorphism $\mathbf{S}$ on $\Lambda_N^{\mathbb{C}}$ such that\footnote{For this situation, such an endomorphism has been in fact constructed explicitly in \cite{Bodendorfer:2011pb}, although their definition of the super scalar product differs from the definition chosen here.}
\begin{equation}
    \braket{\,\cdot\,,\,\cdot\,}=\mathscr{S}(\,\cdot\,,\mathbf{S}\,\cdot\,)
    \label{eq:4.11.1}
\end{equation}
Hence, this yields a Hilbert space $\mathcal{H}^N_x:=(\Lambda_N^{\mathbb{C}},\braket{\,\cdot\,,\,\cdot\,})$. On $\mathcal{H}^N_x$ we define the multiplication operators $\widehat{\theta}^A$ as well as odd derivations $\partial_A\equiv\frac{\partial}{\partial\theta^A}$ for $A=1,\ldots,N$ via
\begin{equation}
    \widehat{\theta}^Af:=\theta^A f\quad\text{and}\quad\partial_A\theta^B:=\delta_A^B
\end{equation}
$\forall f\in\Lambda_N^{\mathbb{C}}$. As shown in \cite{Bodendorfer:2011pb}, due to the choice of the scalar product (\ref{eq:4.11}), these operators are indeed self-adjoint on $\mathcal{H}^N_x$. With these ingredients, one can then construct a faithful representation of the CAR *-algebra (\ref{eq:4.8}). Therefore, one takes the tensor product Hilbert space $\mathcal{H}_x:=\mathcal{H}_x^N\otimes\mathcal{H}_x^M$ with $N=12$ and $M=4$ and defines
\begin{equation}
    \widehat{\theta}^{(\rho)\alpha}_i(x):=\mathbf{P}^{\alpha\beta}_{ij}\left[\sqrt{\frac{\hbar}{2}}(\theta^{j}_{\beta}+\partial^{ j}_{\beta})\right]\quad\text{and}\quad\widehat{\theta}^{(\sigma)\alpha}(x):=\frac{3\sqrt{\hbar}}{2}(\theta^{\alpha}+\partial^{\alpha})
    \label{eq:4.11.2}
\end{equation}
on $\mathcal{H}_x^N$ and $\mathcal{H}_x^M$, respectively. By construction, these operators are then self-adjoint as required by the Majorana conditions and moreover satisfy the anticommutation relations
\begin{equation}
    [\widehat{\theta}^{(\rho)}_i(x),\widehat{\theta}^{(\rho)}_j(x)]=\hbar\mathbf{P}_{ij}\quad\text{and}\quad[\widehat{\theta}^{(\sigma)}_i(x),\widehat{\theta}^{(\sigma)}(x)]=\frac{9\hbar}{2}\mathds{1}
\end{equation}
The quantized Rarita-Schwinger field on $\mathcal{H}_x$ is then given by 
\begin{equation}
    \widehat{\theta}_i(x):=\widehat{\theta}^{(\rho)}_i(x)+\frac{1}{3}\gamma_i\widehat{\theta}^{(\sigma)}(x)
\end{equation}
This construction then takes over to a family of points $\{x_1,\ldots,x_k\}$ yielding the tensor product Hilbert space $\mathcal{H}_{\{x_1,\ldots,x_k\}}:=\bigotimes_{i=1}^k\mathcal{H}_{x_i}$. The fermionic Hilbert space $\mathcal{H}_f$ is then obtained as the inductive limit over the corresponding family of Hilbert spaces $\mathcal{H}_{\{x_1,\ldots,x_k\}}$.   
\subsection{Quantization of the SUSY constraint}\label{section 5.2}
\subsubsection{Part I}
Having derived the compact expression (\ref{eq:3.22}) of the classical Supersymmetry constraint with half-densitized fermionic fields, we next want to find an implementation in the quantum theory. As stated in \cite{Sawaguchi:2001wi}, the Poisson bracket of the SUSY constraint with itself should be proportional to the Hamiltonian constraint modulo Gauss and diffeomorphism constraint. Hence, in the quantum theory, it is expected that, on the subspace of gauge and diffeomorphism invariant states, the commutator of the SUSY constraint operator reproduces the Hamiltonian constraint operator. This is in fact a very interesting and important feature in canonical supergravity theories as this provides a very strong relationship between both operators and thus serves as a consistency condition in the quantum theory. This may also fix some of the quantization ambiguities. In fact, in the framework of self-dual loop quantum cosmology, for a certain subclass of symmetry reduced models, it was shown explicitly in \cite{Eder:2020okh} that this strong relationship even holds exactly in the quantum theory. More precisely, it is shown that the (graded) commutator between the SUSY constraints exactly reproduces the classical Poisson relation.\\ 
Another point of view is that the SUSY constraint is superior to the Hamiltonian constraint in the sense that once the SUSY constraint is quantized (or even solved) this immediately yields the quantization (or solution) of the Hamiltonian constraint by computing the commutator. For this reason, it is desirable to quantize the SUSY constraint in a way that does not involve the Hamiltonian constraint. For instance, it should not depend on the extrinsic curvature as this, via Thiemann's proposal, would involve commutators with the Euclidean part of the Hamiltonian.         
On the other hand, in order to be able to compare it with the Hamiltonian constraint, it is desirable to find an as compact expression as possible.\\
In the following, we will propose a specific quantization scheme of the SUSY constraint that does not involve the Hamiltonian constraint. As a first step, let us therefore consider the first part in the classical expression (\ref{eq:3.22}) depending on the covariant derivative of the fermionic fields\footnote{in order to simplify our notation in what follows, the prime indicating the transformed Ashtekar connection in case of half-densitized fermionic variables will be dropped.} 
\begin{equation}
    S^{(1)}[\eta]:=\int_{\Sigma}{\mathrm{d}^3x\,\bar{\eta}i\tensor{\epsilon}{^{abc}}e_a^i\gamma_i\gamma_{*}D_b^{(A)}\left(\frac{1}{\sqrt[4]{q}}e^j_c\phi_j\right)}
    \label{eq:3.1.1}
\end{equation}
This expression looks quite similar to the Dirac Hamiltonian studied for instance in \cite{Thiemann:1997rt} with the crucial difference that in (\ref{eq:3.1.1}) the conjugate spinor $\bar{\eta}$ now plays the role a smearing function and thus is not a dynamical variable. Hence, in contrast to \cite{Thiemann:1997rt}, we cannot change its density weight going over to half-densities for the regularization as this will change the density weight of the constraint operator as a whole. Moreover, changing the density weight of the smearing function may change the constraint algebra which should be avoided. Hence, particular attention is required for its regularization.

\begin{figure}
     \centering
     \includegraphics[height=6cm]{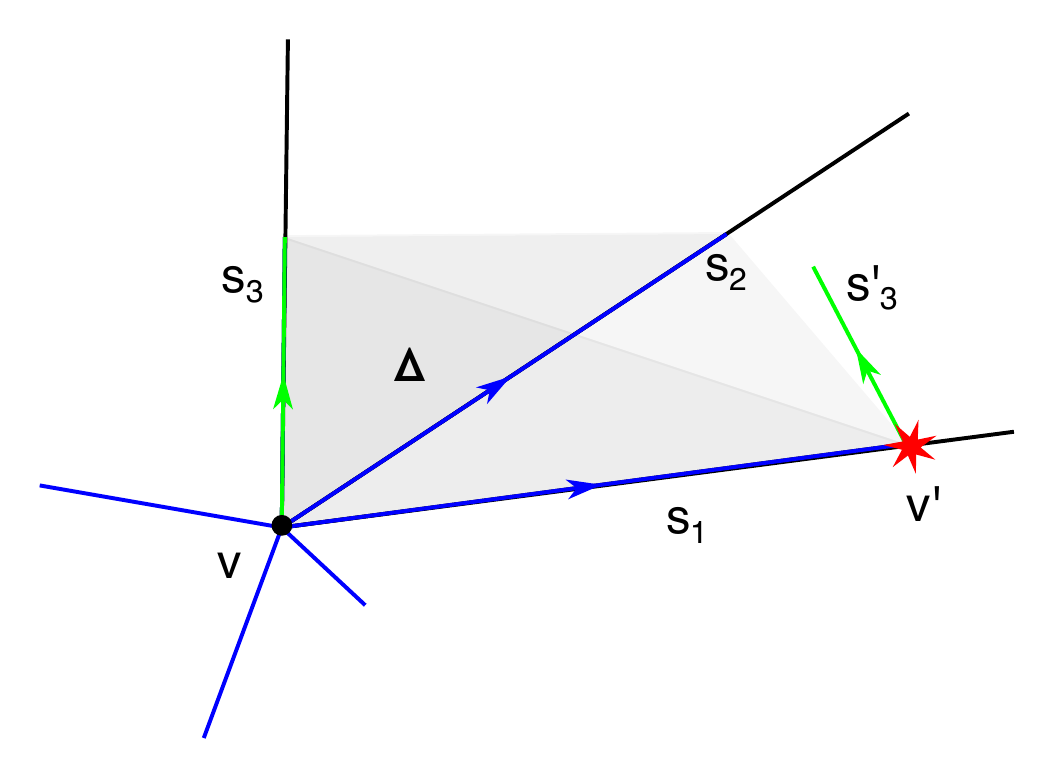}
     \caption{A tetrahedron $\Delta$ with the edges used for the regularization. The star marks the location of the fermion operator.}
     \label{fig:tet}
 \end{figure}

We will proceed in analogy with \cite{Thiemann:1996aw}, i.e., we will consider triangulations adapted to a graph $\gamma$. First, we describe triangulations of the neighborhood of a vertex $v$ of $\gamma$ that are labeled by a triplet of edges $(e_I,e_J,e_K)$ at $v$. We will keep track of the of fineness of these triangulations, measured in a fixed fiducial metric around the vertex, in terms of a parameter $\delta>0$.   
%
%
\begin{enumerate}[label=(\roman*)]
 	\item All edges of the graph are assumed be outgoing in the sense that if $e$ is an edge with vertices $v,v'$ as endpoints, subdivide it into two new edges $e_1$ and $e_2$ such that $e=e_1\circ e_2$ and $e_1$ and $e_2$ are outgoing at $v$ and $v'$. respectively.
	\item Given an edge $e_I$ incident at a vertex $v$, choose a segment $s_I:\,[0,1]\rightarrow \Sigma$ of $e_I$ such that $s_I$ is also incident and outgoing at $v$ and such that it does not include any other endpoint of the edge $e_I$.
 	\item In order to treat all edges of the graph equally, at each vertex $v$, let $(e_I,e_J,e_K)$ be an arbitrary triple of mutually distinct edges incident at the common vertex $v$.\footnote{If the vertex is two-valent, one can adjoin a third edge in an arbitrary manner. However, it will become clear below that the action of the operator on such vertices is trivial.} For each triple, we chose  corresponding segments $(s_I,s_J,s_K)$ shorter than $\delta$. They  span a tetrahedron $\Delta$ with basepoint $v(\Delta)=v$ (see figure \ref{fig:tet}), where the missing three edges of $\Delta$ are chosen in a diffeomorphism covariant way\cite{Thiemann:1996aw}. Furthermore, we assume that the triple is ordered in such a way that the tangents of the segments are positively oriented, i.e., $\mathrm{det}(\dot{s}_I,\dot{s}_J,\dot{s}_K)>0$.
	\item  Let $(e_I,e_J,e_K)$ be a positively oriented triple of edges as in (iii) with corresponding segments $(s_I,s_J,s_K)$. For any $\delta>0$, we introduce another segment $s'_K:\,[0,1]\rightarrow \Sigma$ which is incident and at outgoing at $s_I(1)$ in such a way, that in the limit $\delta\rightarrow 0$, $s'_K$  converges to the segment $s_K$ (see figure \ref{fig:tet}). As it will become clear in what follows, the end result will not depend on the specific choice of such an additional edge provided it satisfies the requirements just mentioned.     
	\item To obtain a triangulation $T(\gamma, v,\delta,IJK)$ of a neighborhood of $v$, we proceed as in \cite{Thiemann:1996aw} and construct seven additional (``mirror'') tetrahedra.
\end{enumerate}
We will now write down a regularization of the classical expression (\ref{eq:3.1.1}), using some triangulation $T(\delta)$ of fineness $\delta$. Let $\Delta_i$ be a tetrahedron from this triangulation spanned by some triplet $(s_I,s_J,s_K)$ of edges. We will additionally assume that edges $s'_I$ have been chosen according to (iv) above. As usual, we apply Thiemann's trick and replace the frame fields $e_i^a$ by the Poisson bracket of the connection with the volume 
\begin{equation}
    2e_a^i=\frac{1}{\kappa}\{A_a^i,V\}=\frac{1}{\kappa}\{A_a^i,V(x,\delta)\}
    \label{eq:3.1.2}
\end{equation}
where
\begin{equation}
    V(x,\delta):=\int_{\Sigma}\mathrm{d}^3y\,\chi_{\delta}(x,y)\sqrt{q(y)}
    \label{eq:3.1.3}
\end{equation}
is the volume of the tetrahedron $\Delta$ containing $x\in\Sigma$, with $\chi_{\delta}$ its characteristic function, such that in the limit $\delta\rightarrow 0$ one has $\lim_{\delta\rightarrow 0}\frac{6}{\delta^3}V(x,\delta)=\sqrt{q(x)}$. For $\delta> 0$ small enough, the holonomy  $h_{s}[A]$ along any segment $s$ in triple can approximately be written as $h_{s}[A]=\mathds{1}+\delta\dot{s}^a A_a^i\tau_i+\mathcal{O}(\delta^2)$ such that, using $\mathrm{tr}(\tau_i\tau_j)=-\frac{1}{2}\delta_{ij}$, it follows
\begin{equation}
   2\mathrm{tr}(\tau_i h_s[A]\{h_s[A]^{-1},V(x,\delta)\})=\delta_{ij}\delta \dot{s}^a\{A_a^j(x),V(x,\delta)\}
   \label{eq:3.1.4}
\end{equation}
This enables one to express (\ref{eq:3.1.2}) in terms of holonomies and fluxes with the latter implicitly contained in the definition of the volume.\\
Finally, in order regulate the covariant derivative in (\ref{eq:3.2.2}), for any segment $s$, let 
\begin{equation}
    H_s[A]:=\mathcal{P}\exp\left(\int_{s}\kappa_{\mathbb{R}*}(A)\right)
    \label{eq:3.1.5}
\end{equation}
be the holonomy of $A$ in the $\mathfrak{su}(2)$ sub representation of the real Majorana representation $\kappa_{\mathbb{R}*}$ which, according to (\ref{eq:2.28}), in the chiral representation consists of a direct sum of two spin-$\frac{1}{2}$ representations such that, w.r.t. this representation, $H_s[A]=\mathrm{diag}(h_s[A],h_s[A])$ is in fact block-diagonal. Again, in the limit of small $\delta>0$, the holonomy can approximately be written in the form $H_s[A]=\mathds{1}+\delta\dot{s}^a \frac{i}{2}\gamma_*\gamma_{0i}A_a^i+\mathcal{O}(\delta^2)$ which yields
\begin{equation}
    H_s[A](0,\delta)\Psi(s(\delta))-\Psi(s(0))=\delta\dot{s}^a(0)(D^{(A)}_a\Psi)(s(0))
    \label{eq:3.1.6}
\end{equation}
where $\Psi$ stands for an arbitrary spinor-valued field defined on $\Sigma$. With these preparations, we are now ready to write down a regularization of (\ref{eq:3.1.1}). Given the triangulation $T(\delta)$ 
of fineness $\delta >0$, we set 
\begin{align}
    S^{(1)}_{\delta}[\eta]:=\frac{1}{6\kappa^2}\sum_{\Delta_i\in T(\gamma,\delta)}
    \bar{\eta}(x_i)i\epsilon^{IJK}\mathrm{tr}(\tau_j h_{s_I(\Delta_i)}\{h_{s_I(\Delta_i)}^{-1},V(x_i,\delta)\})\gamma_j\gamma_*[\mathscr{X}_K(s_J(\Delta_i))-\mathscr{X}_K(x_i)]
    \label{eq:3.1.7}
\end{align}
with
\begin{equation}
    \mathscr{X}_K(s_J(\Delta_i)):=\frac{\mathrm{tr}(\tau_k h_{s'_{K}(\Delta_i)}\{h_{s'_K(\Delta_i)}^{-1},V(s_J(\Delta_i),\delta)\})}{\sqrt{V(s_J(\Delta_i),\delta)}}H_{s_J(\Delta_i)}\theta^{\delta}_k(s_J(\Delta_i)(\delta))
    \label{eq:3.1.8}
\end{equation}
and
\begin{equation}
    \mathscr{X}_K(x_i):=\frac{\mathrm{tr}(\tau_k h_{s_{K}(\Delta_i)}\{h_{s_{K}(\Delta_i)}^{-1},V(x_i,\delta)\})}{\sqrt{V(x_i,\delta)}}\theta^{\delta}_k(x_i)
    \label{eq:3.1.9}
\end{equation}
where in (\ref{eq:3.1.7}) for any basepoint $x_i\equiv v(\Delta_i)$, we have chosen a particular triple of segments $(s_I(\Delta_i),s_J(\Delta_i),s_K(\Delta_i))$ incident at $x_i$ and an additional segment $s'_K$ such that the above requirements are satisfied. First, let us show that (\ref{eq:3.1.7}) indeed provides a regularization of (\ref{eq:3.1.1}). Therefore, we use the fact that, by property (iv), $s_K'$ converges to $s_K$ in the limit $\delta\rightarrow 0$ such that for small $\delta$, due to (\ref{eq:3.1.6}), we can approximately write
\begin{align}
    \mathscr{X}_K(s_J(\Delta_i))-\mathscr{X}_K(x_i)\approx \delta^2\dot{s}_J^{b}(\Delta_i)\dot{s}_{K}^{c}(\Delta_i)D^{(A)}_{b}\left(\frac{\{A_{c}^k,V(x_i,\delta)\}}{\sqrt{V(x_i,\delta)}}\theta_k^{\delta}(x_i)\right)
    \label{eq:3.1.10}
\end{align} 
Recall that, by (\ref{eq:4.9.1}), $\theta^{\delta}_i$ is defined as
\begin{equation}
    \theta_i^{\delta}(x)=\int{\mathrm{d}^3y\,\frac{\chi_{\delta}(x-y)}{\sqrt{\frac{\delta^3}{6}}}\phi_i(y)}
    \label{eq:3.1.11}
\end{equation}
so that, using $\partial_{x^a}\chi_{\delta}(x-y)=-\partial_{y^a}\chi_{\delta}(x-y)$ \cite{Thiemann:1997rt}, it follows
\begin{equation}
    \partial_{x^a}\theta_i^{\delta}(x)=-\int_{\Sigma}{\mathrm{d}^3y\,\frac{\partial_{y^a}\chi_{\delta}(x-y)}{\sqrt{\frac{\delta^3}{6}}}\phi_i(y)}=\int_{\Sigma}{\mathrm{d}^3y\,\frac{\chi_{\delta}(x-y)}{\sqrt{\frac{\delta^3}{6}}}\partial_{y^a}\phi_i(y)}
\end{equation}
Hence, if $\mathcal{B}^k(x_i)$ denotes the term inside the covariant derivative of (\ref{eq:3.1.10}) depending on the volume $V(x_i,\delta)$, we can rewrite (\ref{eq:3.1.10}) as 
\begin{align}
    D_{a}^{(A)}\left(\mathcal{B}^k(x_i)\theta^{\delta}_k(x_i)\right)=(\partial_{x^{a}}\mathcal{B}^k)(x_i)\theta^{\delta}_k(x_i)+\mathcal{B}^k(x_i)\partial_{x^{a}}\theta^{\delta}_k(x_i)+\mathcal{B}^k(x_i)\frac{i}{2}\gamma_*\gamma_{0i}A_a^i(x_i)\theta^{\delta}_k(x_i)\nonumber\\
    =\int{\mathrm{d}^3y\,\frac{\chi_{\Delta}(x_i-y)}{\sqrt{\frac{\delta^3}{6}}}\left((\partial_{x^{a}}\mathcal{B}^k)(x_i)\theta^{\delta}_k(y)+\mathcal{B}^k(x_i)\partial_{x^{a}}\theta^{\delta}_k(y)+\mathcal{B}^k(x_i)\frac{i}{2}\gamma_*\gamma_{0i}A_a^i(x_i)\partial_{y^a}\phi_i(y)\right)}
    \label{eq:3.1.12}
\end{align}
By definition, for small $\delta$ we have  $V(x_i,\delta)\approx\frac{\delta^3}{6}\sqrt{q(x_i)}$. Hence, approximating the denominator in $\mathcal{B}^k(x_i)$ by $\sqrt{\delta^3/{6}}\sqrt[4]{q(x_i)}$ and inserting it into equation (\ref{eq:3.1.12}) and finally using the fact that in the limit $\delta\rightarrow 0$ one has $\chi_{\delta}(x_i-y)/\frac{\delta^3}{6}\rightarrow \delta(x_i-y)$, (\ref{eq:3.1.7}) becomes
\begin{align}
    \frac{1}{24\kappa^2}\lim_{\delta\rightarrow 0}\sum_{\Delta_i\in T(\gamma,\delta)}
    \bar{\eta}(x_i)&i\{A^j_{a}(x_i),V(x_i,\delta)\})\gamma_j\gamma_*D^{(A)}_{b}\left(\frac{\{A_{c}^k,V(x_i,\delta)\}}{\sqrt[4]{q(x_i)}}\phi_k(x_i)\right)\times\label{eq:3.1.13}\\
    &\times\epsilon^{IJK}\delta^3{s}_{I}^a(\Delta_i)\dot{s}_J^{a_1}(\Delta_i)\dot{s}_{K}^{a_2}(\Delta_i)\dot\nonumber
\end{align}
Hence, if we finally use
\begin{equation}
    \epsilon^{IJK}\delta^3{s}_{I}^a(\Delta_i)\dot{s}_J^{a_1}(\Delta_i)\dot{s}_{K}^{a_2}(\Delta_i)=\epsilon^{abc}\delta^3\mathrm{det}(\dot{s}_I,\dot{s}_J,\dot{s}_K)(\Delta_i)=6\epsilon^{abc}\mathrm{vol}(\Delta_i)
    \label{eq:3.1.12.1}
\end{equation}
equation (\ref{eq:3.1.13}) takes the form of a Riemann sum which in the limit $\delta\rightarrow 0$ converges to a Riemann integral which precisely coincides with expression (\ref{eq:3.1.1}). That is, we found
\begin{equation}
    \lim_{\delta\rightarrow 0}\,S^{(1)}_{\delta}[\eta]=S^{(1)}[\eta]
\end{equation}
Hence, we can use \eqref{eq:3.1.7} as a staring point for the quantization. Therefore, we apply the identity 
\begin{equation}
    \{A_a^i,\sqrt{V(x,\delta)}\}=\frac{1}{2\sqrt{V(x,\delta)}}\{A_a^i,V(x,\delta)\}
\end{equation}
in order to express (\ref{eq:3.1.7}) resp. (\ref{eq:3.1.8}) purely in terms of Poisson brackets between holonomies and volume. The corresponding quantum operator is then obtained by replacing the classical phase space variables by their respective quantum counterparts and replacing the Poisson bracket by the commutator $\{\cdot\,,\,\cdot\}\rightarrow \frac{1}{i\hbar}[\cdot\,,\,\cdot]$. 

At this point we have to pause, however, since we have to specify the triangulation $T(\delta)$ in adaptation to the graph $\gamma$. To do this, We follow precisely the procedure from \cite{Thiemann:1996aw}: Triangulations around the vertices are chosen as $T(\gamma, v,\delta,IJK)$, and the rest of the space triangulated arbitrarily. Finally an averaging over $I,J,K$ at each vertex is carried out. To write out this averaging, we denote by $E(v)$ the number of triples at the given vertex. With this procedure, we end up with 
\begin{align}
    \widehat{S}^{(1)}_{\delta}[\eta]:=-\frac{1}{3\hbar^2\kappa^2}\sum_{v\in\gamma}
    \frac{8}{E(v)}
    \bar{\eta}(x_i)i\epsilon^{IJK}\gamma_j\gamma_*[\widehat{\mathscr{X}}_K(s_J(\Delta))-\widehat{\mathscr{X}}_K(x)]\mathrm{tr}(\tau_j h_{s_K(\Delta)}[h_{s_I(\Delta)}^{-1},\widehat{V}_v])
    \label{eq:3.1.14}
\end{align}
with
\begin{equation}
    \widehat{\mathscr{X}}_K(s_J(\Delta)):=\mathrm{tr}(\tau_k h_{s'_{K}(\Delta)}[h_{s'_K(\Delta)}^{-1},\sqrt{\widehat{V}}_{s_J(\Delta)}])H_{s_J(\Delta)}\widehat{\theta}_k(s_J(\Delta))
    \label{eq:3.1.15}
\end{equation}
and
\begin{equation}
    \widehat{\mathscr{X}}_K(x):=\mathrm{tr}(\tau_k h_{s_{K}(\Delta)}[h_{s_{K}(\Delta)}^{-1},\sqrt{\widehat{V}}_v])\widehat{\theta}_k(v)
    \label{eq:3.1.16}
\end{equation}
where for reasons that will become clear below, the first factor in the classical expression (\ref{eq:3.1.7}) depending on the volume has been ordered to the right.\\
Note that in (\ref{eq:3.1.14}) we have implicitly assumed that the discrete sum over all tetrahedra in the triangulation collapses to a sum over the vertices of the underlying spin network graph $\gamma$. This is permissible in case of the Ashtekar-Lewandowski volume operator as this operator acts trivially on planar vertices. However, this also implies that the operator $\widehat{\mathscr{X}}_K(s_J(\Delta))$ in (\ref{eq:3.1.15}) becomes trivial as $\sqrt{\widehat{V}}_{s_J(\Delta)}$ acts on a vertex with coplanar tangent vectors. But then $\widehat{\mathscr{X}}_K(s_J(\Delta))-\widehat{\mathscr{X}}_K(x)$ is not a difference operator and therefore this would not resemble a quantization of a regularized covariant derivative. A resolution would be to quantize a different classical quantity in which the covariant derivative operator acts directly on the Rarita-Schwinger field. The regularization can then be performed as described above. However, we would like to keep the SUSY constraint operator as simple as possible. For this reason, we consider another possibility ensuring nontriviality of the action of $\widehat{\mathscr{X}}_K(s_J(\Delta))$. Therefore, let us choose instead the Rovelli-Smolin variant of the volume operator \cite{Rovelli:1994ge,DePietri:1996tvo,Lewandowski:1996gk}. This operator is defined on cylindrical functions $\Psi_{\gamma}$ according to \cite{Lewandowski:1996gk,Ashtekar:1997fb}
\begin{equation}
    \widehat{V}_v\Psi_{\gamma}:=\sum_{v\in\gamma}\sqrt{|\widehat{q}_v|}\Psi_{\gamma}
    \label{eq:3.2.1.0}
\end{equation}
with $|\widehat{q}_v|$ defined as
\begin{equation}
    |\widehat{q}_v|:=\frac{1}{48}\sum_{I\neq J\neq K\neq I}|\widehat{q}_{IJK}|:=\frac{1}{48}\sum_{I\neq J\neq K\neq I}|\epsilon_{ijk}J^i_I J^j_J J^k_K|
    \label{eq:3.2.1.1}
\end{equation}
where the sum is taken over all possible triples $(e_I,e_J.e_K)$ of mutually distinct edges at $v$. The operator $\widehat{q}_{IJK}$ can also be written in the form
\begin{equation}
    q_{IJK}=\epsilon^{ijk}J^i_IJ^j_JJ^k_K=\frac{i}{4}[(J_{IJ})^2,(J_{JK})^2]
    \label{eq:3.2.1}
\end{equation}
with $(J_{IJ})^2:=(J_I+J_J)^2$ the Casimir operator corresponding to the total angular momentum $J_{IJ}:=J_I+J_J$. Note that the modulus appears inside the sum. For this reason, the action of the Rovelli-Smolin volume operator on vertices with coplanar tangent vectors is in general nontrivial. At first sight, this seems to be a problem as then the sum in (\ref{eq:3.1.14}) would also include basepoints of tetrahedra located inside a given edge of a spin network graph, i.e., the sum would be a priori infinite. However, due to our choice of the factor ordering, we will see that this indeed not the case. 
\begin{figure}
    \centering
    \includegraphics[height=3cm]{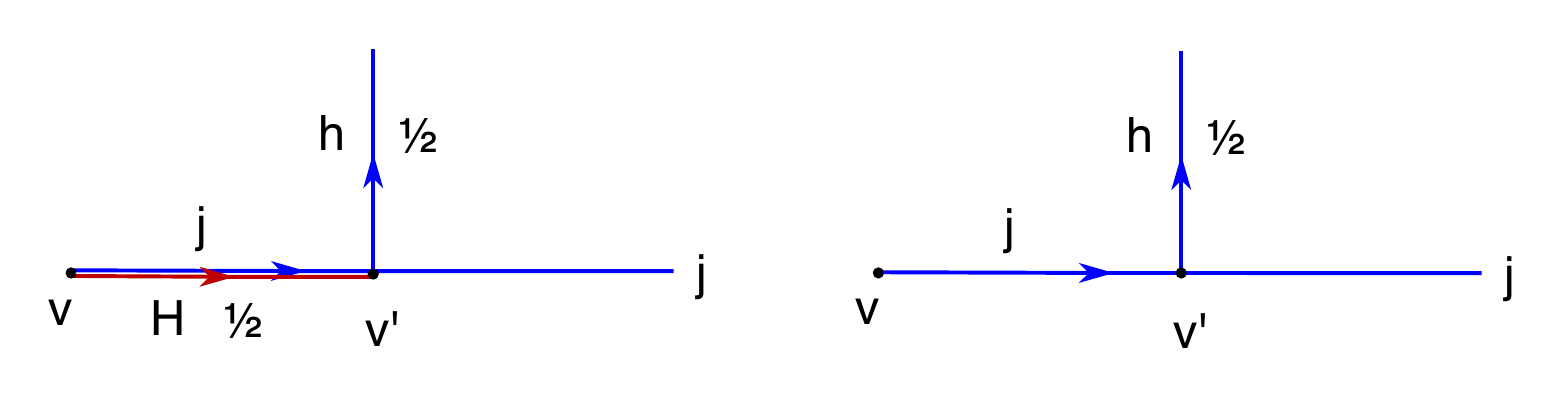}
    \caption{Illustrations of the action of $\widehat{S}^{(1)}[\eta]$ on spin network states. The picture on the right shows the action of the trace operator $\widehat{\mathcal{O}}$ defined in (\ref{eq:3.2.2}) creating a new vertex $v'$ by adding a new edge labeled with spin-$1/2$. The picture on the left illustrates the action of $\widehat{\mathscr{X}}_K(s_J(\Delta))$ in (\ref{eq:3.1.14}) which, in contrast to $\widehat{\mathcal{O}}$, creates two new spin-$1/2$ edges at $v'$, one parallel and one transversal to the spin-network edge $j$. }
    \label{fig:volume_sketch}
\end{figure}
Therefore, let us consider the operator
\begin{align}
    \widehat{\mathcal{O}}:=\mathrm{tr}\left(\tau_i h_e[h_e^{-1},\sqrt{\widehat{V}}]\right)
    \label{eq:3.2.2}
\end{align}
appearing for instance to the right in (\ref{eq:3.1.14}) where the holonomy $h_e$ is taken along an edge $e$ incident at a vertex sitting inside an spin network edge and which is transversal to that particular edge (see figure \ref{fig:volume_sketch}). Given a spin network state $\Psi_{\gamma}$, this operator will take the form
\begin{align}
    \widehat{\mathcal{O}}\Psi_{\gamma}&=\mathrm{tr}\left(\tau_ih_e[h_e^{-1},\sqrt{\widehat{V}}]\right)\Psi_{\gamma}=\mathrm{tr}(\tau_i)\sqrt{\widehat{V}}\Psi_{\gamma}-\mathrm{tr}\left(\tau_ih_e\sqrt{\widehat{V}}h_e^{-1}\right)\Psi_{\gamma}\nonumber\\
    &=-\tensor{{\tau_i}}{^k_l}\tensor{{h_e}}{^l_m}\sqrt{\widehat{V}}\tensor{{h_e^{-1}}}{^m_k}\Psi_{\gamma}
    \label{eq:3.2.3}
\end{align}
where the first term in second equation vanishes due the trace freeness of the Pauli matrices. Since the the matrix components of a holonomy $\tensor{{h_e[A]}}{^m_k}=\tensor{{\pi_{\frac{1}{2}}(h_e[A])}}{^m_k}$ can be identified with the matrix components of the spin-$\frac{1}{2}$ representation, it follows
\begin{align}
    (\widehat{\mathcal{O}}\Psi_{\gamma})[A]=-\tensor{{\tau_i}}{^k_l}\tensor{{\pi_{\frac{1}{2}}(h_e^{-1}[A])}}{^l_m}\sqrt{\widehat{V}}\left(\tensor{{\pi_{\frac{1}{2}}(h_e[A])}}{^m_k}\Psi_{\gamma}[A]\right)
    \label{eq:3.2.4}
\end{align}
Hence, according to (\ref{eq:3.2.4}), the holonomy $h_e$ adds a new edge to the spin network graph $\gamma$ with spin quantum number $j=\frac{1}{2}$ (see figure \ref{fig:volume_sketch}). To evaluate the action of the volume operator, note that, effectively, the state located at the new created vertex can symbolically be written in the form
\begin{equation}
   \Psi_{j_{12}}:=\ket{(j_1j_2)j_{12},\frac{1}{2};\underline{j}\underline{m}}
   \label{eq:3.2.4.1}
\end{equation}
with $j_1=j_2=j$ the spin quantum number of the original spin network edge with $j_{12}=0$ (for divalent spin network vertices) and $j_3=\frac{1}{2}$ the spin quantum number of the new created edge. For later purposes, it is worthwhile to keep the computation a bit more general and assume that $j_1$ and $j_2$ are not necessarily equal (therefore $j_{12}$ does not have be zero). For the vertex under consideration, the operator (\ref{eq:3.2.1}) takes the form
\begin{equation}
    \widehat{q}_{123}=:\widehat{q}=\frac{i}{4}[(J_{12})^2,(J_{23})^2]
    \label{eq:3.2.4.2}
\end{equation}
Hence, in order to determine its action on (\ref{eq:3.2.4.1}), we have to perform a recoupling of angular momenta by coupling $j_2$ and $j_3$. This can be done using the Wigner 6-$j$ symbols which yields
\begin{align}
    \Psi_{j_{12}}=\sum_{j_{23}}{(-1)^{j_1+j_2+\frac{1}{2}+\underline{j}}\sqrt{(2j_{12}+1)}\sqrt{(2j_{23}+1)}\sj{\frac{1}{2}}{j_{12}}{\underline{j}}{j_1}{j_{23}}{j_2}}\ket{j_1,(j_2\,\frac{1}{2})j_{23};\underline{j}\underline{m}}
    \label{eq:3.2.5}
\end{align}
In this form, it is particularly easy to compute the action of $(J_{12})^2$ which gives
\begin{align}
    (J_{23})^2\Psi_{j_{12}}=&\sum_{j_{23}}{(-1)^{j_1+j_2+\frac{1}{2}+\underline{j}}j_{23}(j_{23}+1)\sqrt{(2j_{12}+1)}\sqrt{(2j_{23}+1)}\sj{\frac{1}{2}}{j_{12}}{\underline{j}}{j_1}{j_{23}}{j_2}}\ket{j_1,(j_2\,\frac{1}{2})j_{23};\underline{j}\underline{m}}\nonumber\\
    =&\sum_{j_{23}}(-1)^{j_1+j_2+\frac{1}{2}+\underline{j}}j_{23}(j_{23}+1)\sqrt{(2j_{12}+1)}\sqrt{(2j_{23}+1)}\sj{\frac{1}{2}}{j_{12}}{\underline{j}}{j_1}{j_{23}}{j_2}\times\nonumber\\
    &\times\sum_{j'_{12}}(-1)^{j_1+j_2+\frac{1}{2}+\underline{j}}\sqrt{(2j'_{12}+1)}\sqrt{(2j_{23}+1)}\sj{\frac{1}{2}}{j'_{12}}{\underline{j}}{j_1}{j_{23}}{j_2}\ket{(j_1j_2)j'_{12}.\frac{1}{2};\underline{j}\underline{m}}\nonumber\\
    =&\sqrt{(2j_{12}+1)}\sum_{j_{23},j'_{12}}j_{23}(j_{23}+1)(2j_{23}+1)\sqrt{(2j'_{12}+1)}\sj{\frac{1}{2}}{j_{12}}{\underline{j}}{j_1}{j_{23}}{j_2}\sj{\frac{1}{2}}{j'_{12}}{\underline{j}}{j_1}{j_{23}}{j_2}\Psi_{j'_{12}}
    \label{eq:3.2.6}
\end{align}
where in the last line we have again performed a recoupling by coupling $j_1$ with $j_2$. This immediately yields
\begin{align}
    (J_{12})^2[(J_{23})^2\Psi_{j_{12}}]=&\sqrt{(2j_{12}+1)}\sum_{j'_{12}}j'_{12}(j'_{12}+1)\sqrt{(2j'_{12}+1)}\times\nonumber\\
    &\times\sum_{j_{23}}j_{23}(j_{23}+1)(2j_{23}+1)\sj{\frac{1}{2}}{j_{12}}{\underline{j}}{j_1}{j_{23}}{j_2}\sj{\frac{1}{2}}{j'_{12}}{\underline{j}}{j_1}{j_{23}}{j_2}\Psi_{j'_{12}}
    \label{eq:3.2.7}
\end{align}
Remains to evaluate the last term in the commutator of (\ref{eq:3.2.4.2}). In a similar way as above, one finds
\begin{align}
    (J_{23})^2[(J_{12})^2\Psi_{j_{12}}]=&j_{12}(j_{12}+1)(J_{23})^2\Psi_{j_{12}}=j_{12}(j_{12}+1)\sqrt{(2j_{12}+1)}\sum_{j'_{12}}\sqrt{(2j'_{12}+1)}\times\nonumber\\
    &\times\sum_{j_{23}}j_{23}(j_{23}+1)(2j_{23}+1)\sj{\frac{1}{2}}{j_{12}}{\underline{j}}{j_1}{j_{23}}{j_2}\sj{\frac{1}{2}}{j'_{12}}{\underline{j}}{j_1}{j_{23}}{j_2}\Psi_{j'_{12}}
    \label{eq:3.2.8}
\end{align}
Hence, we found
\begin{align}
    \widehat{q}\Psi_{j_{12}}=&\frac{i}{4}[(J_{12})^2,(J_{23})^2]\Psi_{j_{12}}=\frac{i}{4}\sqrt{(2j_{12}+1)}\sum_{j'_{12}}\sqrt{(2j'_{12}+1)}\left(j'_{12}(j'_{12}+1)-j_{12}(j_{12}+1)\right)\times\nonumber\\
    &\times\sum_{j_{23}}j_{23}(j_{23}+1)(2j_{23}+1)\sj{\frac{1}{2}}{j_{12}}{\underline{j}}{j_1}{j_{23}}{j_2}\sj{\frac{1}{2}}{j'_{12}}{\underline{j}}{j_1}{j_{23}}{j_2}\Psi_{j'_{12}}
    \label{eq:3.2.9}
\end{align}
In fact, this expression can be further simplified using the identity \cite{Brunnemann:2004xi}
\begin{align}
\begin{split}
&\sum_{j_{23}}(2j_{23}+1)j_{23}(j_{23}+1)\sj{j_1}{j_{12}}{j_{2}}{j_3}{j_{23}}{j_{4}}\sj{j_1}{j'_{12}}{j_{2}}{j_3}{j_{23}}{j_{4}}\\
&=\frac{1}{2}(-1)^{j_1+j_2+j_3+j_4+j_{12}+j_{12}'+1}X(j_1,j_4)^{\frac{1}{2}}\sj{j_2}{j_1}{j_{12}}{1}{j_{12}'}{j_1}\sj{j_3}{j_4}{j_{12}}{1}{j_{12}'}{j_4}\\
&\quad+\frac{j_1(j_1+1)+j_4(j_4+1)}{2j_{12}+1}\delta_{j_{12}j_{12}'}
\end{split}
\label{eq:3.2.10}
\end{align}
with $X(j_1,j_4):=2j_1(2j_1+1)(2j_1+2)2j_4(2j_4+1)(2j_4+2)$. Due to the difference appearing in (\ref{eq:3.2.9}), it is immediate that the matrix representation of $\widehat{q}$ is purely off-diagonal, i.e., only entries with $j_{12}\neq j'_{12}$ are nonzero. In this case, (\ref{eq:3.2.10}) becomes
\begin{align}
\begin{split}
&\sum_{j_{23}}(2j_{23}+1)j_{23}(j_{23}+1)\sj{\frac{1}{2}}{j_{12}}{\underline{j}}{j_1}{j_{23}}{j_2}\sj{\frac{1}{2}}{j'_{12}}{\underline{j}}{j_1}{j_{23}}{j_2}\\
&=\frac{1}{2}(-1)^{j_1+j_2+\underline{j}+j_{12}+j_{12}'+\frac{3}{2}}X(\frac{1}{2},j_2)^{\frac{1}{2}}\sj{\underline{j}}{\frac{1}{2}}{j_{12}}{1}{j_{12}'}{\frac{1}{2}}\sj{j_1}{j_2}{j_{12}}{1}{j'_{12}}{j_2}
\end{split}
\label{eq:3.2.11}
\end{align}
with $X(\frac{1}{2},j_2)^{\frac{1}{2}}=2\sqrt{6}\sqrt{j_2(j_2+1)}\sqrt{(2j_2+1)}$. Furthermore, by the properties of the 6-$j$ symbols, in order for (\ref{eq:3.2.11}) to be nonzero $j'_{12}$ has to appear in the decomposition of the tensor product representation $j_{12}\otimes 1\cong(j_{12}-1)\otimes j_{12}\otimes(j_{12}+1)$, that is
$j'_{12}\in\{j_{12}-1,j_{12}+1\}$. Thus, Inserting (\ref{eq:3.2.11}) into (\ref{eq:3.2.9}), we finally obtain
\begin{align}
    \widehat{q}\Psi_{j_{12}}=&-\frac{i\sqrt{6}}{4}(-1)^{j_1+j_2+2j_{12}+\underline{j}+\frac{3}{2}}\sqrt{(2j_{12}+1)}\sqrt{j_2(j_2+1)}\sqrt{(2j_2+1)}\times\label{eq:3.2.10.1}\\
    &\times\sum_{k\in\{\pm1\}}k(2j_{12}+k+1)\sqrt{2j_{12}+2k+1}\sj{\underline{j}}{\frac{1}{2}}{j_{12}}{1}{j_{12}+k}{\frac{1}{2}}\sj{j_1}{j_2}{j_{12}}{1}{j_{12}+k}{j_2}\Psi_{j_{12}+k}\nonumber
\end{align}
This is the most general form for the action of $\widehat{q}$ on a planar vertex with an additional decoupled edge labeled by spin-$\frac{1}{2}$. Applying (\ref{eq:3.2.10.1}) to our situation, i.e., $j_1=j_2=:j$ and $j_{12}=0$, this yields
\begin{align}
    \widehat{q}\Psi_{0}=&\frac{3i\sqrt{2}}{2}(-1)^{2j+1}\sqrt{(2j+1)}\sqrt{j(j+1)}\sj{\frac{1}{2}}{\frac{1}{2}}{1}{0}{1}{\frac{1}{2}}\sj{j}{j}{1}{0}{1}{j}\Psi_{1}\nonumber\\
    =&\frac{3i\sqrt{2}}{2}(-1)^{2j+1}\sqrt{(2j+1)}\sqrt{j(j+1)}\frac{1}{\sqrt{6}}\frac{(-1)^{2j+1}}{\sqrt{2j+1}\sqrt{3}}\Psi_1\nonumber\\
    =&\frac{i}{2}\sqrt{j(j+1)}\Psi_1
    \label{eq:3.2.12}
\end{align}
where we used that
\begin{align}
    \sj{a}{b}{c}{0}{c}{b}=\frac{(-1)^{a+b+c}}{\sqrt{(2b+1)}\sqrt{(2c+1)}}
    \label{eq:3.2.13}
\end{align}
Similarly, for $j_{12}=1$ one obtains
\begin{align}
    \widehat{q}\Psi_{1}=-\frac{i}{2}\sqrt{j(j+1)}\Psi_0
    \label{eq:3.2.14}
\end{align}
Hence, w.r.t. the subspace spanned by the orthonormal basis $\Psi_0$ and $\Psi_1$, the operator $\widehat{q}$ has the following matrix representation
\begin{equation}
    \widehat{q}=\frac{i}{2}\sqrt{j(j+1)}\begin{pmatrix}
0 & 1 \\
-1 & 0  
\end{pmatrix}
\label{eq:3.2.15}
\end{equation}
from which we can directly deduce that
\begin{equation}
    |\widehat{q}|=\sqrt{\widehat{q}^{\dagger}\widehat{q}}=\frac{1}{2}\sqrt{j(j+1)}\mathds{1}
    \label{eq:3.2.16}
\end{equation}
Hence, the Rovelli-Smolin volume operator (\ref{eq:3.2.1.0}) acts via multiplication with the constant factor $C$ on the subspace spanned by $\Psi_0$ and $\Psi_1$. This immediately implies that the action of (\ref{eq:3.2.2}) is given by
\begin{align}
    (\widehat{\mathcal{O}}\Psi_{\gamma})[A]=&-\tensor{(\tau_i)}{^k_l}\tensor{{\pi_{\frac{1}{2}}(h_e[A])}}{^l_m}\sqrt{\widehat{V}}\left(\tensor{{\pi_{\frac{1}{2}}(h_e^{-1}[A])}}{^m_k}\Psi_{\gamma}[A]\right)\nonumber\\
    =&-C^{\frac{1}{4}}\mathrm{tr}(\tau_i h_e[A]h_e^{-1}[A])\Psi_{\gamma}[A]=-C^{\frac{1}{4}}\mathrm{tr}(\tau_i)\Psi_{\gamma}[A]=0
    \label{eq:3.2.17}
\end{align}
that is, $\widehat{\mathcal{O}}$ simply vanishes on these type of edges and therefore is only nonzero in case of spin network vertices proving that (\ref{eq:3.1.14}) is indeed finite also justifying or choice of the factor ordering. This is in fact different to the situation of the standard regularization of the Hamiltonian constraint \cite{Thiemann:1996aw} as, e.g, the Euclidean part contains a term of the form $\mathrm{tr}(h_{\alpha}h_{e}[h_{e}^{-1},\widehat{V}])$ where $\alpha$ is a closed loop. In contrast to (\ref{eq:3.2.2}), the action of this operator will then, in general, be nonzero (in fact, as observed in (\ref{eq:3.2.17}), the triviality of the action of $\widehat{\mathcal{O}}$ mainly arose due to the appearence of the Pauli matrix inside the trace). At first sight, this may look like a contradiction, as the the commutator of the SUSY constraint should reproduce the Hamiltonian constraint. However, as already explained in the beginning of this section, the SUSY constraint is superior to the Hamiltonian constraint, i.e., once the SUSY constraint is quantized, this yields a quantization of the Hamiltonian constraint by computing its commutator. Hence, our proposal of the quantum SUSY constraint provides, at least in principle, another possibility for the quantization of the Hamiltonian constraint.\\
\\
It finally remains to the check that the action of the operator $\widehat{\mathscr{X}}_K(s_J(\Delta))$ in (\ref{eq:3.1.15}) is nontrivial such that $\widehat{\mathscr{X}}_K(s_J(\Delta))-\widehat{\mathscr{X}}_K(x)$ can indeed be viewed as a quantization of a regularized covariant derivative. Therefore, we have to study the action of $\widehat{q}$ on decoupled product states of the form
\begin{equation}
    \ket{(j\,j)0}\otimes\ket{\frac{1}{2},m}\otimes\ket{\frac{1}{2},m'}
    \label{eq:3.2.18}
\end{equation}
where $\ket{(j\,j)0}$ is again the gauge invariant divalent vertex located inside a spin network edge and $\ket{\frac{1}{2},m}$ resp. $\ket{\frac{1}{2},m'}$ are the additional edges with spin-$\frac{1}{2}$ arising from the holonomies $h_{s'_{K}(\Delta)}$ resp. $H_{s_J(\Delta)}$ contained in (\ref{eq:3.1.15}) (see figure \ref{fig:volume_sketch}). Note  that for the Ansatz (\ref{eq:3.2.18}) we have implicitly chosen the chiral represenation of the gamma matrices so that the holonomy $H_e$ is indeed block diagonal according to the decomposition of the restricted Majorana representation into a direct sum of two spin-$\frac{1}{2}$ representations. Hence, this operator does not mix between the two chiral sub representations so that it suffices to restrict to one particular chiral sector. However, note that for the quantization of the Rarita-Schwinger field in section \ref{section 5.1} a representation was chosen in which the gamma matrices are explicitly real. But, since both representations are related via a similarity transformations, one can map from one representation to the other.\\
In order to compute the action of (\ref{eq:3.2.4.2}) on the state (\ref{eq:3.2.18}), we first need to couple the angular momentum $j$ corresponing to the one part of the spin network edge $e$ that is incident at the vertex $v\in\gamma$ under consideration with the spin-$\frac{1}{2}$ quantum number corresponding to the segment $s'_{K}(\Delta)$ that is parallel to that edge. Using again Wigner 6-$j$ symbols, we find
\begin{align}
    \ket{(j\,j)0}\otimes\ket{\frac{1}{2},m}\otimes\ket{\frac{1}{2},m'}=&\ket{(j\,j)0,\frac{1}{2};\frac{1}{2}\,m}\otimes\ket{\frac{1}{2},m'}\nonumber\\
    =&\left(\sum_{j_{23}}(-1)^{2j+1}\sqrt{2j_{23}+1}\sj{j_{23}}{j}{\frac{1}{2}}{0}{\frac{1}{2}}{j}\ket{j,(j\,\frac{1}{2})j_{23},\frac{1}{2}\,m}\right)\otimes\ket{\frac{1}{2},m'}\nonumber\\
    =&\frac{(-1)^{2j+1}}{\sqrt{2}\sqrt{2j+1}}\sum_{j_{23}}(-1)^{j+\frac{1}{2}+j_{23}}\sqrt{2j_{23}+1}\ket{j,(j\,\frac{1}{2})j_{23},\frac{1}{2}\,m}\otimes\ket{\frac{1}{2},m'}\nonumber\\
    =&\sqrt{\frac{j+1}{2j+1}}\ket{j,(j\,\frac{1}{2})j+\frac{1}{2},\frac{1}{2}\,m}\otimes\ket{\frac{1}{2},m'}\nonumber\\
    &-\sqrt{\frac{j}{2j+1}}\ket{j,(j\,\frac{1}{2})j-\frac{1}{2},\frac{1}{2}\,m}\otimes\ket{\frac{1}{2},m'}
    \label{eq:3.2.19}
\end{align}
This can then be coupled with the remaining spin-$\frac{1}{2}$ quantum number using the well-known identities
\begin{align}
    \ket{\frac{1}{2},\frac{1}{2}}\otimes\ket{\frac{1}{2},\frac{1}{2}}=\ket{1,1},\quad\ket{\frac{1}{2},-\frac{1}{2}}\otimes\ket{\frac{1}{2},-\frac{1}{2}}=\ket{1,-1}
    \label{eq:3.2.20}
\end{align}
and 
\begin{align}
    \ket{\frac{1}{2},\pm\frac{1}{2}}\otimes\ket{\frac{1}{2},\mp\frac{1}{2}}=\frac{1}{\sqrt{2}}\ket{1,0}\pm\frac{1}{\sqrt{2}}\ket{0,0}
    \label{eq:3.2.21}
\end{align}
Hence, we have to determine the action of (\ref{eq:3.2.4.2}) on states of the form
\begin{equation}
    \Psi^{\pm}_{\frac{1}{2},\frac{1}{2},\underline{j}}:=\ket{(j\pm\frac{1}{2}\,\,j)\frac{1}{2},\frac{1}{2},\underline{j}\,\underline{m}},\quad\text{with}\quad\underline{j}\in\{0,1\}
    \label{eq:3.2.22}
\end{equation}
The action of $\widehat{q}$ on (\ref{eq:3.2.22}) now follows directly from the general formula (\ref{eq:3.2.10.1}) setting $j_1=j\pm\frac{1}{2}$ and $j_2=j$. Since $j_{12}=\frac{1}{2}$ in this case, only the $k=+1$-term in the sum of (\ref{eq:3.2.10.1}) remains yielding
\begin{align}
    \widehat{q}\Psi^{\pm}_{\frac{1}{2},\frac{1}{2},\underline{j}}=-3i\sqrt{3}(-1)^{2j+\frac{1}{2}\pm\frac{1}{2}+\underline{j}}\sqrt{j(j+1)}\sqrt{(2j+1)}\sj{\underline{j}}{\frac{1}{2}}{\frac{1}{2}}{1}{\frac{3}{2}}{\frac{1}{2}}\sj{j\pm\frac{1}{2}}{j}{\frac{1}{2}}{1}{\frac{3}{2}}{j}\Psi_{\frac{3}{2},\frac{1}{2},\underline{j}}
    \label{eq:3.2.23}
\end{align}
which, according to the first $6j$-symbol appearing in (\ref{eq:3.2.23}), will be nonzero if and only if $\underline{j}\in\frac{3}{2}\otimes\frac{1}{2}\cong1\oplus 2$. Hence, in particular, for $\underline{j}=0$ this immediately implies
\begin{equation}
    \widehat{q}\Psi^{\pm}_{\frac{1}{2},\frac{1}{2},0}=0
    \label{eq:3.2.24}
\end{equation}
On the other hand, for $\underline{j}=1$, one obtains
\begin{align}
    \widehat{q}\Psi^{\pm}_{\frac{1}{2},\frac{1}{2},1}=3i\sqrt{3}(-1)^{2j+\frac{1}{2}\pm\frac{1}{2}}\sqrt{j(j+1)}\sqrt{(2j+1)}\sj{1}{\frac{1}{2}}{\frac{1}{2}}{1}{\frac{3}{2}}{\frac{1}{2}}\sj{j\pm\frac{1}{2}}{j}{\frac{1}{2}}{1}{\frac{3}{2}}{j}\Psi_{\frac{3}{2},\frac{1}{2},1}
    \label{eq:3.2.25}
\end{align}
Using the general formula
\begin{align}
\sj{a}{j}{\frac{1}{2}}{1}{\frac{3}{2}}{j}=&\sj{a}{j}{\frac{3}{2}}{1}{\frac{1}{2}}{j}\nonumber\\
=&\frac{(-1)^{a+\frac{3}{2}+j}}{4\sqrt{3}\sqrt{2j+1}\sqrt{j(j+1)}}\left(\left(a+j+\frac{5}{2}\right)\left(\frac{3}{2}+j-a\right)\left(\frac{3}{2}+a-j\right)\left(a-\frac{1}{2}+j\right)\right)^{\frac{1}{2}}
\label{eq:3.2.26}
\end{align}
it follows for $a=1$ and $j=\frac{1}{2}$
\begin{align}
\sj{1}{\frac{1}{2}}{\frac{1}{2}}{1}{\frac{3}{2}}{\frac{1}{2}}=&-\frac{1}{3}
\label{eq:3.2.27}
\end{align}
For $a=j+\frac{1}{2}$ one finds
\begin{align}
\sj{j+\frac{1}{2}}{j}{\frac{1}{2}}{1}{\frac{3}{2}}{j}=&\frac{(-1)^{2j}}{2\sqrt{3}}\frac{\sqrt{j(2j+3)}}{\sqrt{2j+1}\sqrt{j(j+1)}}
\label{eq:3.2.28}
\end{align}
and finally for $a=j-\frac{1}{2}$
\begin{align}
\sj{j-\frac{1}{2}}{j}{\frac{1}{2}}{1}{\frac{3}{2}}{j}=&\frac{(-1)^{2j+1}}{2\sqrt{3}}\frac{\sqrt{(j+1)(2j-1)}}{\sqrt{2j+1}\sqrt{j(j+1)}}
\label{eq:3.2.29}
\end{align}
Thus, inserting (\ref{eq:3.2.27}), (\ref{eq:3.2.28}) and (\ref{eq:3.2.29}) into (\ref{eq:3.2.25}) this yields
\begin{align}
    \widehat{q}\Psi^{\pm}_{\frac{1}{2},\frac{1}{2},1}=\frac{ia_{\pm}}{2}\Psi_{\frac{3}{2},\frac{1}{2},1}
    \label{eq:3.2.30}
\end{align}
with $a_+:=\sqrt{j(2j+3)}$ and $a_-:=\sqrt{(j+1)(2j-1)}$. Since $\widehat{q}$ is Hermitian, its matrix representation in the subspace spanned by the orthonormal basis $\Psi^{\pm}_{\frac{1}{2},\frac{1}{2},1}$ and $\Psi^{\pm}_{\frac{3}{2},\frac{1}{2},1}$ thus takes the form
\begin{equation}
    \widehat{q}=\frac{ia_{\pm}}{2}\begin{pmatrix}
0 & 1 \\
-1 & 0  
\end{pmatrix}
\label{eq:3.2.31}
\end{equation}
As a consequence, the Rovelli-Smolin volume operator is diagonal on this subspace so that, in particular,
\begin{equation}
    \sqrt{\widehat{V}}=\sqrt[4]{|\widehat{q}|}=\sqrt[4]{\frac{a_{\pm}}{2}}\mathds{1}=:C_{\pm}\mathds{1}
    \label{eq:3.2.32}
\end{equation}
i.e. $\sqrt{\widehat{V}}$ acts a multiplication operator with the constant factor $C_{\pm}$. In order to simply our notation, we define
\begin{align*}
    \ket{(j\,j)0,\frac{1}{2};\frac{1}{2}\,m}\otimes\ket{\frac{1}{2},m'}=\ket{(j\,j)0}\otimes\ket{\frac{1}{2},m}\otimes\ket{\frac{1}{2},m'}:=\left\{\begin{array}{lr}
        \ket{0,\upuparrows}, & \text{for } m=m'=\frac{1}{2}\\
        \ket{0,\updownarrows}, & \text{for } m=\frac{1}{2},\,m'=-\frac{1}{2}\\
        \ket{0,\downuparrows}, & \text{for } m=-\frac{1}{2},\,m'=\frac{1}{2}\\
        \ket{0,\downdownarrows}, & \text{for } m=m'=-\frac{1}{2}
        \end{array}\right.
\end{align*}
Using then (\ref{eq:3.2.19}), (\ref{eq:3.2.20}) and (\ref{eq:3.2.21}) as well as (\ref{eq:3.2.32}), we find
\begin{align}
    \sqrt{\widehat{V}}\ket{0,\upuparrows}=&\sqrt{\frac{j+1}{2j+1}}\sqrt{\widehat{V}}\ket{j,(j\,\frac{1}{2})j+\frac{1}{2};\frac{1}{2},\frac{1}{2}}\otimes\ket{\frac{1}{2},\frac{1}{2}}\nonumber\\
    &-\sqrt{\frac{j}{2j+1}}\sqrt{\widehat{V}}\ket{j,(j\,\frac{1}{2})j-\frac{1}{2};\frac{1}{2},\frac{1}{2}}\otimes\ket{\frac{1}{2},\frac{1}{2}}\nonumber\\
    =&\sqrt{\frac{j+1}{2j+1}}C_+\ket{(j\,j+\frac{1}{2})\frac{1}{2},\frac{1}{2};1,1}-\sqrt{\frac{j}{2j+1}}C_-\ket{(j\,j-\frac{1}{2})\frac{1}{2},\frac{1}{2};1,1}\nonumber\\
    =&\sqrt{\frac{j+1}{2j+1}}C_+\ket{j,(j\,\frac{1}{2})j+\frac{1}{2};\frac{1}{2},\frac{1}{2}}\otimes\ket{\frac{1}{2},\frac{1}{2}}\nonumber\\
    &-\sqrt{\frac{j}{2j+1}}C_-\ket{j,(j\,\frac{1}{2})j-\frac{1}{2};\frac{1}{2},\frac{1}{2}}\otimes\ket{\frac{1}{2},\frac{1}{2}}\nonumber\\
    =:&A\ket{+,\uparrow}\otimes\ket{\uparrow}-B\ket{-,\uparrow}\otimes\ket{\uparrow}
    \label{eq:3.2.33}
\end{align}
and similarly
\begin{align}
    \sqrt{\widehat{V}}\ket{0,\downdownarrows}
    =&\sqrt{\frac{j+1}{2j+1}}C_+\ket{j,(j\,\frac{1}{2})j+\frac{1}{2};\frac{1}{2},-\frac{1}{2}}\otimes\ket{\frac{1}{2},-\frac{1}{2}}\nonumber\\
    &-\sqrt{\frac{j}{2j+1}}C_-\ket{j,(j\,\frac{1}{2})j-\frac{1}{2};\frac{1}{2},-\frac{1}{2}}\otimes\ket{\frac{1}{2},-\frac{1}{2}}\nonumber\\
    =&A\ket{+,\downarrow}\otimes\ket{\downarrow}-B\ket{-,\downarrow}\otimes\ket{\downarrow}
    \label{eq:3.2.34}
\end{align}
Finally, using (\ref{eq:3.2.21}) and the fact that the action of the volume operator on states with vanishing total angular momentum $\underline{j}=0$ is zero (cf. (\ref{eq:3.2.24})), we find for the mixed spin-components
\begin{align}
    \sqrt{\widehat{V}}\ket{0,\downuparrows}=&\sqrt{\frac{j+1}{2j+1}}\sqrt{\widehat{V}}\ket{j,(j\,\frac{1}{2})j+\frac{1}{2};\frac{1}{2},-\frac{1}{2}}\otimes\ket{\frac{1}{2},\frac{1}{2}}\nonumber\\
    &-\sqrt{\frac{j}{2j+1}}\sqrt{\widehat{V}}\ket{j,(j\,\frac{1}{2})j-\frac{1}{2};\frac{1}{2},-\frac{1}{2}}\otimes\ket{\frac{1}{2},\frac{1}{2}}\nonumber\\
    =&\frac{1}{\sqrt{2}}\sqrt{\frac{j+1}{2j+1}}C_+\ket{(j\,j+\frac{1}{2})\frac{1}{2},\frac{1}{2};1,0}-\frac{1}{\sqrt{2}}\sqrt{\frac{j}{2j+1}}C_-\ket{(j\,j-\frac{1}{2})\frac{1}{2},\frac{1}{2};1,0}\nonumber\\
    =&\sqrt{\frac{j+1}{2j+1}}\frac{C_+}{2}\left(\ket{+,\uparrow}\otimes\ket{\downarrow}+\ket{+,\downarrow}\otimes\ket{\uparrow}\right)-\sqrt{\frac{j}{2j+1}}\frac{C_-}{2}\left(\ket{-,\uparrow}\otimes\ket{\downarrow}+\ket{-,\downarrow}\otimes\ket{\uparrow}\right)\nonumber\\
    =&\frac{A}{2}\ket{+,\uparrow}\otimes\ket{\downarrow}+\frac{A}{2}\ket{+,\downarrow}\otimes\ket{\uparrow}-\frac{B}{2}\ket{-,\uparrow}\otimes\ket{\downarrow}-\frac{B}{2}\ket{-,\downarrow}\otimes\ket{\uparrow}
    \label{eq:3.2.35}
\end{align}
and analogously
\begin{align}
    \sqrt{\widehat{V}}\ket{0,\updownarrows}=&\frac{A}{2}\ket{+,\uparrow}\otimes\ket{\downarrow}+\frac{A}{2}\ket{+,\downarrow}\otimes\ket{\uparrow}-\frac{B}{2}\ket{-,\uparrow}\otimes\ket{\downarrow}-\frac{B}{2}\ket{-,\downarrow}\otimes\ket{\uparrow}
    \label{eq:3.2.36}
\end{align}
Recall that we want to the determine the action of (\ref{eq:3.1.15}) on the spin network state $\Psi_{\gamma}$. We therefore have already derived all necessary ingredients. It only remains to evaluate the trace appearing in (\ref{eq:3.1.15}). For this, let us recall some basic facts concerning the action of flux operators appearing e.g. in the volume operator.\\
\\
The flux operator $X_n(S)$ smeared over two-dimensional surfaces $S$ with smearing function $n$ acts on holonomies $h_e[A]$ via \cite{Thiemann:2007pyv}
\begin{equation}
    X_n(S)h_{e}[A]=\frac{i\hbar\kappa\beta}{2}\epsilon(e,S)\frac{n(b(e))}{2}h_{e}[A]
    \label{eq:3.2.37}
\end{equation}
Since $\{E_n(S),h_e[A]^{-1}\}=-h_e[A]^{-1}\{E_n(S),h_e[A]\}h_e[A]^{-1}$, this yields in case of a single edge $e$ ingoing at $S\cap e$
\begin{equation}
    X_n(S)h_{e}[A]^{-1}=-h_{e}[A]^{-1}(X_n(S)h_e[A])h_{e}[A]^{-1}=\frac{i\hbar\kappa\beta}{4}h_{e}[A]^{-1}n(b(e))
    \label{eq:3.2.38}
\end{equation}
such that
\begin{align}
    X_n(S)f_{\gamma}(h_e[A]^{-1})&=\frac{\partial f_{\gamma}}{\partial\tensor{\left(h_{e}[A]^{-1}\right)}{^k_l}}(h_{e}[A]^{-1})\tensor{\left(\frac{i\hbar\kappa\beta}{4}h_{e}[A]^{-1}n(b(e))\right)}{^k_l}\nonumber\\
    &=\frac{i\hbar\kappa\beta}{4}n(b(e))^j\frac{\mathrm{d}}{\mathrm{d}t}\bigg{|}_{t=0}f_{\gamma}(h_e[A]^{-1}e^{t\tau_j})=\frac{\kappa\beta}{4}n(b(e))^j(i\hbar L_j f_{\gamma})(h_e[A]^{-1})
    \label{eq:3.2.39}
\end{align}
with $L_j$ the left-invariant vector field generated by $\tau_j\in\mathfrak{su}(2)$, $j\in1,\ldots,3$, which is related to the pushforward representation of the right regular representation
\begin{equation}
    \rho_R:\,\mathrm{SU}(2)\rightarrow B(L^2(\mathrm{SU}(2))),\quad g\mapsto (\rho_R(g):\,f\mapsto f(\,\cdot\,g))
    \label{eq:3.2.40}
\end{equation}
according to
\begin{equation}
    (L_jf)(h)=\frac{\mathrm{d}}{\mathrm{d}t}\bigg{|}_{t=0}f(he^{t\tau_j})=\frac{\mathrm{d}}{\mathrm{d}t}\bigg{|}_{t=0}\rho_R(e^{t\tau_j})(f)(h)=\rho_{R*}(\tau_j)f(h)
    \label{eq:3.2.41}
\end{equation}
$\forall f\in C^{\infty}(\mathrm{SU}(2))$ and $h\in\mathrm{SU}(2)$ and extended uniquely to a (unbounded) self-adjoint operator on $L^2(\mathrm{SU}(2))$, that is,
\begin{equation}
    J^j:=i\hbar\rho_{R*}(\tau_j)=i\hbar L_j
    \label{eq:3.2.42}
\end{equation}
In our case $f_{\gamma}$ corresponds to the matrix components of the spin-$\frac{1}{2}$ representation of $\mathrm{SU}(2)$, i.e. $f_{\gamma}=\tensor{\pi_{\frac{1}{2}}(h_{e}[A]^{-1})}{^k_l}$ for any $k,l\in\{0,1\}$. As it is very well-known, these matrix components generate a proper invariant subrepresentation of the right regular representation on $L^2(\mathrm{SU}(2))$. In fact, since for general spin-$j$
\begin{align}
    \rho_R(g)\tensor{(\pi_j)}{^k_l}(h)=\tensor{\pi_j(hg)}{^k_l}=\tensor{\pi_j(h)}{^k_m}\tensor{\pi_j(g)}{^m_l}
    \label{eq:3.2.43}
\end{align}
for any $g\in\mathrm{SU}(2)$, it follows that $\rho_R(g)V_k\subseteq V_k$ with $V_k:=\mathrm{span}_{\mathbb{R}}\left\{\tensor{(\pi_j)}{^k_m}|\,m\in\{0,1\}\right\}$ and thus, in particular,
\begin{equation}
    J^jV_k\subseteq V_k,\quad\forall k=0,1
    \label{eq:3.2.44}
\end{equation}
Moreover, for $j=\frac{1}{2}$, it follows
\begin{align}
    J^3\tensor{(\pi_{\frac{1}{2}})}{^k_m}(h)&=i\hbar\frac{\mathrm{d}}{\mathrm{d}t}\bigg{|}_{t=0}\tensor{\pi_{\frac{1}{2}}(he^{t\tau_3})}{^k_m}=i\hbar\tensor{\pi_{\frac{1}{2}}(h)}{^k_n}\frac{\mathrm{d}}{\mathrm{d}t}\bigg{|}_{t=0}\tensor{\left(e^{t\tau_3}\right)}{^n_m}\nonumber\\
    &=\frac{\hbar}{2}\tensor{\pi_{\frac{1}{2}}(h)}{^k_n}\tensor{(\sigma_3)}{^n_m}
    \label{eq:3.2.45}
\end{align}
so that
\begin{equation}
    J^3\tensor{(\pi_{\frac{1}{2}})}{^k_0}=\frac{\hbar}{2}\tensor{(\pi_{\frac{1}{2}})}{^k_0}\quad\text{and}\quad J^3\tensor{(\pi_{\frac{1}{2}})}{^k_1}=-\frac{\hbar}{2}\tensor{(\pi_{\frac{1}{2}})}{^k_1}
    \label{eq:3.2.46}
\end{equation}
To summarize, we have
\begin{equation}
    \pi_{\frac{1}{2}}=\begin{pmatrix}
    \ket{\frac{1}{2},\frac{1}{2}} & \ket{\frac{1}{2},-\frac{1}{2}}\\
    \ket{\frac{1}{2},\frac{1}{2}} & \ket{\frac{1}{2},-\frac{1}{2}}
    \end{pmatrix}
    \label{eq:3.2.47}
\end{equation}
and, due to (\ref{eq:3.2.43}), the rows in (\ref{eq:3.2.47}) define 2-dimensional invariant subspaces w.r.t. the angular momentum operator $J^j$ and thus, in particular, w.r.t. the fluxes $X_n(S)$.\\
\\
With these observations, let us now compute the action of (\ref{eq:3.1.15}) on the spin network state $\Psi_{\gamma}$ which we take as a product state $\Psi_{\gamma}=\psi_b\otimes\psi_f$ with $\psi_b$ a proper spin network function and $\psi_f$ an element of the fermionic part of the Hilbert space. Using (\ref{eq:3.2.33}) as well as (\ref{eq:3.2.47}) and (\ref{eq:3.2.43}), we then immediately find 
\begin{align}
    \sqrt{\widehat{V}}\tensor{{h[A]^{-1}}}{^m_0}\tensor{H}{^0_0}\widehat{\theta}^0_i\Psi_{\gamma}[A]&\equiv\sqrt{\widehat{V}}\ket{0,\upuparrows}\otimes\widehat{\theta}_i^0\psi_f\nonumber\\
    &=\left(A\tensor{{h^{-1}}}{^m_0}\ket{+,\uparrow}-B\tensor{{h^{-1}}}{^m_0}\ket{-,\uparrow}\right)\otimes\widehat{\theta}_i^0\psi_f
    \label{eq:3.2.48}
\end{align}
On the other hand, we have
\begin{align}
    \sqrt{\widehat{V}}\tensor{{h[A]^{-1}}}{^m_1}\tensor{H}{^0_0}\widehat{\theta}_i^0\Psi_{\gamma}[A]\equiv&\sqrt{\widehat{V}}\ket{0,\downuparrows}\otimes\widehat{\theta}_i^0\psi_f\nonumber\\
    =&\left(\frac{A}{2}\tensor{{h^{-1}}}{^m_0}\ket{+,\downarrow}+\frac{A}{2}\tensor{{h^{-1}}}{^m_1}\ket{+,\uparrow}\right.\nonumber\\
    &\left.-\frac{B}{2}\tensor{{h^{-1}}}{^m_0}\ket{-,\downarrow}-\frac{B}{2}\tensor{{h^{-1}}}{^m_1}\ket{-,\uparrow}\right)\otimes\widehat{\theta}_i^0\psi_f
    \label{eq:3.2.49}
\end{align}
as well as
\begin{align}
    \sqrt{\widehat{V}}\tensor{{h[A]^{-1}}}{^m_0}\tensor{H}{^0_1}\widehat{\theta}_i^1\Psi_{\gamma}[A]\equiv&\sqrt{\widehat{V}}\ket{0,\updownarrows}\otimes\widehat{\theta}_i^1\psi_f\nonumber\\
    =&\left(\frac{A}{2}\tensor{{h^{-1}}}{^m_0}\ket{+,\downarrow}+\frac{A}{2}\tensor{{h^{-1}}}{^m_1}\ket{+,\uparrow}\right.\nonumber\\
    &\left.-\frac{B}{2}\tensor{{h^{-1}}}{^m_0}\ket{-,\downarrow}-\frac{B}{2}\tensor{{h^{-1}}}{^m_1}\ket{-,\uparrow}\right)\otimes\widehat{\theta}_i^1\psi_f
    \label{eq:3.2.50}
\end{align}
and finally
\begin{align}
    \sqrt{\widehat{V}}\tensor{{h[A]^{-1}}}{^m_1}\tensor{H}{^0_1}\widehat{\theta}_i^1\Psi_{\gamma}[A]&\equiv\sqrt{\widehat{V}}\ket{0,\downdownarrows}\otimes\widehat{\theta}_i^1\psi_f\nonumber\\
    &=\left(A\tensor{{h^{-1}}}{^m_1}\ket{+,\downarrow}-B\tensor{{h^{-1}}}{^m_1}\ket{-,\downarrow}\right)\otimes\widehat{\theta}_i^1\psi_f
    \label{eq:3.2.51}
\end{align}
If we write for the holonomy
\begin{equation}
    h^{-1}:=\begin{pmatrix}
    \alpha & \beta\\
    \gamma & \delta
    \end{pmatrix}
    \label{eq:3.2.52}
\end{equation}
this yields for the action of (\ref{eq:3.1.15})
\begin{align}
    (\widehat{\mathscr{X}}\Psi_{\gamma})[A]=&\mathrm{tr}(\tau_ih[A]\sqrt{\widehat{V}}h[A]^{-1})\tensor{H}{^0_{\beta}}\widehat{\theta}^{\beta}\psi[A]=\tensor{{\tau_i}}{^k_l}\tensor{h}{^l_m}\sqrt{\widehat{V}}\tensor{{h^{-1}}}{^m_k}\tensor{H}{^0_{\beta}}\psi_b\otimes\widehat{\theta}_i^{\beta}\psi_f\nonumber\\
    =&\mathrm{tr}(\tau_i h\begin{pmatrix}
    A\alpha & \frac{A}{2}\beta\\
    A\gamma & \frac{A}{2}\delta
    \end{pmatrix})\ket{+,\uparrow}\otimes\widehat{\theta}_i^{0}\psi_f+\mathrm{tr}(\tau_i h\begin{pmatrix}
    0 & \frac{A}{2}\alpha\\
    0 & \frac{A}{2}\gamma
    \end{pmatrix})\ket{+,\downarrow}\otimes\widehat{\theta}_i^{0}\psi_f\nonumber\\
    -&\mathrm{tr}(\tau_i h\begin{pmatrix}
    B\alpha & \frac{B}{2}\beta\\
    B\gamma & \frac{B}{2}\delta
    \end{pmatrix})\ket{-,\uparrow}\otimes\widehat{\theta}_i^{0}\psi_f-\mathrm{tr}(\tau_i h\begin{pmatrix}
    0 & \frac{B}{2}\alpha\\
    0 & \frac{B}{2}\gamma
    \end{pmatrix})\ket{-,\downarrow}\otimes\widehat{\theta}_i^{0}\psi_f\nonumber\\
    +&\mathrm{tr}(\tau_i h\begin{pmatrix}
    \frac{A}{2}\beta & 0\\
    \frac{A}{2}\delta & 0
    \end{pmatrix})\ket{+,\uparrow}\otimes\widehat{\theta}_i^{1}\psi_f+\mathrm{tr}(\tau_i h\begin{pmatrix}
    \frac{A}{2}\alpha & A\beta\\
    \frac{A}{2}\gamma & A\delta
    \end{pmatrix})\ket{+,\downarrow}\otimes\widehat{\theta}_i^{1}\psi_f\nonumber\\
    -&\mathrm{tr}(\tau_i h\begin{pmatrix}
    \frac{B}{2}\beta & 0\\
    \frac{B}{2}\delta & 0
    \end{pmatrix})\ket{-,\uparrow}\otimes\widehat{\theta}_i^{1}\psi_f-\mathrm{tr}(\tau_i h\begin{pmatrix}
    \frac{B}{2}\alpha & B\beta\\
    \frac{B}{2}\gamma & B\delta
    \end{pmatrix})\ket{-,\downarrow}\otimes\widehat{\theta}_i^{1}\psi_f
    \label{eq:3.2.53}
\end{align}
This can be further simplified using that
\begin{align}
    \begin{pmatrix}
    A\alpha & \frac{A}{2}\beta\\
    A\gamma & \frac{A}{2}\delta
    \end{pmatrix}=\begin{pmatrix}
    \alpha & \beta\\
    \gamma & \delta
    \end{pmatrix}\begin{pmatrix}
    A & 0\\
    0 & \frac{A}{2}
    \end{pmatrix}=h^{-1}\begin{pmatrix}
    A & 0\\
    0 & \frac{A}{2}
    \end{pmatrix}
    \label{eq:3.2.54}
\end{align}
and
\begin{align}
    \begin{pmatrix}
    0 & \frac{A}{2}\alpha\\
    0 & \frac{A}{2}\gamma
    \end{pmatrix}=\begin{pmatrix}
    \alpha & \beta\\
    \gamma & \delta
    \end{pmatrix}\begin{pmatrix}
    0 & \frac{A}{2}\\
    0 & 0
    \end{pmatrix}=h^{-1}\begin{pmatrix}
    0 & \frac{A}{2}\\
    0 & 0
    \end{pmatrix}
    \label{eq:3.2.55}
\end{align}
as well as
\begin{align}
    \begin{pmatrix}
    \frac{A}{2}\beta & 0\\
    \frac{A}{2}\delta & 0
    \end{pmatrix}=h^{-1}\begin{pmatrix}
    0 & 0\\
    \frac{A}{2} & 0
    \end{pmatrix}
    \label{eq:3.2.56}
\end{align}
such that, for instance,
\begin{align}
    \mathrm{tr}(\tau_i\begin{pmatrix}
    A\alpha & \frac{A}{2}\beta\\
    A\gamma & \frac{A}{2}\delta
    \end{pmatrix})=\left\{\begin{array}{lr}
        0, & \text{for } i=1\\
        0, & \text{for } i=2\\
        \frac{A}{4i}, & \text{for } i=3
        \end{array}\right.
        \label{eq:3.2.57}
\end{align}
and similar for the other traces. Hence, we finally end up with
\begin{align}
    (\widehat{\mathscr{X}}\Psi_{\gamma})[A]=&\frac{A}{4i}\ket{+,\uparrow}\otimes\widehat{\theta}_3^{0}\psi_f+\frac{A}{4i}\ket{+,\downarrow}\otimes(\widehat{\theta}_1^{0}+i\widehat{\theta}_2^{0})\psi_f\nonumber\\
    -&\frac{B}{4i}\ket{-,\uparrow}\otimes\widehat{\theta}_3^{0}\psi_f-\frac{B}{4i}\ket{-,\downarrow}\otimes(\widehat{\theta}_1^{0}+i\widehat{\theta}_2^{0})\psi_f\nonumber\\
    +&\frac{A}{4i}\ket{+,\uparrow}\otimes(\widehat{\theta}_1^{1}-i\widehat{\theta}_2^{1})\psi_f-\frac{A}{4i}\ket{+,\downarrow}\otimes\widehat{\theta}_3^{1}\psi_f\nonumber\\
    -&\frac{B}{4i}\ket{-,\uparrow}\otimes(\widehat{\theta}_1^{1}-i\widehat{\theta}_2^{1})\psi_f+\frac{B}{4i}\ket{-,\downarrow}\otimes\widehat{\theta}_3^{1}\psi_f
    \label{eq:3.2.60}
\end{align}
and thus
\begin{align}
    (\widehat{\mathscr{X}}\Psi_{\gamma})[A]=&\frac{A}{4i}\ket{+,\uparrow}\otimes(\widehat{\theta}_3^{0}+\widehat{\theta}_1^{1}-i\widehat{\theta}_2^{1})\psi_f+\frac{A}{4i}\ket{+,\downarrow}\otimes(\widehat{\theta}_1^{0}+i\widehat{\theta}_2^{0}-\widehat{\theta}_3^{1})\psi_f\nonumber\\
    -&\frac{B}{4i}\ket{-,\uparrow}\otimes(\widehat{\theta}_3^{0}+\widehat{\theta}_1^{1}-i\widehat{\theta}_2^{1})\psi_f-\frac{B}{4i}\ket{-,\downarrow}\otimes(\widehat{\theta}_1^{0}+i\widehat{\theta}_2^{0}-\widehat{\theta}_3^{1})\psi_f
    \label{eq:3.2.61}
\end{align}
where
\begin{equation}
    A=\sqrt{\frac{j+1}{2(2j+1)}}\left(j(2j+3)\right)^{\frac{1}{4}}\quad\text{and}\quad B=\sqrt{\frac{j}{2(2j+1)}}\left((j+1)(2j-1)\right)^{\frac{1}{4}}
    \label{eq:3.2.62}
\end{equation}
As we see, the action of (\ref{eq:3.1.15}) is indeed nontrivial as required and, moreover, creates a new vertex coupled to a to fermion. In particular, we see that (\ref{eq:3.2.61}) is completely independent on the additional segment $s'_K(\Delta)$ which was needed for the regularization. This is indeed a good thing as the choice of such an additional segment would be completely arbitrary and not bases on any fundamental principles justifying the assumption made in (iv) above.
Let us make two final remarks about the quantization chosen here. 
\begin{remark}
We have seen that the properties of the additional edge added at the new vertex, in the definition of (\ref{eq:3.2.61}) are irrelevant for the end result. This property can have some side effects, however. Consider the situation depicted in figure, \ref{fig:tet}, and additionally consider a second tetrahedron spanned by the edge segments $s_1, s_2$ and a third segment $t_3$ along an edge different from $s_1,s_2,s_3$. 
Depending on the orientation of the tangent vectors, the triplet $(s_1,s_2,t_3$ may be either positively or negatively oriented. However, the action of (\ref{eq:3.2.61}) will otherwise be exactly the same in both cases. The relative orientation of the two triplets enters through the $\epsilon$ tensor and gives a relative minus sign in one of the cases. 
If the orientations differ, the two contributions to the operator $\widehat{S}^{(1)}$ cancel after all. This runs counter to the intuition from the classical theory. Thus one might consider defining a variant of this operator in which an additional sign depending on the orientation is introduced in (\ref{eq:3.2.61}).   
\end{remark}
\begin{remark}
Another possibility in quantizing the first term in the SUSY constraint (\ref{eq:3.22}) would be to choose a different variant in which the covariant derivative acts directly on the Rarita-Schwinger field involving of course additional contributions due to the derivation property. That is, one could instead consider an expression of the form
\begin{equation}
    S'^{(1)}[\eta]:=\int_{\Sigma}{\mathrm{d}^3x\,\bar{\eta}\frac{i}{\sqrt[4]{q}}\tensor{\epsilon}{^{abc}}e_a^i\gamma_i\gamma_{*}e^j_c D_b^{(A)}\phi_j}
    \label{remark:1}
\end{equation}
Following the standard procedure, it is then immediate to see that a regularization of (\ref{remark:1}) is given by (see also Part II below)
\begin{align}
    S'^{(1)}_{\delta}[\eta]=&\sum_{\Delta_i\in T(\gamma,\delta)}\bar{\eta}(x_i)\frac{1}{\kappa^2\sqrt{V(x_i,\delta)}}\epsilon^{IJK}\mathrm{tr}(\tau_l h_{s_I(\Delta_i)}[A]\{h_{s_I(\Delta_i)}[A]^{-1},V(x_i,\delta)\})\gamma_l\gamma_{*}\times\label{remark:2}\\
   &\times\mathrm{tr}(\tau_j h_{s_J(\Delta_i)}[A]\{h_{s_J(\Delta_i)}[A]^{-1},V(x_i,\delta)\})\left(H(A)(s_K(\Delta_i)(\delta))\theta^{\delta}_j(s_K(\Delta_i)(\delta))-\theta^{\delta}_j(x_i)\right)\nonumber
\end{align}
For the quantization of (\ref{remark:2}), one can now use either the Ashtekar-Lewandowski or Rovelli-Smolin volume operator. In both cases, based on our observations above, the resulting operator will be finite, i.e., only terms involving spin-network vertices contribute. Moreover, one obtains a nontrivial action for the difference operator resulting from the last term in (\ref{remark:2}) which is consistent for a regularization of a covariant derivative. 
\end{remark}

\subsubsection{Part II}
Next, let us turn to the quantization of the second term in the SUSY constraint (\ref{eq:3.22}) depending on the covariant derivative of the frame field
\begin{equation}
    S^{(2)}[\eta]:=\int_{\Sigma}{\mathrm{d}^3x\,\bar{\eta}\frac{1}{\sqrt[4]{q}}\tensor{\epsilon}{^{abc}}e_c^l\frac{\mathds{1}+i\beta\gamma_{*}}{2\beta}\gamma_k(D_a^{(A)}e_b^k)\phi_l}
    \label{PartII:1}
\end{equation}
We want to quantize this expression by similar means as in the foregoing section. As we have recently observed, the implementation of the regularized covariant derivative in (\ref{eq:3.1.7}) yields an operator that creates new vertices. However, according to (\ref{eq:3.2.61}),  this new vertex is strongly coupled with the fermion. Hence, in order for this additional contribution to be nonzero, the presence of a fermion is crucial. One may therefore expect that the quantization of the covariant derivative in (\ref{PartII:1}) by similar means will lead to vanishing contributions of the operator acting apart from the spin network vertex which seems to be inconsistent for the regularization of a covariant derivative. For this reason, let us introduce the total covariant derivative $\nabla_a^{(A)}$ which acts on both internal indices and spinor indices. With respect to this covariant derivative, we can write  
\begin{equation}
    (D_a^{(A)}e_b^k)\phi_l=\nabla_a^{(A)}(e_b^k\phi_l)-e_b^k \nabla^{(A)}_a\phi_l
    \label{PartII:2}
\end{equation}
In the quantum theory, this then has the advantage of creating vertices coupled to fermion fields and therefore, based on our previous observations, yields nontrivial contributions. Inserting (\ref{PartII:2}) into (\ref{PartII:1}) yields two terms, one which is very similar to expression (\ref{eq:3.1.1}) replacing the covariant derivative acting on purely spinor indices with the new total covariant derivative which also acts on internal indices. The implementation of this quantity can be performed in analogy to the foregoing section. For this reason, we will not explain the steps in detail. Concerning the second contribution, one arrives at an expression of the form 
\begin{equation}
    S'^{(2)}[\eta]:=\int_{\Sigma}{\mathrm{d}^3x\,\bar{\eta}\frac{1}{\sqrt[4]{q}}\tensor{\epsilon}{^{abc}}e_c^l\frac{\mathds{1}+i\beta\gamma_{*}}{2\beta}\gamma_k e_b^k \nabla^{(A)}_a\phi_l}
    \label{PartII:3}
\end{equation}
We make the following Ansatz for a regularization of (\ref{PartII:3})
\begin{align}
    S'^{(2)}_{\delta}[\eta]=&\sum_{\Delta_i\in T(\gamma,\delta)}\bar{\eta}(x_i)\frac{-1}{6\kappa^2\sqrt{V(x_i,\delta)}}\epsilon^{IJK}\mathrm{tr}(\tau_l h_{s_K(\Delta_i)}[A]\{h_{s_K(\Delta_i)}[A]^{-1},V(x,\delta)\})\frac{\mathds{1}+i\beta\gamma_{*}}{2\beta}\gamma_k\times\label{PartII:3.1}\\
    &\times\mathrm{tr}(\tau_k h_{s_J(\Delta_i)}[A]\{h_{s_J(\Delta_i)}[A]^{-1},V(x,\delta)\})\left(\mathscr{Y}^{\delta}_l(s_{I}(\Delta))-\mathscr{Y}^{\delta}_l(x_i)\right)\nonumber
\end{align}
where
\begin{align}
   \mathscr{Y}^{\delta}_l(s_{I}(\Delta)):=\underline{H}(A)(s_I(\Delta_i)(\delta))\theta^{\delta}_l(s_I(\Delta_i)(\delta))
\end{align}
and
\begin{align}
   \mathscr{Y}_l(x_i):=\theta^{\delta}_l(x_i)
\end{align}
Here, $\underline{H}(A)$ denotes the holonomy induced by the total covariant derivative $\nabla^{(A)}$ which, in the limit of small $\delta$, satisfies
\begin{equation}
    \underline{H}(A)(s_I(\Delta_i)(\delta))\Psi_l(s_I(\Delta_i)(\delta))-\Psi_l(x_i)=\delta \dot{s}_I(\Delta_i)^a\nabla_a^{(A)}\Psi_l(x_i)
\end{equation}
where $\Psi$ is some spinor-valued co-vector field (w.r.t. internal indices) defined on $\Sigma$. Following the same steps as in the previous section, it can be shown immediately that for $\delta\rightarrow 0$, one obtains
\begin{align}
   \lim_{\delta\rightarrow 0}S'^{(2)}_{\delta}[\eta]=\lim_{\delta\rightarrow 0}\sum_{\Delta_i\in T(\gamma,\delta)}\frac{-1}{32\kappa^2\sqrt[4]{q(x_i)}}&\{A_c^l,V(x_i,\delta)\}\frac{\mathds{1}+i\beta\gamma_{*}}{2\beta}\gamma_k\{A_b^k,V(x_i,\delta)\}\nabla_a^{(A)}\phi_l(x_i)\times\nonumber\\
   &\times\epsilon^{IJK}\delta^3\dot{s}^a_I(\Delta_i)\dot{s}^b_J(\Delta_i)\dot{s}^c_K(\Delta_i)
\end{align}
so that, together with (\ref{eq:3.1.12.1}) and (\ref{eq:3.1.2}), this yields a Riemann sum so that in the limit $\delta\rightarrow 0$ one finally arrives at 
\begin{equation}
    \lim_{\delta\rightarrow 0} S'^{(2)}_{\delta}[\eta]=S'^{(2)}[\eta]
    \label{PartII:4}
\end{equation}
For the quantization of the regularized expression (\ref{PartII:3.1}), we use 
\begin{equation}
    \frac{1}{\sqrt{V(x,\delta)}}\{A_c^l,V(x,\delta)\}\{A_b^k,V(x,\delta)\}=\frac{16}{9}\{A_c^l,V(x,\delta)^{\frac{3}{4}}\}\{A_b^k,V(x,\delta)^{\frac{3}{4}}\}
\end{equation}
and replace Poisson brackets by the respective commutator yielding
\begin{align}
    \widehat{S}'^{(2)}_{\delta}[\eta]:=&\frac{8}{27\hbar^2\kappa^2}\sum_{v\in \gamma}\frac{8}{E(v)}\bar{\eta}(v)\epsilon^{IJK}\frac{\mathds{1}+i\beta\gamma_{*}}{2\beta}\gamma_k\mathrm{tr}(\tau_k h_{s_J(\Delta)}[A][h_{s_J(\Delta)}[A]^{-1},\widehat{V}_v^{\frac{3}{4}}])\times\label{eq:PartII:5}\\
    &\times\left(\widehat{\mathscr{Y}}_l(s_{I}(\Delta))-\widehat{\mathscr{Y}}_l(x_i)\right)\mathrm{tr}(\tau_l h_{s_K(\Delta)}[A][h_{s_K(\Delta)}[A]^{-1},\widehat{V}_v^{\frac{3}{4}}])\nonumber
\end{align}
with
\begin{equation}
    \widehat{\mathscr{Y}}_l(s_{I}(\Delta)):=\underline{H}(A)(s_I(\Delta)(\delta))\widehat{\theta}_l(s_I(\Delta)(\delta))\quad\text{and}\quad\widehat{\mathscr{Y}}_l(v):=\widehat{\theta}_l(v)
\end{equation}
In the infinite sum of (\ref{eq:PartII:5}) we were again allowed to restrict to the sum over the vertices of the underlying spin network graph since one of the trace-terms was ordered to the right. By (\ref{eq:3.2.17}), this yields vanishing contributions in case the Rovelli-Smolin volume operator does not act on a spin network vertex.   
\subsubsection{Part III}
Finally, we need to quantize the last three terms in the SUSY constraint (\ref{eq:3.22}). These terms are all of very similar structure and, in particular, do not contain any covariant derivatives. Hence, it suffices for instance to consider the last one which we write in the form
\begin{equation}
    S^{(3)}[\eta]:=\int_{\Sigma}{\mathrm{d}^3x\,\bar{\eta}\frac{\kappa}{4\sqrt[4]{q}}\gamma_0\phi^{i}\left(\tensor{\epsilon}{^{jkl}}\bar{\phi}_{j}\gamma_0\frac{\mathds{1}+i\beta\gamma_{*}}{2\beta}\gamma_k\gamma_{i}\phi_{l}\right)}
    \label{PartIII:1}
\end{equation}
For its regularization, we make the Ansatz
\begin{equation}
    S^{(3)}_{\delta}[\eta]:=\sum_{\Delta_i\in T(\gamma,\delta)}\bar{\eta}(x_i)\frac{\kappa}{4\sqrt{V(x_i,\delta)}}\gamma_0\theta^{\delta}_i(x_i)\left(\tensor{\epsilon}{^{jkl}}\bar{\theta}^{\delta}_{j}(x_i)\gamma_0\frac{\mathds{1}+i\beta\gamma_{*}}{2\beta}\gamma_k\gamma^{i}\theta^{\delta}_{l}(x_i)\right)
    \label{PartIII:2}
\end{equation}
Due to (\ref{eq:4.9.1}), we have
\begin{align}
    &\tensor{\epsilon}{^{jkl}}\bar{\theta}^{\delta}_{j}(x_i)\gamma_0\frac{\mathds{1}+i\beta\gamma_{*}}{2\beta}\gamma_k\gamma^{i}\theta^{\delta}_{l}(x_i)\\
    =&\int{\mathrm{d}^3y\int{\mathrm{d}^3z\,\frac{\chi_{\delta}(x_i-y)\chi_{\delta}(x_i-z)}{\delta^3/6}\tensor{\epsilon}{^{jkl}}\bar{\phi}_{j}(y)\gamma_0\frac{\mathds{1}+i\beta\gamma_{*}}{2\beta}\gamma_k\gamma^{i}\phi_{l}(z)}}\nonumber
\end{align}
and on the other hand
\begin{equation}
    \frac{\kappa}{4\sqrt{V(x_i,\delta)}}\gamma_0\theta^{\delta}_i(x_i)=\int{\mathrm{d}^3x\,\frac{\chi_{\delta}(x_i-x)}{\delta^3/6}\frac{\kappa}{4\sqrt[4]{q(x_i)}}\gamma_0\phi_i(x)}
\end{equation}
In the limit $\delta\rightarrow 0$, it follows  $\chi_{\delta}(x_i-x)/\frac{\delta^3}{6}\rightarrow\delta(x_i-x)$ and moreover $\chi_{\delta}(x_i-z)/\frac{\delta^3}{6}\rightarrow\delta(x_i-z)$ and $\chi_{\delta}(x_i-y)$ can be replaced by the Kronecker delta $\delta_{x_i,y}$. Therefore, in this limit, (\ref{PartIII:2}) finally becomes
\begin{align}
    \lim_{\delta\rightarrow 0} S^{(3)}_{\delta}[\eta]&=\lim_{\delta\rightarrow 0}\sum_{\Delta_i\in T(\gamma,\delta)}\bar{\eta}(x_i)\frac{\kappa}{4\sqrt[4]{q(x_i)}}\gamma_0\phi_i(x_i)\left(\tensor{\epsilon}{^{jkl}}\bar{\phi}_{j}\gamma_0\frac{\mathds{1}+i\beta\gamma_{*}}{2\beta}\gamma_k\gamma^{i}\phi_{l}(x_i)\right)\mathrm{vol}(\Delta_i)\nonumber\\
    &=\int_{\Sigma}{\mathrm{d}^3x\,\bar{\eta}\frac{\kappa}{4\sqrt[4]{q}}\gamma_0\phi^{i}\left(\tensor{\epsilon}{^{jkl}}\bar{\phi}_{j}\gamma_0\frac{\mathds{1}+i\beta\gamma_{*}}{2\beta}\gamma_k\gamma_{i}\phi_{l}\right)}=S^{(3)}[\eta]
\end{align}
and therefore (\ref{PartIII:2}) indeed provides an appropriate regularization of (\ref{PartIII:1}). Its implementation in the quantum theory is now straightforward yielding
\begin{equation}
    \widehat{S}^{(3)}[\eta]:=\frac{\kappa}{4}\sum_{v\in V(\gamma)}\frac{8}{E(v)}\bar{\eta}(v)\sqrt{\widehat{V_v^{-1}}}\gamma_0\widehat{\theta}_i(v)\left(i\tensor{\epsilon}{^{jkl}}\widehat{\theta}^T_{j}(v)\frac{\mathds{1}+i\beta\gamma_{*}}{2\beta}\gamma_k\gamma^{i}\widehat{\theta}_{l}(v)\right)
\end{equation}
where, in the real representation of the gamma matrices, we used that the charge conjugation matrix is given by $C=i\gamma^0$. There exist various possibilities for the implementation of the inverse volume operator $\widehat{V^{-1}}$ such that this operator is well-defined and non-singular. For instance, one can re-express it in terms of a product of Poisson brackets of the form (\ref{eq:3.1.2}). However, for sake of simplicity, let us choose a quantization as proposed in \cite{Assanioussi:2015gka}. There, one quantizes the inverse volume via 
\begin{equation}
    \widehat{V^{-1}}:=\mathrm{lim}_{t \rightarrow 0}(\widehat{V}^2+t^2l_{p}^6)^{-1}\widehat{V}
    \label{PartIII:3}
\end{equation}
with $l_p$ the Planck-length. This operator then simply vanishes while acting on vertices with zero volume and therefore provides a suitable regularization.  
\subsection{Solutions of the quantum SUSY constraint}\label{section 5.3}
\begin{figure}
    \centering
    \includegraphics[height=5cm]{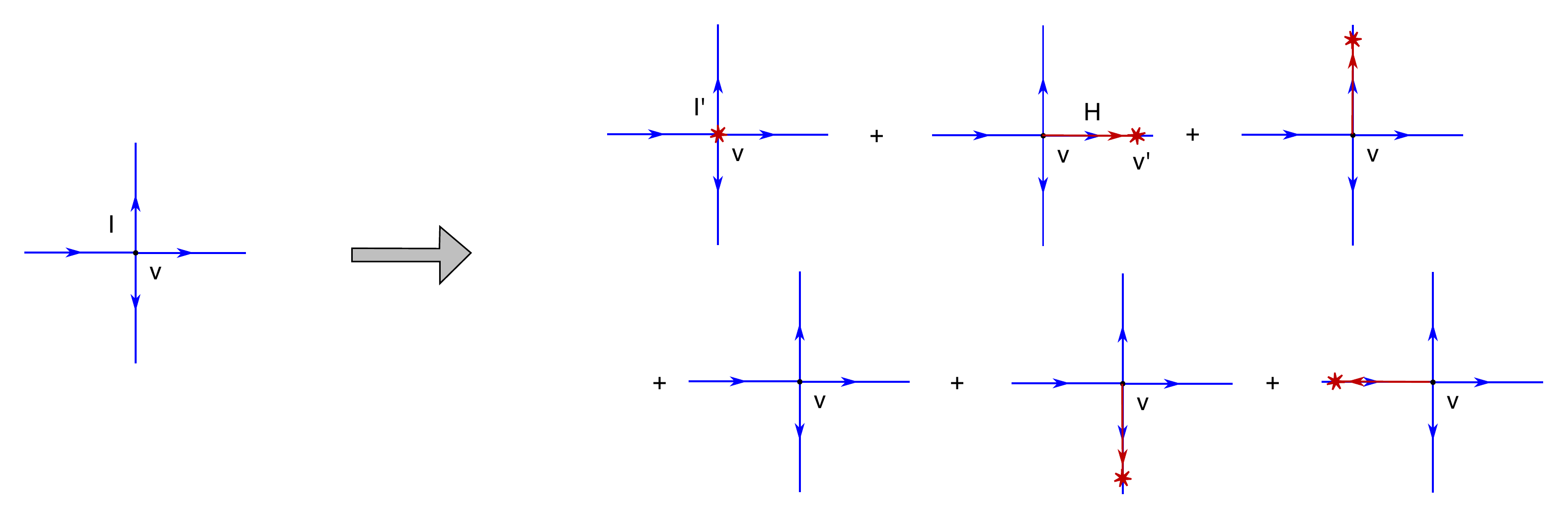}
    \caption{Schematic depiction of the action of the supersymmetry constraint on a 4-valent vertex $v$ with intertwiner $I$. Each sub diagram on the right side of the arrow represents a type of term that is appearing in the result. The star symbol represents a vertex containing a fermion, and $H$ is the new holonomy that connects a new vertex $v'$ to the intertwiner at $v$.}
    \label{fig:action}
\end{figure}

In this last section, we would like to sketch possible solutions of the quantum SUSY constraint. Going over to the sector of diffeomorphism-invariant states, we are thus looking for vectors $\Psi_{\mathrm{phys}}\in\mathcal{D}_{\mathrm{diff}}^*$ such that\footnote{Actually, working on the dual requires an antilinear representation of the constraint algebra involving rather the adjoint $\widehat{S}[\eta]^{\dagger}$ of the SUSY constraint. However, since the classical theory, the SUSY constraint is a real function and thus we could equally quantize the complex conjugate $\bar{S}[\eta]$ which then yields $\widehat{S}[\eta]^{\dagger}$.}
\begin{equation}
    (\Psi_{\mathrm{phys}}\ket{\widehat{S}[\eta]\psi}=0,\quad\forall\psi\in\mathcal{H}_{\mathrm{kin}}=\mathcal{H}_{\mathrm{grav}}\otimes\mathcal{H}_f,\,\eta\in\Gamma(E_{\mathbb{R}})
    \label{eq:5.3.1}
\end{equation}
where $\Gamma(E_{\mathbb{R}})$ denotes the space of smooth sections of the spinor bundle $E_{\mathbb{R}}:=P_{\mathrm{spin}}\times_{\kappa_{\mathbb{R}}}\Delta_{\mathbb{R}}$ induced by the Majorana representation on $\Delta_{\mathbb{R}}$.\\
Considering the first part (\ref{eq:3.1.14}) of the quantum SUSY constraint studied in section \ref{section 5.1}, this operator creates new vertices coupled to a fermion. A qualitative description of the action is depicted in figure \ref{fig:action}. 
Each diagram on the right side of the arrow represents a type of term that is appearing in the result. Fermions are created both, at the original vertex $v$ and at new vertices $v'$ that lie on the edges incident at $v$. The creation of fermions is a generic feature of the quantum SUSY constraint because the conjugate spinor plays the role of smearing function. In case of an ordinary Dirac fermion, this would mean that even if, on the right hand side of (\ref{eq:5.3.1}), one initially started with a state $\psi$ in the pure gravitational sector of the Hilbert state, i.e., an ordinary spin network state without any fermions, this operator would always create states with nontrivial fermionic degrees of freedom. But then, any pure gravitational state $\Psi_{\mathrm{phys}}$ would be a solution of (\ref{eq:5.3.1}) as the inner product between a pure bosonic and fermionic state is always zero by (\ref{eq:4.11}) (or (\ref{eq:4.11.1})). \\
This is however no longer true in case of Majorana fermions. In fact, as seen in section \ref{section 5.1} (see formula (\ref{eq:4.11.2})), due to the Majorana condition, it follows that the quantization of the Rarita-Schwinger field necessarily involves both multiplication operators and derivations, i.e., creation and annihilation operators. Therefore, the quantum SUSY constraint generically both creates and annihilates fermionic degrees of freedom. As a consequence, pure gravitational states cannot be a solution of (\ref{eq:5.3.1}). \\
For purely fermionic states, the situation is less clear, we can not immediately rule out their existence. In any case, such solutions of (\ref{eq:5.3.1}) would seem to be unphysical. 


\section{Conclusions}
In this paper, we have studied the canonical theory of $\mathcal{N}=1$ Poincaré and anti-de Sitter supergravity in four spacetime dimensions based on the Holst action of supergravity as first introduced by Tsuda in \cite{Tsuda:1999bg}. In this framework we considered half-densitized fermion fields as suggested by Thiemann \cite{Thiemann:1997rq} in order to simplify the reality conditions for the Rarita-Schwinger field. We then derived a compact expression for the classical SUSY constraint which then served as a starting point for its implementation in the quantum theory. Therefore, following \cite{Bodendorfer:2011pb}, we quantitzed the Rarita-Schwinger field by appropriately extending the classical phase space.\\
With these prerequisites, we turned to the quantitzation of the supersymmetry constraint which so far has not been considered in the literature. This is important because the quantum SUSY constraint in canonical supergravity theories is as important as the quantum Hamiltonian constraint in quantum gravity theories without local supersymmetry. We therefore first need derive a suitable regularization of the continuum expression guided by the principle that the resulting operator should be as compact as possible. For the regularization, special care was required. This is mainly due to the fact that, although the SUSY constraint looks similar to the Dirac Hamiltonian constraint, there is a crucial difference: The conjugate spinor plays the role of a Lagrangian multiplier. As a result, one cannot simply follow the standard regularization procedure as the density weight of the smearing function should be kept fixed in order not the change the density weight of the SUSY constraint as a whole. Changing its density weight may change the resulting quantum algebra and thus its strong relationship to the Hamiltonian constraint as indicated in the classical regime in \cite{Sawaguchi:2001wi} in case of real Ashtekar-Barbero variables. We succeeded in finding an appropriate regularization such the density weight is maintained. \\
The resulting operator consists of various different terms one of which arose from the quantization of the covariant derivative on the fermion field considered in section \ref{section 5.2}. Requiring consistency with the classical theory forced us to choose the Rovelli-Smolin variant of the volume operator for the quantization of the triads via Thiemann's trick. Based on an explicit calculation, it was shown, choosing an appropriate factor ordering, that the resulting operator was still finite as the sum over the tetrahedra in the triangulation again restricts on the sum over vertices of the underlying graph. Different implementations in the quantum theory involving the Ashtekar-Lewandowski volume operator have also been discussed. For this, a different but equivalent form of the classical SUSY constraint needs to be considered. \\
As it turns out, the operator thus obtained has an interesting feature as it creates new vertices strongly coupled to fermions. This was shown via explicit computation evaluating its action on generic spin-network states. Due to this fact, it is expected that solutions of the quantum SUSY constraint need to contain both, gravity and matter degrees of freedom, as required for supersymmetry. We have see that the reality condition enforced on Majorana spinors is important. Whether these solutions indeed contain the same number of bosons and fermions, however, is still unclear so far and remains a question for the future. Also it would be highly desirable to study the commutator algebra of the quantum SUSY constraint. In, particular, it would be very interesting in which sense the commutator on diffeomorphism and gauge invariant states is related to the Hamiltonian constraint. As a first step, one could try to evaluate the commutator of the terms involving the quantization of the covariant derivative and investigate whether this can be related to the quantization of the curvature of the connection along loops. 
\enlargethispage{10\baselineskip}
\section*{Acknowledgments}
We thank Norbert Bodendorfer, Klaus Liegener and Thomas Thiemann for helpful comments and discussions on quantizing supergravity with methods from loop quantum gravity, and Jorge Pullin and Lee Smolin for communications at an early stage of this work. KE thanks the German Academic Scholarship Foundation (Studienstiftung des Deutschen
Volkes) for financial support.

\end{document}